\documentclass[reprint,amsmath,amssymb,aps,floatfix]{revtex4-2}

\usepackage{hyperref}
\hypersetup{
    colorlinks=true,
    linkcolor=blue,
    citecolor=blue,
    urlcolor=cyan
}

\usepackage{graphicx}
\usepackage{mathdots}
\usepackage[normalem]{ulem}
\usepackage[usenames,svgnames]{xcolor}
\usepackage[toc,page]{appendix}
\usepackage{bm}
\usepackage{multirow}

\newcommand{\Eq}[1]{Eq.~(\ref{#1})}
\newcommand{\Fig}[1]{Fig.~\ref{#1}}

\newcommand{\Sec}[1]{Sec.~\ref{#1}}
\newcommand{\cRef}[1]{Ref.~\cc{#1}}
\newcommand{\cRefs}[1]{Refs.~\cc{#1}}
\newcommand{\cc}[1]{~\cite{#1}}

\newcommand{\App}[1]{Appendix~\ref{#1}}

\newcommand{\norm}[1]{{\| #1 \|}}  
\newcommand{\ket}[1]{{ |{#1} \rangle }}  

\newcommand{\av}[1]{{ \langle {#1} \rangle }}

\newcommand{\ketbra}[2]{{ |{#1} \rangle\!\,\langle {#2} | }}

\newcommand{\EqDef}{\stackrel{\mathrm{def}}{=}}
\newcommand{\Id}{\mathbb{I}}
\newcommand{\BBC}{\mathbb{C}}

\DeclareMathOperator*{\Tr}{Tr}
\DeclareMathOperator{\poly}{poly}
\newcommand{\ux}{\underline{x}}
\newcommand{\uy}{\underline{y}}

\newcommand{\mcU}{\mathcal{U}}
\newcommand{\mcI}{\mathcal{I}}
\newcommand{\mcE}{\mathcal{E}}
\newcommand{\mcF}{\mathcal{F}}
\newcommand{\mcN}{\mathcal{N}}
\newcommand{\mcL}{\mathcal{L}}
\newcommand{\rhoss}{{\rho_\infty}}
\newcommand{\rhossI}{{\rho_{\infty}^{I}}}
\newcommand{\rhossII}{{\rho_{\infty}^{II}}}
\newcommand{\vtheta}{{\bm{\theta}}}
\DeclareMathOperator{\CX}{CX}
\DeclareMathOperator{\CR}{CR}
\DeclareMathOperator{\RESCX}{RESCX}
\DeclareMathOperator{\RESU}{RESU}
\DeclareMathOperator{\supp}{supp}
\DeclareMathOperator{\RESET}{RESET}
\DeclareMathOperator{\CNOT}{CNOT}
\DeclareMathOperator{\nRESET}{R\widetilde{ESE}T}
\newcommand{\Balpha}{{\bm{\alpha}}}
\newcommand{\Bbeta}{{\bm{\beta}}}
\newcommand{\bP}{{\bar{P}}}

\begin{document}

\title{Learning a quantum channel from its steady-state}
  
\author{Yigal Ilin$^{1}$}
\email{yigal.ilin@gmail.com}
\email{yigali@campus.technion.ac.il}

\author{Itai Arad$^{2,3,4}$}

\affiliation{$^1$Andrew and Erna Viterbi Department of Electrical
and Computer Engineering, Technion – Israel Institute of Technology,
3200003, Haifa, Israel}

\affiliation{$^2$Physics Department, Technion – Israel Institute of
Technology, 3200003, Haifa, Israel}

\affiliation{$^3$The Helen Diller Quantum Center, Technion – Israel
Institute of Technology, 3200003, Haifa, Israel}

\affiliation{$^4$Centre for Quantum Technologies, National
University of Singapore, 117543 Singapore, Singapore}

\begin{abstract} 
  We present a scalable method for learning local quantum channels
  using local expectation values measured on a \emph{single} state
  --- their steady state. Our method is inspired by the algorithms
  for learning local Hamiltonians from their ground states. For it
  to succeed, the steady state must be non-trivial, and therefore
  the channel needs to be non-unital. Such non-unital channels are
  readily implementable on present day quantum computers using
  mid-circuit measurements or RESET gates. We demonstrate that the
  full structure of such channels is encoded in their steady states,
  and can be learned efficiently using only the expectation values
  of local observables on these states. We emphasize two immediate
  applications to illustrate our approach: (i) Using engineered
  dissipative dynamics, we offer a straightforward way to assess the
  accuracy of a given noise model in a regime where all qubits are
  actively utilized for a significant duration. (ii) Given a
  parameterized noise model for the entire system, our method can
  learn its underlying parameters. We demonstrate both applications
  using numerical simulations and experimental trials conducted on
  an IBMQ machine.
\end{abstract}

{\hypersetup{hidelinks}
\maketitle
}

\section{Introduction}\label{introduction}

In recent years, with the advances in quantum information and
quantum computation, there have been a growing interest in methods
for learning many-body quantum systems. These methods include, for
example, learning a many-body Hamiltonian based on its
dynamics\cc{ref:Burgarth2009, ref:DiFranco2009, ref:Shabani2011a,
ref:DaSilva2011, ref:Zhang2014, ref:DeClercq2016, ref:Wang2018,
ref:Zubida2021, ref:Kokail2021, ref:Yu2022, ref:Wilde2022}, its
steady states\cc{ref:Kappen2020, ref:Anshu2021, ref:Haah2021,
ref:Lifshitz2021}, or using a trusted quantum
simulator\cc{ref:Granade2012, ref:Wiebe2014, ref:Wiebe2014a,
ref:Wiebe2015, ref:Wang2017}.

Among these methods, a promising line of research tries to learn the
underlying generators of dynamics using local measurements on a
steady state of the system\cc{ref:Sone2017, ref:Greiter2018,
ref:Chertkov2018, ref:Qi2019, ref:Bairey2019, ref:Bairey2020,
ref:Evans2019}. It can be shown that if the underlying Hamiltonians
or Lindbladians are sufficiently generic, they can be uniquely
determined by the expectation values of local observables in these
steady states. Specifically, \cRef{ref:Bairey2019} showed that for a
system governed by a local Hamiltonian, there exists a set of linear
constraints between expectation values of local observables in a
steady state of the system, from which an unknown local Hamiltonian
can be uniquely determined. Following that work,
\cRef{ref:Bairey2020} showed that a similar set of constraints can
be obtained for the case of open quantum systems, allowing
reconstruction of unknown local Lindbladians by performing local
measurements on a steady state. Together, \cRefs{ref:Bairey2019,
ref:Bairey2020} imply that by measuring local expectation values
from a single state within a given patch of the system, we can infer
the underlying Hamiltonian or Lindbladian terms in that region
\footnote{Strictly speaking, this is true up to an overall factor.}.

In some respect, at least in the case of local Hamiltonians, the
method relies on the non-commutativity of the local Hamiltonian
terms, and is therefore a `purely quantum' method, with no classical
equivalent. It gives a surprisingly simple and efficient method for
learning the generators of the local dynamics, and can also be used
to test how well a particular model of the Hamiltonian or
Lindbladian agrees with the empirical measurements. All this is done
\emph{without having to classically calculate local expectation
values}, which is crucial for the scalablity of the approach.

The simplicity and efficiency of the methods developed in
\cRefs{ref:Bairey2019, ref:Bairey2020} raises the question of
whether and how they can be applied to discrete quantum channels,
particularly those describing (or engineered on) quantum computers.
Specifically, we consider an engineered dissipative quantum channel,
realized on a quantum computer. It consists of a series of known
quantum gates that we apply, together with an unavoidable (and
unknown) noise channel. Can we learn the unknown noise parameters
solely by measuring local expectation values on its steady state?
Can we do all that without classically simulating the
system?

To generalize the aforementioned techniques to the domain of quantum
computers, several points have to be
addressed. At the abstract level, the evolution of local
Hamiltonians and Lindbladians is \emph{continuous}, while quantum
computers evolve \emph{discretely} through gate-based operations.
Furthermore, the introduction of noise disrupts locality, as the noise channel can act
simultaniously on all qubits.

In this work we propose a method that generalizes the approach of \cRefs{ref:Bairey2019,
ref:Bairey2020} to discrete channels that describe gate-based
quantum computers. Our main idea is to engineer a dissipative
channel that can be implemented on a quantum computer using
non-unitary operations, such as the $\RESET$ gates, available on
current day machines. These non-unitary operations lead to a
dynamic that is described by a \emph{non-unital} channel, i.e.,
channels $\mcE(\cdot)$ for which $\mcE(\Id)\neq\Id$. By definition,
the steady state of these channels is different from the maximally mixed
state, and may therefore contain fingerprints of the underlying
dynamic. By iteratively applying this channel, we converge to a steady state,
on which we measure local expectation values. These expectation
values satisfy an extensive set of local linear constraints among
themselves, which can be efficiently verified. Efficiency in this context denotes that our method does not require classical estimation of these expectation values. Furthermore, all classical post- and pre-processing procedures exhibit polynomial complexity with respect to the system size. Notably, the number of local linear constraints scales linearly with the system size. Therefore, these
constraints can be used to \emph{efficiently} learn the underlying
channel by measuring only a \emph{single} state --- the steady state
of the channel. 

Like the methods in \cRefs{ref:Qi2019, ref:Bairey2019,
ref:Bairey2020, ref:Evans2019}, our method is highly scalable, and
can be used on systems with a large number of qubits. To demonstrate
it, we introduce two generic ways to construct a dissipative quantum
channel on a quantum computer. The first way constructs a stochastic
evolution that can be described as a local Kraus map (at least in
the absence of noise). It can be implemented on a quantum computer
by randomly applying a gate from a set of gates with a prescribed
set of probabilities. Consequently, the steady state is probed by
running many different random circuits, which realize different
trajectories of the dynamics. The second way applies a deterministic
circuit that is based on a composed 2-local gate, which is both
non-unitary and entangling. Implementing this map on a quantum
computer requires only one circuit, which makes it easier to execute
on currently available hardware such as IBMQ. To test our method we
used numerical simulations of both methods on $5$--$11$ qubits, and
in addition applied the deterministic method to an IBMQ machine
using $5$ qubits.

The structure of this paper is as follows. In
\Sec{sec:preliminaries} we provide brief definitions and basic facts
on quantum channels and gates, and the notation we use throughout
the paper. Then in \Sec{sec:Learning-SS} we introduce the
theoretical framework of learning a local quantum channel from its
steady state, and explain how such framework can be used, for
example, to learn and validate the noisy dynamics of a quantum
computer. In \Sec{sec:optim-and-noise-models} we focus on the
learning and validation task by defining the noise model we will be
using to characterize and validate (benchmark) the noisy dynamics of
IBMQ machines. The optimization method used in the variational
learning part is also described. In \Sec{sec:numerical-simulations}
we describe the results of our numerical simulations, performed on
the two types of engineered dissipative dynamics. In
\Sec{sec:ibmq-implementation} we present the results of applying our
method on an actual IBMQ machine using 5 qubits, which include
benchmarking different noise models, as well as using our method to
learn its noise. Finally, in \Sec{sec:discussion} we present our
conclusions and discuss possible future research directions.

\section{Preliminaries}
\label{sec:preliminaries}

Throughout this work we consider a quantum computer (QC) made of 
$n$ qubits with Hilbert space $(\BBC^2)^{\otimes n}$. Quantum states
will be generally denoted by Greek letters $\rho$, $\sigma$, etc.
The expectation value of an observable $A$ with respect to the
quantum state $\rho$ will be generally denoted by $\av{A}_\rho =
\Tr(\rho A)$.  

Unitary gates will be denoted by $X,Y,Z, \CX, R_x,$ etc. We use
$\CX(i\to j)$ to denote the CNOT (Controlled-NOT) gate between the
control qubit $i$ and the target qubit $j$. We use
$R_{\bm{x}}(\theta)$ to denote a Bloch sphere rotation of angle
$\theta$ around the axis $\bm{x}$.  Quantum channels, i.e.,
completely positive, trace-preserving (CPTP) maps will usually be
denoted by calligraphic letters such as $\mcE(\cdot), \mcN(\cdot)$,
etc. Given a unitary gate $U$, we will denote its corresponding
channel by $\mcU$, i.e., $\mcU(\rho) \EqDef U\rho U^\dagger$. For
any quantum channel $\mcE$, its adjoint is the unique superoperator
$\mcE^*$ that satisfies $( \mcE^*(A),B) = (A, \mcE(B))$ for any
operators $A,B$, where $(A,B)\EqDef \Tr(A^\dagger B)$ is the
Hilbert-Schmidt inner product between two operators. A map $\mcE$ is
trace preserving if and only if $\mcE^*(\Id) = \Id$. A channel
$\mcE(\cdot)$ is called \emph{unital} if it also holds that
$\mcE(\Id) =\Id$.  In other words, the maximally mixed state is a
fixed point of the channel.

An important example of a non-unital channel that can be applied on
a QC is formed by the $\RESET$ gate, which performs an active
mid-circuit reset of a qubit to the $\ket{0}$ state. Ideally, it
corresponds to the channel
\begin{align*}
  \mcE_\text{RESET}(\rho) = \ketbra{0}{1}\rho\ketbra{1}{0}
   + \ketbra{0}{0}\rho\ketbra{0}{0} .
\end{align*}
It can also be realized by measuring the qubit in the standard
basis, and applying $X$ gate whenever the result is $\ket{1}$.

\begin{figure*}
  \includegraphics[width=\linewidth]{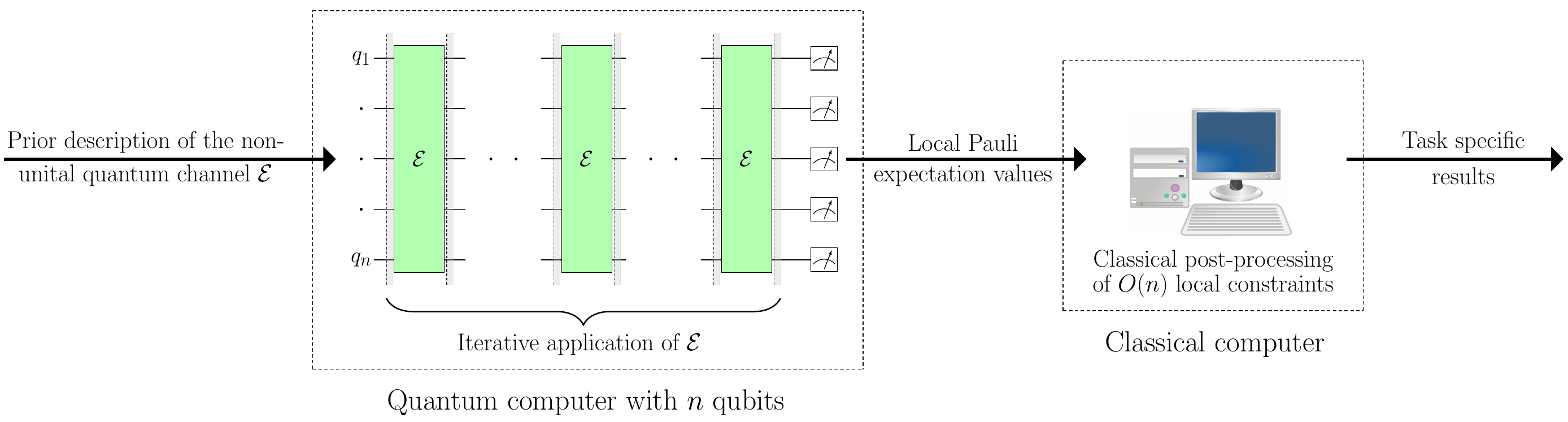}
    \caption{Block diagram summarizing our method. Given a
    description of the non-unital channel $\mcE(\cdot)$ (e.g. its
    quantum gates, activation probabilities, etc), we iteratively
    apply it on a QC to reach an approximate steady state. The local
    Pauli expectation values are sampled from the steady state, and
    used as input to the classical post-processing routine. They
    satisfy a set of local constraints with coefficients that depend
    on the noise parameters of the underlying quantum device. The
    local constraints can be verified efficiently and their number
    grows linearly with the system size. The classical
    post-processing routine is readily tailored to learn specific
    channel properties such as rotation angles, coherence times,
    gate activation probabilities and so on.}
    \label{fig:block-diagram}
\end{figure*}

\section{Learning quantum channels from their steady state}
\label{sec:Learning-SS}

In this section we introduce our method for learning quantum
channels using local measurements on their steady state. We shall
consider channels that can be realized on a quantum computer. A
summary of the main steps of our method is given in
\Fig{fig:block-diagram}. Two examples to keep in mind are (i) the
application of quantum gate drawn from a set of gates according to a
fixed probability distribution, and (ii) a low-depth quantum
circuit. In both examples, we assume that the overall channel is
non-unital because of the noise and/or because of actively using the
RESET gate. Unlike unital channels, whose fixed point is typically
the maximally mixed state, the fixed points of non-unital channels
are typically non-trivial in the sense that two different channels
from a relevant set of channels will have different fixed points. We
shall further assume that $\mcE(\cdot)$ has a \emph{unique} fixed
point $\mcE(\rhoss)=\rhoss$, which we call the \emph{steady state}.
We shall also assume that $\rhoss$ can be approximately reached by
the QC using few applications of $\mcE$, starting from an initial
state $\rho_0 = \ketbra{0}{0}^{\otimes n}$. As we shall see,
our protocol uses expectation values of local Pauli operators in the
steady state. Therefore, a practical way to verify that we are close
enough to this state would be to demand that these local
expectation values converge, i.e., that the difference between
the expectation values we get in $T$ steps and in $T+1$ steps is of
the order of the statistical error. For the channels that we
consider in this work, convergence was always attained after few
tens of steps. For brevity, we will denote
expectation values with respect to $\rhoss$ by $\av{\cdot}_\infty$
instead of $\av{\cdot}_\rhoss$.

Given a channel $\mcE$, assume that the system is in its steady
state $\rhoss$. Then the expectation value of any observable $A$
must satisfy
\begin{align}
\label{eq:general_steady_state}
  \av{A}_\infty &= \Tr(A\rhoss) = \Tr(A\mcE(\rhoss)) \\
  &= \Tr(\mcE^*(A)\rhoss)
    =\av{\mcE^*(A)}_\infty.
    \nonumber
\end{align}
The above equation is the basis of our method. For the channels that
will be discussed below, when $A$ is a local observable, the
observable $\mcE^*(A)$ is also local, or at least can be estimated
using local measurements. In such case, we can use the equality
between the LHS and RHS to constrain, or even learn $\mcE$, without
having to calculate the expectation values in $\rho_\infty$ ---
which might be hard classically. Moreover, as we are measuring the
steady state of the system, our results are independent from the
initial state of the system, and therefore insensitive to state
preparation errors.

Below, we describe two methods to construct such $\mcE$, which we
call the stochastic method and the deterministic method.  In
\Sec{sec:stochastic-ideal} and \Sec{sec:stochastic-noisy} we
describe the first method, and in \Sec{sec:deterministic} describe
the second.

\subsection{A stochastic map: the strictly local case}
\label{sec:stochastic-ideal}

Let $\{\mcE_k\}$ be a set of local channels implementable by a QC.
We can think of every $\mcE_k$ as being the application of a 
quantum gate or a small circuit acting on a few contiguous qubits.
We assume that the $\mcE_k$ channels are strictly local in the sense
that they act trivially outside the support of the small circuit
that defines them. This assumption breaks down in realistic quantum
hardware, where additional noise channels act \emph{simulatiniously
on all} qubits. In the next subsection we will show how to extend
our method to many of these cases (where the noise is local and
Markovian), but for the sake of keeping the presentation clear, we
start by assuming that the $\mcE_k$ channels are strictly local,
acting on at most two neighboring qubits. In addition, we allow some
of the $\mcE_k$ to be non-unital by containing RESET gates. 

Together with $\{\mcE_k\}$, we will use a probability distribution
$\{p_k\}$ and define our quantum channel by
\begin{align}
\label{eq:stochastic-E}
  \mcE \EqDef \sum_k p_k \mcE_k .
\end{align}
Physically, $\mcE$ can be implemented on a QC in a stochastic way:
we choose a $k$ with probability $p_k$ and apply $\mcE_k$. 

For example, the three qubits non-unital quantum channel
\begin{align*}
  \mcE(\rho) &= 0.2\cdot X^{1/2}_1 \rho {X^{{1/2}}_1}^{\dagger}
    + 0.2\cdot \CX(1\to 2) \rho \CX(1\to 2)\\
  &+ 0.4\cdot \CX(2\to 3) \rho \CX(2\to 3) \\
  &+ 0.2\cdot \RESET_{3}(\rho)
\end{align*}
can be implemented on a QC by randomly applying one of the gates
$\{X^{1/2}_1, \CX(1\to 2), \CX(2\to~3), \RESET_3\}$
according to the probabilities $\{0.2, 0.2, 0.4, 0.2\}$. A
simple numerical check shows that the steady state of the this
channel is a non-product state which can be approximated to within a
trace distance of $10^{-5}$ using few tens of steps. 

Plugging \Eq{eq:stochastic-E} in \Eq{eq:general_steady_state}, we
see that for any local observable $A$, 
\begin{align}
\label{eq:stoch-vanilla}
  \av{A}_\infty = \sum_k p_k\av{\mcE_k^*(A)}_\infty .
\end{align}
Importantly, if $\mcE_k$ is a $2$-local channel, then $\mcE_k^*$
also acts non-trivially only on the two-qubits on which $\mcE_k$ is
defined. Consequently, if $A$ is a $t$-local observable, then
$\mcE_k^*(A)$ is at most a $(t+2)$-local observable. In fact, it is
at most $(t+1)$-local for the following reason. The only case where
$\mcE_k^*(A)$ might be $(t+2)$-local is when the supports of
$\mcE_k$ and $A$ are disjoint. But in such case, $\mcE_k^*(A) =
A\mcE_k^*(\Id) = A$, which is $t$-local. 

Using this observation, we may write \Eq{eq:stoch-vanilla} as
\begin{align*}
  \av{A}_\infty = \sum_{k\in \supp(A)} p_k\av{\mcE_k^*(A)}_\infty
   + \sum_{k\notin \supp(A)} p_k\av{A}_\infty,
\end{align*}
where we use the notation $k\in \supp(A)$ to denote enumeration over
all $k$ indices for which $\mcE_k$ acts non trivially on $A$.
Writing the LHS as $\sum_{k\in \supp(A)} p_k\av{A}_\infty +
\sum_{k\notin \supp(A)} p_k\av{A}_\infty$ and re-organizing the
equation, we obtain the following equation, which holds for every
local observable $A$:
\begin{align}
\label{eq:stoch-noiseless-local}
  \av{\mcE^*(A)}_\infty-\av{A}_\infty 
   = \!\!\!\!\!\!\sum_{k\in\supp(A)} p_k\big(\av{\mcE_k^*(A)}_\infty 
    - \av{A}_\infty\big) = 0 .
\end{align}

The above equation can be used to \emph{validate} a channel $\mcE$. Given a channel $\mcE$, we
shall now let $S$ denote the set of local observables. Throughout this
paper, we shall always take $S$ to be the set of  geometrically $t$-local
Pauli operators for some fixed $t$ --- but at large, our method is
independent of that choice. Using $S$, we define a
\emph{cost function} $\Phi$ as a normalized sum of differences between the LHS and
RHS of \Eq{eq:stoch-noiseless-local} for all local observables $A\in S$:
\begin{align}
\label{eq:cost-Kraus-ideal-validation}
  \Phi &\EqDef  \frac{1}{C_S}\sum_{A\in S} \Big( \av{\mcE^*(A)}_\infty
      -  \av{A}_\infty\Big)^2 , \\
  C_S &\EqDef \sum_{A\in S} \av{A}^2_\infty .      
\label{def:C_S}
\end{align}

Assuming the observables in $S$ are orthonormal with respect
to the Hilbert-Schmidt norm, the expression $\sum_{A\in
S}\av{A}^2_\rho = \sum_{A\in S} \big(\Tr(\rho A)\big)^2$ is equal to
the square of the Frobenius norm of the projection of $\rho$ to the
operator subspace spanned by observables in $S$. Denoting this norm
by $\norm{\cdot}_S$, we can compactly write the cost function $\Phi$
by
\begin{align}
  \Phi \EqDef \norm{\mcE(\rho)-\rho}^2_S \Big/ \norm{\rho}^2_S .
\end{align}
With this definition, the cost function $\Phi$ for a state $\rho$ for which $\mcE(\rho)$ is very different from $\rho$, (i.e., $\rho$ is far from being a steady state of $\mcE$), will be $\approx1$. Admittedly, it would have been more natural to define the cost
function using the trace norm instead of the Frobenious norm, i.e.,
by $\norm{\mcE(\rho)-\rho}_1$. By definition, $\norm{\rho}_1=1$ for
every quantum state and the trace distance serves as an upperbound
to the difference between expectation values of arbitrary normalized
observables over two states. However, since we want to be able to
estimate the cost function using \emph{local} expectation values,
and we want to optimize over it, the projected Frobenious norm
$\norm{\cdot}_S$ was chosen. Unlike the trace norm, the Frobenious
norm of a quantum state is only upperbounded by 1, and can be as
small as $2^{-n/2}$ for $n$ qubits state. For that reason, we use
the $1/\norm{\rho}_S^2$ normalization \footnote{While it
would also be more natural to normalize by
$\frac{1}{2}(\norm{\rho}_S^2 + \norm{\mcE(\rho)}_S^2)$, such
normalization substantially complicates the optimization routine
because $\norm{\mcE(\rho)}_S^2$ depends on the optimization
parameters. In turn, this normalization does not introduce a
significant change in the accuracy of the optimization as close to
the convergence point $\mcE(\rho)$ has to be close to $\rho$.}.

When the set of observables $S$ consists of local Pauli operators $\{P_i\}$ (which
is case in this study), we can explicitly write
$\mcE^*(P_i)$ in terms of other Pauli operators (which are
often less local and therefore not
necessarily in $S$):
\begin{align}
  \mcE^*(P_i) - P_i \EqDef \sum_j C_{ij}P_j .
\end{align}
Here $C_{ij}$ is a matrix that fully describes the channel. 
With this matrix, the cost function is given by
\begin{align}
\label{eq:cost-Pauli-Kraus-validation}
  \Phi = \frac{1}{C_S} 
    \sum_i \Big( \sum_j C_{ij}\av{P_j}_\infty\Big)^2 .
\end{align}

When $\mcE$ is given by \Eq{eq:stochastic-E} with $2$-local channels
$\mcE_k$, and $S$ is a set of $t$-local Pauli operators, then 
$\mcE^*(P_i)$ is at most a $(t+1)$-local operator, represented in the $(t+1)$-local Pauli basis. Therefore, by
measuring the $(t+1)$-local Pauli expectation values in the steady
state, we can calculate $\Phi$ and see how well the model fits the
actual quantum hardware.

This procedure can also be made local by considering local cost
functions $\Phi_q$ for qubit $q$.  Denoting by $S_q$ the set of observables $A$ for which
$\mcE^*(A)$ acts non-trivially
on $q$, the local cost function is given by
\begin{align}
\begin{split}
\label{eq:loss-Kraus-validation}
  \Phi_q &\EqDef \frac{1}{C_{S_q}}
  \sum_{A\in S_q}
    \Big(\av{\mcE^*(A)}_\infty -
      \av{A}_\infty\Big)^2 \\
  &=\frac{1}{C_{S_q}}
  \sum_{A\in S_q}
    \Big(\sum_{k\in\supp(A)} p_k \big(\av{\mcE_{k}^*(A)}_\infty -
      \av{A}_\infty\big)\Big)^2 ,  
\end{split}
\end{align}
where the $C_{S_q}\EqDef \sum_{A\in S_q} \av{A}^2 =
\norm{\rho}_{S_q}^2$.

The local validation scheme allows us to identify the regions in the
system where the model performs well or badly in a setup in which
possibly all the device qubits are being actively used. Because it
only requires $(t+1)$-local expectation values, it is highly
scalable and has a small computational cost.

The above idea can be pushed further; we can
actually use \Eq{eq:stoch-noiseless-local} to \emph{learn} the
channel $\mcE$. Suppose we have some
parameterization of $\mcE$, denoted by
$\mcE_{\vtheta}$, where $\vtheta$ is a
vector of parameters. We may define the $\vtheta$-dependent cost
function
\begin{align}
\label{eq:cost-Kraus-ideal}
  \Phi(\vtheta) \EqDef \frac{1}{C_S}\sum_{A\in S} 
    \big(\av{\mcE_{\vtheta}^*(A)}_\infty
    - \av{A}_\infty\big)^2 ,
\end{align}
and find the particular $\vtheta$ that best describes the system
by \emph{minimizing} $\Phi(\vtheta)$. When $S$ consists of
local Pauli operators $\{P_i\}$, we can repeat the derivation that
led to \Eq{eq:cost-Pauli-Kraus-validation} 
and write the cost function as
\begin{align}
\label{eq:cost-Pauli-Kraus}
  \Phi(\vtheta) = \frac{1}{C_S} 
    \sum_i \Big( \sum_j C_{ij}(\vtheta)\av{P_j}_\infty\Big)^2 .
\end{align}

Obtaining expectation values $\av{P_j}_\infty$ in the steady state allows us to learn the best
$\vtheta$ that describes the quantum hardware. Note that once
expectation values are obtained, the minimization of $\Phi(\vtheta)$ can
be done without the $1/C_S$ normalization, as it is
$\vtheta$-independent. In such case, the cost function is a simple
quadratic expression in $\av{P_j}_\infty$ with $\vtheta$-dependent
coefficients.

The $\vtheta$ parameterization can be full, in which case we would
need about $16\times 16$ parameters to describe all possible
$2$-qubit channels $\mcE_k$, or it could contain a much smaller
number of parameters if we use prior assumptions on the structure of
$\mcE_k$. For example, we may
model the channel that corresponds to an $R_{\bm{x}}(\alpha)$
rotation using one parameter --- $\alpha$, in order to see if the
hardware gate under-rotates or over-rotates with respect to the $\hat{x}$
axis.

For a system with $n$ qubits, there are exactly $N_t =
\sum_{k=1}^{t} 3^{k}\left(n-k+1\right)$ $t$-local Pauli operators
with contiguous support, not counting the trivial identity operator.
By selecting $t$-local Pauli operators as the observables $A$ in
\Eq{eq:cost-Kraus-ideal}, we effectively minimize over $N_t$
constraints, and therefore it is desirable that the number of free
parameters will be smaller than $N_t$\cc{ref:Bairey2020}. Although this requirement may seem restrictive, it still permits the consideration of highly detailed models. For instance, with a system size of $n=6$ and observables' locality $t=2$, the number of constraints amounts to $N_t=63$, enabling the incorporation of models with up to 10 parameters per qubit. If a greater number of parameters per qubit is required, one can simply increase the locality $t$ to generate additional constraints for the same system size.

\subsection{A stochastic map: the non strictly-local case}
\label{sec:stochastic-noisy}

In the previous section we derived our main equation,
\Eq{eq:stoch-noiseless-local}, under the assumption that $\mcE_k$
are \emph{strictly local} channels, acting non trivially on at most
$O(1)$ qubits. This assumption is no longer
true in realistic noise models, in which also idle qubits, far away
from the support of $\mcE_k$, experience noise such as dephasing and
amplitude damping. In this section we show how our method can be
generalized to account also for (some of) these cases. Specifically,
we shall assume a local Markovian noise model without cross talks,
with the following properties:
\begin{enumerate}
  \item Each qubit $j$ has its own idle noise channel $\mcN_j$, 
    which acts independently of what happens to the other qubits.
    Thus the noise on the set of idle qubits $I_\text{idle}$ factors 
    into a tensor product of single-qubit noise channels
    $\bigotimes_{j\in I_\text{idle}} \mcN_j$.

  \item When a unitary gate or a RESET gate acts on one or two
    qubits, the actual channel applied by the QC is a noisy version
    of these gates that acts only on the qubits in the support of
    the ideal gates (i.e., no spillover or cross-talks to other
    qubits), \emph{together} with the idle noise channel on the rest
    of the qubits.
\end{enumerate}
Together, these two assumptions imply that the noisy channel can be written by
\begin{align}
  \mcE = \sum_k p_k \tilde{\mcE}_k
    \otimes\Big(\bigotimes_{j\notin \supp(k)} \mcN_j\Big),
\end{align}
where $\tilde{\mcE}_k$ is the noisy version of $\mcE_k$ and
$\supp(k)$ denotes the qubits in the support of $\mcE_k$ (which,
under our assumptions, also define the support of $\tilde{\mcE}_k$).

To proceed, we define superoperators $\mcF_k$
\begin{align}
\label{def:Fk}
  \mcF_k \EqDef \Big(\bigotimes_{j\in\supp(k)}\mcN^{-1}_j\Big)
    \tilde{\mcE}_k .
\end{align}
We will use $\mcF_k$ as a mathematical tool to simplify the
equations we derive below. Note that $\mcF_k$ is not necessarily a
channel, since $\mcN^{-1}_j$ by itself is not necessarily a channel.
Nevertheless, it is a \emph{local} superoperator which satisfies
$\mcF_k(A) = A$ when $A$ is outside its support. The point of using
$\mcF_k$ is that it allows us to write the global action of the
$k$th local channel without the $j\notin \supp(k)$ condition: 
\begin{align*}
  \tilde{\mcE}_k\otimes\Big(\bigotimes_{j\notin \supp(k)} \mcN_j\Big)
   = \Big(\bigotimes_j\mcN_j\Big) \cdot \mcF_k .
\end{align*}
Therefore,
\begin{align}
\label{eq:pre-noisyE}
  \mcE^* = \big(\sum_k p_k\mcF_k^*\big)\cdot \bigotimes_j\mcN^*_j
\end{align}
By the same argument used in the previous section, we note that the
local noise channels act on $A$ non-trivially only if they are in
its support, and therefore we can define a ``noisy version'' of $A$
by
\begin{align}
\label{def:noisy-A}
  \Big(\bigotimes_j \mcN^*_j\Big) (A) 
    = \Big(\bigotimes_{j\in\supp(A)} 
    \mcN^*_j\Big)(A) \EqDef \tilde{A} .
\end{align}
Plugging this back into \Eq{eq:pre-noisyE}, the formula for $\mcE^*(A)$ becomes
\begin{align}
  \mcE^*(A) = \sum_k p_k \mcF_k^*(\tilde{A})
\end{align}
Following the treatment in \Eq{eq:stoch-noiseless-local}, we can
further simplify the formula, by noting that $\mcF_k^*(\tilde{A}) =
\tilde{A}$ whenever $\mcF_k$ acts outside the support of $A$, and
therefore
\begin{align}
  \mcE^*(A) = 
  \sum_{k\in \supp(A)} p_k 
    \big(\mcF_k^*(\tilde{A})-\tilde{A}\big) + \tilde{A} .
\end{align}
Note that in the above equation only $k$ for which $\mcF_k$
intersects the support of $A$ are summed over.
Consequently, \Eq{eq:general_steady_state} becomes
\begin{align}
\label{eq:local-Kraus-cond-noisy}
  \av{A-\tilde{A}}_\infty =
    \sum_{k\in \supp(A)} p_k\big(\av{\mcF_k^*(\tilde{A})}_\infty 
      - \av{\tilde{A}}_\infty\big) .
\end{align}

As in the previous section, we can use the above set of constraints
as a validation tool for particular noise models by defining local
cost functions, as done in \Eq{eq:loss-Kraus-validation}. We can
also use it to learn the best noise model from a family of noise
models parameterized by $\vtheta$, by using the global cost function
\begin{widetext}
\begin{align}
\label{eq:cost-Kraus-noisy}
  \Phi(\vtheta) \EqDef \frac{1}{C_S}\sum_{A\in S} \Big(
    \sum_{k\in \supp(A)}
      p_k\big(\av{\mcF_{k,\vtheta}^*(\tilde{A}_\vtheta)}_\infty 
        - \av{\tilde{A}_\vtheta}_\infty\big)
    - \av{A-\tilde{A}_\vtheta}_\infty \Big)^2 ,
\end{align}  
\end{widetext}
where $\tilde{A}_\vtheta$ is defined in \Eq{def:noisy-A}.
As before, to evaluate $\Phi(\vtheta)$ from local measurements, we
expand the operators $\tilde{A}_\vtheta$,
$\mcF_{k,\vtheta}^*(\tilde{A}_\vtheta)$ in terms of $(t+1)$-local
Pauli strings with $\vtheta$-dependent coefficients and write the
cost function $\Phi(\theta)$ as a quadratic expression of these
expectations (see \Eq{eq:cost-Pauli-Kraus}).

\subsection{Deterministic map}
\label{sec:deterministic}
 
Realizing the method presented in \Sec{sec:stochastic-noisy} on a QC requires the execution of a large
number of \emph{different} quantum circuits, or trajectories, in
order to properly sample the steady state. This makes it challenging
to implement on currently available devices, which are often limited
by the number of distinct circuits that can be executed in a single
experimental batch.

\begin{figure}
  \includegraphics[width=\linewidth]{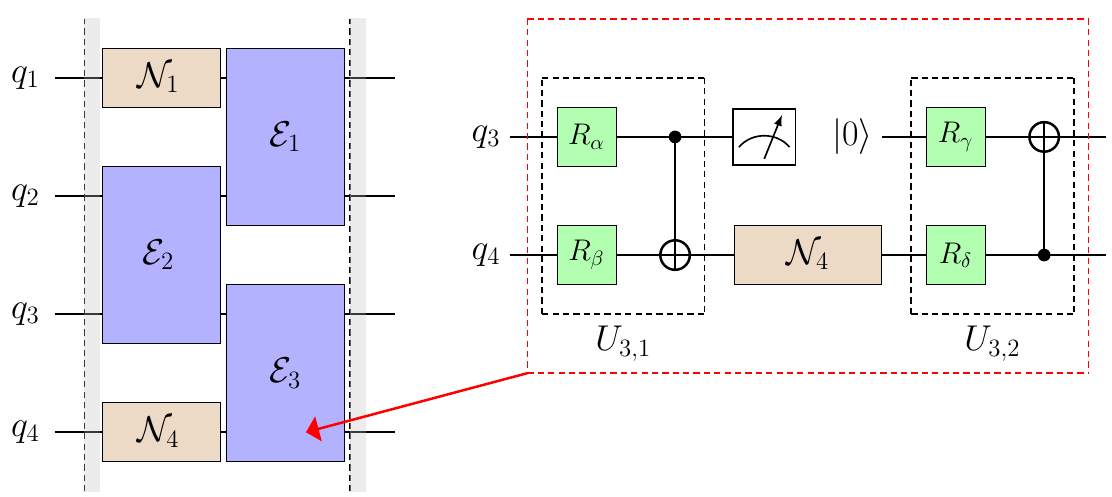}
    \caption{Example of the deterministic channel implementation on
    a QC. Left: Odd channel $\mcE_I$ applied on four qubits, the
    even layer $\mcE_\text{even}$ is followed by the odd layer
    $\mcE_\text{odd}$. During the action of the $\mcE_2$ on qubits
    $2,3$ in the even layer, we assume that the idle noise channels
    $\mcN_{1},\mcN_{4}$ are acting on the qubits $1,4$. Right: Inner
    structure of the 2-local channel $\mcE_{3}$. The 2-local
    non-unital channel $\mcE_3$ consists of applying a two-qubit
    unitary gate $U_{3,1}$ on qubits $3, 4$, followed by a RESET
    gate on qubit $3$ along with idle noise channel $\mcN_{4}$ on
    qubit $4$, and ends with applying a two-qubit unitary gate
    $U_{3,2}$. The two-qubit unitary gate $U_{3,1}$ has two known
    single-qubit rotation gates $R_{\bm{\alpha}}, R_{\bm{\beta}}$
    about axes $\hat{\alpha}$ and $\hat{\beta}$ together with a CX
    gate. The exact details of our realization of the noisy
    non-unital channels $\{\mcE_k\}$ are given in
    \Sec{sec:deterministic-resu-details}.} 
    \label{fig:determinist-example}
\end{figure}

To overcome this limitation, we propose an alternative
way to construct a non-unital channel, which relies
on a \emph{deterministic} map that can be implemented on a QC using
a \emph{single} circuit. For simplicity, we shall describe our
construction in the 1D case with open boundary conditions, but
generalization to other geometries and higher dimensions is
straightforward. In such case, our map $\mcE$ is defined by a
product of local channels $\{\mcE_k\}$, where $\mcE_k$ acts
non-trivially on the neighboring qubits $k,k+1$. The $\mcE_k$
channels are organized in a two layers brick-wall
structure, as shown in \Fig{fig:determinist-example}.

The key motivations behind this construction arise from scenarios where numerous qubits are simultaneously active, providing a natural way to test the presence of cross-talks. Additionally, reaching an approximate steady state with minimal channel applications is crucial to ensure that this circuit can be executed on the quantum computer within a specified time window.

As explained in \Sec{sec:Learning-SS}, to learn the parameters
of the quantum channel, we want its steady state to be non-trivial
in the sense that it will not be the steady state of other local
channels. To that aim, we want the local $\mcE_k$ channels to be
non-unital and also entangling so that the steady state will be a
non trivially-entangled state.  For example, such $\mcE_k$ may be
realized on a QC by some combination of CX, single-qubit rotations,
and RESET gates, as shown in \Fig{fig:determinist-example} and
discussed below.

\begin{figure}
  \includegraphics[width=0.7\linewidth]{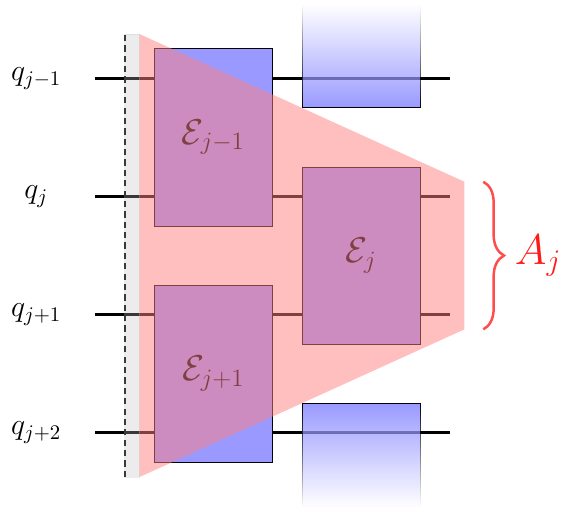}
  \caption{A light cone originating from a 2-local observable $A_j$
  (red shaded area) has at most 4-local support. For the case of
  finite system with open boundary conditions the light cone is
  truncated from 4-local to 3-local at each end of the system.}
  \label{fig:determinist-lightcone}
\end{figure}

To derive our constraints, we view $\mcE$ as the product of two
layers. We define the \emph{odd layer} by $\mcE_\text{odd}
\EqDef\bigotimes_\text{odd $k$}\mcE_k$ and \emph{even layer} by
$\mcE_\text{even}\EqDef \bigotimes_\text{even $k$}\mcE_k$. There are
two possible choices for the overall channel $\mcE$, based on the
order in which the layers are applied: $\mcE_I\EqDef
\mcE_\text{odd}\cdot\mcE_\text{even}$ and $\mcE_{II}\EqDef
\mcE_\text{even}\cdot\mcE_\text{odd}$. An example of the $\mcE_I$
channel is given in \Fig{fig:determinist-example}. The steady state
of $\mcE_I$ is defined by
$\mcE_I(\rho^{I}_{\infty})=\rho^{I}_{\infty}$, and the steady state
of $\mcE_{II}$ by $\mcE_{II}(\rho^{II}_\infty)=\rho^{II}_\infty$. In
general, they will differ from each other. For brevity, expectation
values calculated under $\rhossI, \rhossII$ are denoted by
$\av{\cdot}_I, \av{\cdot}_{II}$ respectively.

Let us now apply \Eq{eq:general_steady_state} to the case of
$\mcE_I$ and a local observable $A$:
\begin{align}
\label{eq:determinist-vanilla}
  \av{A}_{I} = \av{\mcE_{I}^*(A)}_{I} 
    = \av{\mcE^*_\text{even}\big(\mcE_\text{odd}^*(A)\big)}_I.
\end{align}
To simplify the equation, we restrict our attention to observables
$A_j$ acting on qubits $j,j+1$ for odd $j$. The support $A_j$
coincides with the support of $\mcE_j$ from $\mcE_\text{odd}$. By
the same argument that was used in \Sec{sec:stochastic-ideal},
if the supports of $\mcE_k$ and $A_j$ are disjoint, $\mcE_k^*(A_j) =
A_j$, and so $\mcE_\text{odd}^*(A_j)=\mcE^*_j(A_j)$. Another
simplification comes from the fact that the light cone of $A_j$ has
support only on $\mcE^*_{j-1}$ and $\mcE^*_{j+1}$, as shown in
\Fig{fig:determinist-lightcone}. Therefore,
$\mcE^*_\text{even}\big(\mcE^*_\text{odd}(A_j)\big) =
(\mcE_{j-1}^*\otimes\mcE^*_{j+1})\big(\mcE^*_j(A_j)\big)$. All
together, this allows us to rewrite \Eq{eq:determinist-vanilla} in
terms of 4-local expectation values on the RHS:
\begin{align}
\label{eq:determinist-exp-val}
  \av{A_j}_I = \av{(\mcE^*_{j-1}\otimes \mcE^*_{j+1})
    \mcE^*_{j}(A_j)}_I.
\end{align}
For observables $A_j$ with even $j$, a similar equation holds with
the steady state of $\mcE_{II}$ replacing the steady state of
$\mcE_I$.

Following our previous construction with the stochastic map, we can
use the constraints in \Eq{eq:determinist-exp-val} to validate a
given model for the local channels, or learn the model that best
fits the data from a family of models characterized by a set of
parameters $\vtheta$. For the learning task, we start by defining
the cost function $\Phi_I(\vtheta)$ using expectation values in the
steady state $\rhossI$:
\begin{align}
\label{eq:cost-deterministic-noisy}
  \Phi_I(\vtheta) \EqDef \frac{1}{C_{S_I}}\sum_{A_j\in S_I} \left(
     \av{[\mcE^*_{j-1,\vtheta}\otimes \mcE^*_{j+1,\vtheta}]
      \mcE^*_{j,\vtheta}(A_j)}_{I} - \av{A_j}_{I}\right)^2 .
\end{align}
Above, the sum is over a set $S_I$ of $2$-local
observables $A_j$ defined on $j,j+1$ for odd $j$, which we take to
be Pauli operators. Similarly, we define $\Phi_{II}(\vtheta)$ from
the expectation values at $\rhossII$, and set
\begin{align}
  \label{eq:total-Phi}
  \Phi(\vtheta) \EqDef \Phi_I(\vtheta) + \Phi_{II}(\vtheta) .
\end{align}
For a system with $n$ qubits, there are exactly $3n+9(n-1)=12n-9$
such geometrically $2$-local Pauli operators, excluding the trivial
identity operator. This number provides a rough upper bound to the
number of parameters that can be learned using this method, as discussed earlier at the end of \Sec{sec:stochastic-ideal}.

To validate a given model, we define local cost functions $\Phi_q$,
which for a given $q$ contain all terms in $\Phi_I, \Phi_{II}$ that
overlap $q$. See \Eq{eq:loss-Kraus-validation} for the equivalent
definition in the stochastic case.

As in the case of the stochastic map, we can expand $\mcE^*_I(A),
\mcE^*_{II}(A)$ in terms of $4$-local Pauli operators, and write
$\Phi(\vtheta)$ as a quadratic expression of the expectation values
of these operators, as done in \Eq{eq:cost-Pauli-Kraus}.

There are many ways to realize the non-unital channels $\mcE_k$ on a
QC. In this work we use a composed gate, which we call the $\RESU$
gate ($\RESU$ = $\RESET$ + Unitary). It is defined by two $2$-local
unitaries $U_{k,1}$ and $U_{k,2}$ acting on qubits $k,k+1$ and a
$\RESET$ gate on qubit $k$. Each of the $2$-local unitaries is made
of a $\CX$ gate and two general $1$-qubit rotations, as shown in
\Fig{fig:determinist-example}.  Further details of how we model
this gate in the presence of noise are given in
\Sec{sec:deterministic-resu-details}.

In \Sec{sec:deterministic-numerics} we show numerically that our
protocol can be used to learn a given noise model, and in
\Sec{sec:ibmq-implementation} we demonstrate its performance on
actual quantum hardware.

\section{Noise models and classical optimization}
\label{sec:optim-and-noise-models}

In this section we describe the noise models and the numerical
procedures we used to find $\vtheta$ by optimizing over the cost
functions in Eqs.~(\ref{eq:cost-Kraus-noisy},
\ref{eq:total-Phi}),

\subsection{Local Markovian noise models}
\label{sec:noise-models}

We assume a local Markovian noise model, which can be
generally treated using the Lindbladian master
equation\cc{ref:Head-Marsden2021, ref:Fratus2022, ref:Berg2022}. To model the
evolution of a noisy quantum gate, we write
\begin{align}
\label{eq:general_lindblad}
  \frac{d}{dt}\rho = -\frac{i}{T_0}[H_k(\vtheta_k),\rho] 
    + \frac{1}{T_0}\sum_m\theta_m\mcL_m(\rho).
\end{align}
Above, $T_0$ is a time scale, to be determined later, and the
dimensionless Hamiltonian $H_k(\vtheta_k)$ defines the coherent
evolution of the gate $U_k$, i.e., $U_k=e^{-iH_k(\vtheta_k) T/T_0}$,
where $T$ is the gate time and $\{\vtheta_k\}$ are variational gate
parameters. For example, the $\{\vtheta_k\}$ parameters may describe
a source of coherent error in the two-qubit $\CX$ gate. 

Under our local noise assumption, we take $\mcL_m$ to be
\emph{single-qubit} superoperators that represent different
single-qubit noise processes. They are given in terms of jump
operators $L_m$, $\mcL_m(\rho)\EqDef L_m \rho L^{\dagger}_m -
\frac{1}{2}\{L^{\dagger}_mL_m, \rho\}$, where $\{\cdot,\cdot\}$ is
an anti-commutator.  The variational parameters $\{\theta_m\}$ model
the strength of the different noise processes. For example, for a
single-qubit dephasing noise, we use $\mcL_z(\rho)=Z\rho Z - \rho$,
and the corresponding $\theta/T_0$ is the decoherence rate. 

Given a set of channels $\{\mcE_k\}$ that define either the
stochastic or the deterministic maps, we take $T_0$ to be the
maximum over the running times of $\{\mcE_k\}$. Working with the
IBMQ hardware, this time scale was primarily due to the RESET gate
time, which is typically on the order of a microsecond
\cc{ref:Kjaergaard2020, ref:Jurcevic_2021}. The $T_0$ and gate execution
times we used in our numerical simulations are given in
\App{app-sec:stochastic-channel-details} and
\App{app-sec:deterministic-map-details}.

We allowed the coupling strength (variational) parameters to be
different for each qubit. For a qubit $j$, we use the notation
$\vtheta^{(j)}=\{\theta^{(j)}_m\}$ to denote its relevant noise
parameters.

Using the above assumptions, the noise channel
$\mcN_{\vtheta^{(j)}}$ on an idle qubit $j$ for time $T$ is
represented by:
\begin{align}
\label{eq:numerics-idling-noise}
  \mcN_{\vtheta^{(j)}}(\rho) = e^{\frac{T}{T_0}
    \sum_m \theta^{(j)}_m \mcL_m} (\rho).
\end{align}
Similarly, the noisy version of a unitary gate $U_k$ described by a
Hamiltonian $H_k(\vtheta)$ acting for time $T$ is given by:
\begin{align}
\label{eq:numerics-unitary-gate-noise}
  \tilde{\mcE}_{k,\vtheta^{(j)}}(\rho) 
    = e^{\frac{T}{T_0}\left(-i[H_k(\vtheta_k),\cdot] 
      + \sum_m \theta^{(j)}_m \mcL_m\right)} (\rho).
\end{align}
If $U_k$ is a two-qubits gate, the above equation will contain
the $\theta_m\mcL_m$ operators of both qubits.

Equation~\eqref{eq:numerics-unitary-gate-noise} in its entirety was
in fact only used to model the noisy $\CX$ gate when analyzing the
experimental data from the IBMQ hardware. For the single qubit
gates, or for the $\CX$ gate in the numerical simulations, no
coherent errors were assumed, and $H_k(\vtheta_k)$ was taken to be
the ideal Hamiltonian, which is independent of $\vtheta_k$.  The
full details of the $\CX$ modeling we used are given in
\App{app-sec:CX-details}. 

Another simplification was the use of a first-order Trotter-Suzuki
approximation for the case of single-qubit gates. As the typical
execution time of these gates is much shorter than that of the $\CX$
gate (in the IBMQ hardware their execution time is the range of tens
of nanoseconds\cc{ref:Kjaergaard2020, ref:Jurcevic_2021}, which is about
an order of magnitude shorter than the execution time of $\CX$), we
approximated \Eq{eq:numerics-unitary-gate-noise} by
\begin{align}
\label{eq:numerics-unitary-trotter}
  \tilde{\mcE}_{k,\vtheta^{(j)}}
    \simeq \mcU_k\cdot\mcN_{\vtheta^{(j)}} .
\end{align}
Above, $\mcU_k(\cdot) \EqDef e^{-i\frac{T}{T_0}[H_k,\cdot]}$ is the
ideal single-qubit gate channel. Notice that as the resultant error
of this approximation scales as $O\left((T/T_0)^2\cdot \big\|\big[
[H_k,\cdot],\sum_m \theta^{(j)}_m \mcL_m\big]\big\|\right)$, and as
$\norm{H_k}$ and $\norm{\theta_m\mcL_m}$ are all $O(1)$ (see
\App{app-sec:stochastic-channel-details} and
\App{app-sec:deterministic-map-details} for more details), we deduce
that our Trotter-Suzuki error for single-qubit gates is of the order
of $\big(T/T_0\big)^2\sim10^{-4}$, which is much smaller than all
other error sources (e.g., statistical error) described below.

We conclude this section by describing how we modeled the noisy
$\RESET$ gate. Here we followed \cRef{ref:Rost2021} and modeled the
noisy gate phenomenologically using a Kraus map representation. In
this approach, the noisy $\RESET$ gate on the qubit $j$ is described
by two variational parameters, $\vtheta^{(j)}=
(\theta^{(j)}_0,\theta^{(j)}_1)$. The $\theta^{(j)}_0$ is the
probability to measure 0 given that the qubit $j$ was in the state
0, and the $\theta^{(j)}_1$ is the probability to measure 1 given
that the qubit was in the state 1. All together, the noisy $\RESET$
gate is modeled by
\begin{align}
\label{eq:numerics-reset-noise}
  &\nRESET(\rho) \EqDef \theta_0 \ketbra{0}{0}\,\rho\,\ketbra{0}{0} 
      + \theta_1\ketbra{0}{1}\,\rho\,\ketbra{1}{0} \\        
    &+ (1-\theta_{0})\ketbra{1}{0}\,\rho\, \ketbra{0}{1} 
    + (1-\theta_{1})\ketbra{1}{1}\, \rho\,\ketbra{1}{1}.
\nonumber
\end{align}

\subsection{Classical optimization}
\label{sec:optim-routine}

To define the cost function $\Phi$, we used a set $S$ of all 
$2$-local Pauli operators with contiguous support. The minimization
of the cost function $\Phi(\vtheta)$ in
Eqs.~(\ref{eq:cost-Kraus-noisy}, \ref{eq:total-Phi}) was done using
the stochastic optimization algorithm Adam\cc{ref:Kingma2015},
implemented by PyTorch with an automatic differentiation
engine\cc{ref:NEURIPS2019_9015}. We used the same numerical
optimization procedure for both the numerical simulations and the
experimental results on real quantum hardware. This is because the
cost function depends only on the local expectation values, which
may come either from numerical simulations or actual experiments.
As the cost function normalization constant $1/C_S$ does not
depend on the variational parameters, it can be ignored during the
optimization process, yielding a cost function that is quadratic in
the Pauli expectation values.

The computational cost associated with optimization via a gradient-descent algorithm depends upon specific algorithmic decisions and various hyperparameters, including the learning rate, number of steps, and different regularization techniques. However, as a general rule, the computational resources needed for a single optimization step scale equivalently to those required for evaluating the loss function $\Phi(\boldsymbol{\theta})$. In our scenario, where $\Phi(\boldsymbol{\theta})$ comprises $O(n)$ local constraints, we anticipate the overall runtime to follow a scaling pattern akin to $O(Tn)$, where $T$ denotes the number of optimization steps \cc{ref:Kingma2015}.

Heuristically, we fixed the maximum number of optimization steps to
be $15,000$. We added another heuristic termination criterion for
the optimization process, wherein the optimization ceased if at step
$i$, the difference in the cost function from step $i-500$ was less
than $0.25\%$.

In what follows, we denote by $\vtheta_{est}$ the channel
parameters which achieve the lowest cost function value of
$\Phi(\vtheta_{est})$.

\section{Numerical simulations}
\label{sec:numerical-simulations}

To test our method, we performed several numerical simulations of
the stochastic and deterministic maps, and optimized over the cost
function to learn the underlying noise model. Below, we
briefly describe the technical details of simulations, followed by
their results.

\subsection{Simulation of the dynamics, calculation of the 
  expectation values and validation of the numerical results}
\label{sec:dymanics-expectation-values}

For both maps, we simulated systems of $n=5$ to $n=11$ qubits
arranged on a line, with the initial state $\rho_0 =
\ketbra{0}{0}^{\otimes n}$. The simulations were done by evolving
the full density-matrix on a computer. 

An approximation for steady
state $\rhoss$ of a quantum channel $\mcE$ was obtained numerically
by iteratively applying the quantum channel until the convergence
criteria $D\big(\rho_t, \mcE(\rho_t)\big)\EqDef
\frac{1}{2}\norm{\rho_t-\mcE(\rho_t)}_1<10^{-6}$ was reached. The number of channel applications required to for the convergence may be influenced by various factors such as system size, strength of noise processes, choice of the noise model, and the number of $\RESET$ gates. For both maps considered, the convergence criteria was met within at most few tens of channel applications for all system sizes. The
approximate steady-state was then used to calculate the expectation
values of the local Pauli operators in the different cost functions
$\Phi(\vtheta)$, as given in \Eq{eq:cost-Pauli-Kraus}.

We modeled the statistical noise in the expectation values of the
local observables using a Gaussian random variable that approximates
the binomial distribution of measurement results.  For an observable
$A$ that is a product of Pauli matrices, it is easy to verify that
the statistical noise of $N$ shots is well approximated by a
Gaussian random variable $G\sim\mathcal{N}(0, \sigma)$ with
$\sigma=\sqrt{(1-\av{A}^2_\infty)/N}$. 

Our numerical findings were validated by calculating the diamond distance between the channels $\mcE$ and $\hat{\mcE}$, which represent the original noisy gate and the noisy gate derived from the optimization process, respectively. The diamond distance is defined as:
\begin{align}
\label{eq:diamond-dist}
d_{\diamondsuit}(\mcE,\hat{\mcE})=\max_{\rho}\norm{(\mcE\otimes \mcI)\rho-(\hat{\mcE}\otimes \mcI)\rho}_{1},
\end{align}
where $\mcI$ is the identity channel with the same dimensions as $\mcE$ and $\hat{\mcE}$, and the maximization is taken over all density matrices of dimension $\dim{\{\mcE\otimes\mcI\}}$. Specifically, we focused on evaluating the accuracy of our approach in the presence of statistical noise and across various system sizes.

As detailed in Appendix~\ref{app-sec:stat-error}, the validation process outlined above, along with the statistical noise model employed for numerical simulations, cannot be directly extrapolated to scenarios where expectation values used in the cost functions
$\Phi(\vtheta)$, were acquired from real experiments conducted on quantum computers. Nevertheless, the statistical noise model employed in our numerical simulations provided valuable insights that we utilized in the experimental demonstration.

\subsection{Stochastic map simulations}
\label{sec:stochastic-numerics}

For the stochastic map simulations, our noise model consisted of
dephasing (in the $\hat{z}$ direction) and amplitude damping, as
well as the noisy $\RESET$ gate from \Eq{eq:numerics-reset-noise}.
This amounts to $4$ noise parameters for each qubit: $2$ for the
dephasing and amplitude damping and $2$ for the noisy $\RESET$ gate.
In all unitary gates, we assumed no coherent errors. The exact form
of the Lindbladian dissipators is given in
\App{app-sec:stochastic-noise-model}.

\begin{figure}
\centering
  \includegraphics[width=0.5\linewidth]{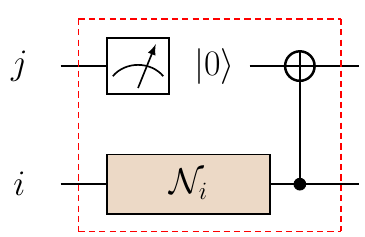}
  \caption{Structure of the $\RESCX(i\to j)$ gate acting two
    neighboring qubits $j,i$. We first break the entanglement
    between $i,j$ by resetting qubit $j$, and then recreate some
    entanglement between them by applying $\CX(i\to j)$. During the action of the $\RESET$ gate on qubit $j$, the idle noise channel $\mcN_{i}$ is acting on qubit $i$.}
\label{fig:rescx-numerics}
\end{figure}

We simulated 3 different stochastic maps, which used the same local
channels but with different probabilities $\{p_k\}$. We refer to
them as probability set 1, 2, and 3. 

The set of gates that defined the stochastic map consisted of the
single-qubits gates $X^{1/2}$, $H$, $R_\Balpha(\psi)$, and a
non-unitary $2$-local gate between neighboring qubits $i,j$, which
we call $\RESCX(i\to j)$. The $R_\Balpha(\psi)$ gate is a
single-qubit rotation gate by an angle $\psi$ around the axis
$\Balpha$, and $\RESCX(i\to j)$ is discussed below.

Following the IBMQ hardware specifications, the
$R_\Balpha(\psi)$ gate was implemented as a combination of two
$X^{1/2}$ and three $R_{\bm{z}}(\phi_i)$ gates, which are native gates in
IBMQ:
\begin{align}
\label{eq:rotation-gate-parameterized}
  R_\Balpha(\psi)= R_{\bm{z}}(\phi_1)X^{1/2}
    R_{\bm{z}}(\phi_2)X^{1/2}R_{\bm{z}}(\phi_3),
\end{align}
In the formula above, the three rotation angles
$\{\phi_i\}_{i=1}^{3}$ implicitly define the rotation angle $\psi$
and the rotation axis $\Balpha$.

The $\RESCX(i\to j)$ gate is a combination of the $\RESET$ gate on
$j$, followed by a $\CX(i\to j)$, as shown in
\Fig{fig:rescx-numerics}. The motivation behind this construction is
that the $\RESET$ gate makes the quantum channel non-unital, but
also breaks entanglement. This might lead to a trivial steady-state,
from which it is hard to learn the underlying noise model. Applying
the $\CX(i\to j)$ gate right after the $\RESET(j)$ was applied
regenerates some entanglement.

The full description of the 3 probability sets, together with the
underlying noise model, the gates and their execution times, are
given in \App{app-sec:stochastic-channel-details}.

\begin{figure}
  \includegraphics[width=0.9\linewidth]{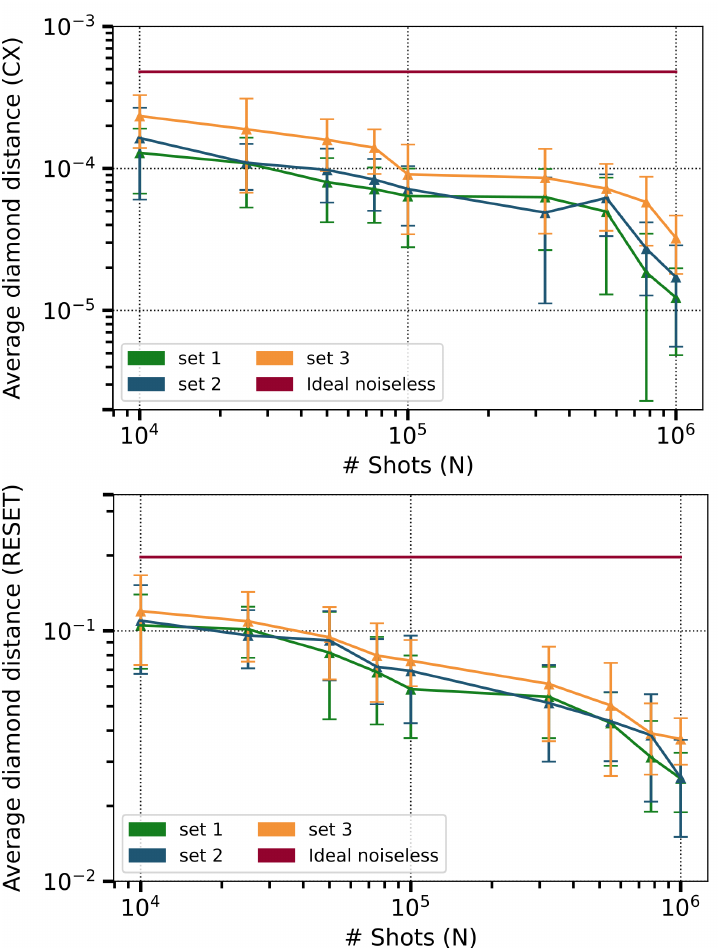}
  \caption{The average diamond distance between the original noisy $\CX$ and $\RESET$ gates and their respective estimations, for a fixed system size $n=6$ qubits. Each data
  point being the average of 10 statistical noise realizations in
  the expectation values. The error bars indicate the uncertainty of
  two standard deviations. (Top): The average diamond distance for
  the $\CX$ gates. (Bottom): The average diamond distance for the
  $\RESET$ gates. Each color represents a different probability set
  $\{p_k\}$. The constant horizontal line (red color) represents the
  diamond distance between the ideal case of the noiseless channel
  and the true noise channel parameters.}
  \label{fig:kraus-numerics-1}
\end{figure}

\begin{figure}
  \includegraphics[width=0.9\linewidth]{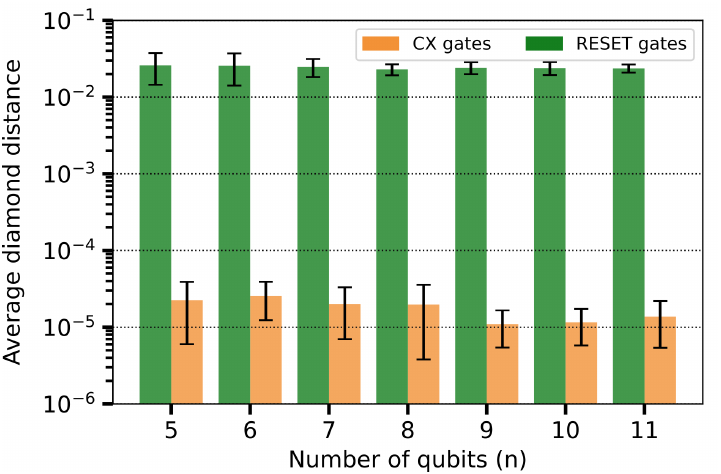}
  \caption{The average diamond distance between the original noisy CX and RESET gates and their respective estimations, for a fixed number of shots $N=10^6$, using the
  probabilities $\{p_k\}$ from the set 2. Each data point being the
  average of 10 statistical noise realizations in the expectation
  values. The error bars indicate the uncertainty of two standard
  deviations.} \label{fig:kraus-numerics-2}
\end{figure}

Our three probability sets were chosen such that the probability of
picking a certain gate is independent of the qubit on which it acts.
Taking the time scale $T_0$ (see also \Sec{sec:noise-models}) to
be the maximum running time over the gate set $\{X^{1/2}, H, 
R_\Balpha(\psi), \RESCX\}$, the implementation of the stochastic
channel from \Sec{sec:stochastic-noisy} on a quantum computer,
is described by the following algorithm:
\begin{enumerate}
  \item Pick a gate $1\leq k\le 4$ from the set $\{X^{1/2}, H, 
  R_\Balpha(\psi), \RESCX\}$ with probability $p_k$.

  \item If the chosen gate is a single-qubit unitary, uniformly 
    choose a qubit and apply it.

  \item If the chosen gate is the $\RESCX$ gate, uniformly 
    choose a neighboring pair of control and target qubits and
    apply it.

  \item If the total running time $T$ of the chosen gate from 
    steps 2 or 3 is smaller than $T_0$, wait $\Delta T = T_0 - T$.
  
\end{enumerate}

Following the discussion in \Sec{sec:stochastic-noisy}, we chose
the observables $A$ in the cost function \eqref{eq:cost-Kraus-noisy}
to be all possible, geometrically 2-local Pauli operators.
Consequently, the optimization procedure used the expectation values
of 3-local Paulis.

To evaluate the accuracy of the numerical optimization, we 
calculated the diamond distance defined in \Eq{eq:diamond-dist}, between the channel representations of the original
noisy $\CX$ and $\RESET$ gates and the ones obtained from
the optimization routine. We focused on these two gates because they
have the longest gate times, and therefore they are the most noisy
gates in the simulation.

\Fig{fig:kraus-numerics-1} and \Fig{fig:kraus-numerics-2}
present the average diamond distance between the original noisy $\CX$ and $\RESET$ gates and their respective estimations. For a system of size $n$, the average is taken over
$n$ possible $\RESET$ gates and $2(n-1)$ possible $\CX$ gates.
\Fig{fig:kraus-numerics-1} shows the average diamond distance as
a function of the number of shots per local Pauli observable, with
the system size fixed at $n=6$ qubits. The results are robust for
the three different probability sets.

With $N=10^6$ shots per Pauli observable, the average diamond
distance for the $\CX$ gate is approximately $2\times 10^{-5}$,
while the average diamond distance for the RESET gates is of the
order $2\times 10^{-2}$. As our method aims to learn the
dynamics of the underlying noise model rather than targeting a
specific gate, the gap between these two distances likely arises
because the noise parameters governing the noisy $\CX$ gate appear
not only in the $\RESCX$ gate, but also in the idle noise channels
and the noisy single qubit gates. In contrast, the noise parameters
associated with the $\RESET$ gate appear only when the $\RESCX$ gate
is applied.

The same behavior can be observed in \Fig{fig:kraus-numerics-2},
where we present the average diamond distance as a function of the
number of qubits in the system using $10^6$ measurements per local
Pauli operator for the probability set 2. 

It would be interesting to find the \emph{optimal} choice of the
gates and probability distribution $\{p_k\}$ for learning a given
model of noise. We leave this for future works.

\subsection{Deterministic maps simulations}
\label{sec:deterministic-numerics}

\subsubsection{The RESU gate}
\label{sec:deterministic-resu-details}

As mentioned in \Sec{sec:deterministic}, we realized the $2$-local
non-unital channels $\{\mcE_k\}$ of the deterministic maps using a
composed gate called the $\RESU$ gate. As shown in 
\Fig{fig:determinist-example}, the $\RESU$ gate on qubits $k,k+1$
consists of two $2$-local unitaries $U_{k,1}$ and $U_{k,2}$ acting
on qubits $k,k+1$ along with a $\RESET$ gate on qubit $k$:
\begin{align}
\label{eq:deterministic-E_k}
  \RESU_k \EqDef U_{k,2} \cdot 
    (\RESET_k\otimes \Id_{k+1}) \cdot U_{k,1} .
\end{align}
Each of the unitaries $U_{k,i=1,2}$ is made of a CX gate and two
arbitrary single-qubit rotation gates. Therefore, different $\RESU$
gates are characterized by four single-qubit rotations
$R_\Balpha(\psi)$, which we denote by
$\{\Balpha_k,\psi_k\}_{k=1}^{4}$. As in
\Sec{sec:stochastic-numerics}, each single-qubit rotation
$R_\Balpha(\psi)$ is implemented as a product of two $X^{1/2}$ and
three $R_{\bm{z}}(\phi_i)$ gates, as described in
\Eq{eq:rotation-gate-parameterized}. 

To model the noise in the deterministic map, we used the framework
of \Sec{sec:noise-models}. For each unitary $U_{k,i=1,2}$ we
defined its noisy version $\tilde{U}_{k,i=1,2}$ by replacing its
noiseless composing gates by their noisy versions. The noisy $\CX$
gates in $\tilde{U}_{k,i=1,2}$ were modeled by
\Eq{eq:numerics-unitary-gate-noise}, and the noisy single-qubit
rotations $R_\Balpha(\psi)$ were modeled by
\Eq{eq:numerics-unitary-trotter}. Finally, the noisy $\RESET$ gate
is modeled by \Eq{eq:numerics-reset-noise}. Together, the
expression of the noisy $\RESU$ gate becomes:
\begin{align}
\label{eq:deterministic-E_k-noisy} \RESU_k \EqDef
  \tilde{U}_{k,2} \cdot (\nRESET_k\otimes \mcN_{k+1}) \cdot
  \tilde{U}_{k,1} ,
\end{align}
where $\mcN_{k+1}$ is an idle noise channel on qubit $k+1$ for the
duration of the noisy $\RESET$ gate.

For the time scale $T_0$, as defined in \Sec{sec:noise-models},
we took the maximum running time over all $\{\RESU_k\}$ gates. For
any $\RESU_k$ gate with the running time $T_k<T_0$, a two-qubit idle
noise channel $\mcN_{k}\otimes\mcN_{k+1}$ is applied for the
duration $\Delta T_k = T_0 - T_k$. On a real QC the idle noise
channel is replaced by the \emph{delay} instruction.

\subsubsection{Numerical results}
\label{sec:deterministic-numerics-results}

We simulated three deterministic maps, which we refer to as map-1,
map-2, and map-3. Each map used two different $\RESU$ gates: one for
the even layer, and one for the odd layer. In each layer, the same
$\RESU$ gate was used on all qubits. We note that while the general
$\RESU$ gate is defined by four single-qubit rotations $R_\Balpha
(\psi)$, the actual gates used in the simulations and experiments
only used two single qubits rotations in each gate by setting
$R_\Balpha = R_\Bbeta$ and $R_{\bm{\gamma}}=R_{\bm{\delta}}$ (see
\Fig{fig:determinist-example}). This was due to a technical
reason: limiting the number of parameters helped us to find sets of
angles for which the convergence to the steady state was rapid, at most within 10 steps (as
required by the IBMQ hardware). All together, each deterministic map
was defined by a set of $4$ single-qubit rotations ($2$ for the even
layer $\RESU$ and $2$ for the odd layer $\RESU$). The full details
of the maps, including the rotation axes and angles, as well the
different gate times, are described in
\App{app-sec:deterministic-map-details}.

\begin{figure}
  \includegraphics[width=0.9\linewidth]{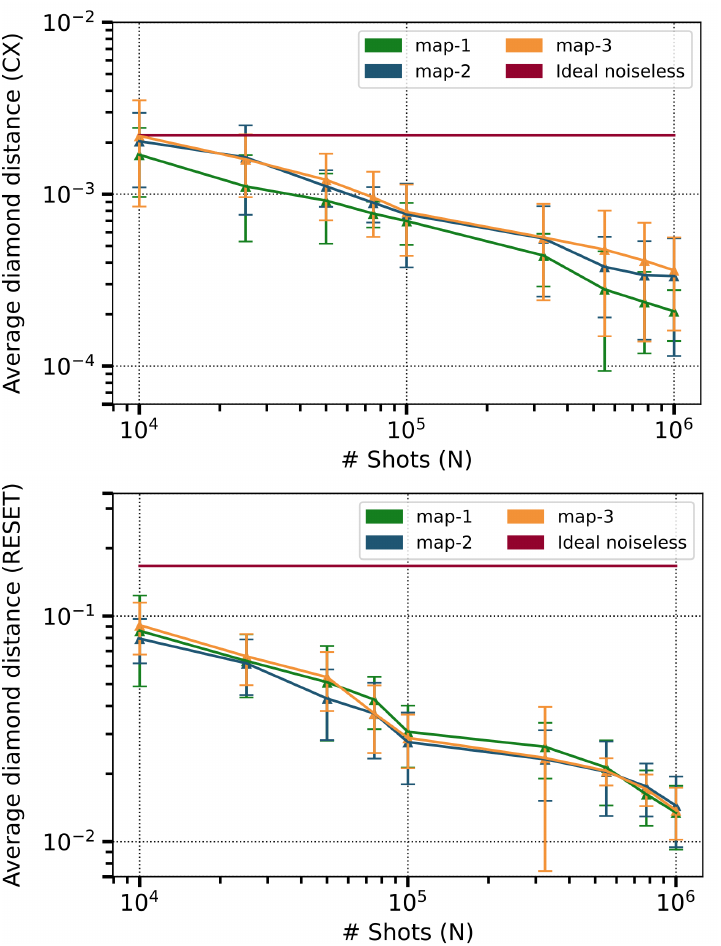}
  \caption{The average diamond distance between the original noisy CX and RESET gates and their respective estimations, for a fixed system size $n=8$ qubits. Each data
  point being the average of 10 statistical noise realizations in
  the expectation values. The error bars indicate the uncertainty of
  two standard deviations. (Top): The average diamond distance for
  the $\CX$ gates. (Bottom): The average diamond distance for the
  $\RESET$ gates. Each color represents a different map defined by
  the rotation axes and angles,
  $\{\Balpha_k,\psi_k\}_{k=1}^{4}$. For a specific map, the
  pairs $\{\Balpha_k,\psi_k\}$ were randomly generated, and kept
  constant. The constant horizontal line (red color) represents the
  diamond distance between the ideal case of the noiseless channel
  and the true noise channel parameters.}
  \label{fig:RESU-numerics-1}
\end{figure}

\begin{figure}
  \includegraphics[width=0.9\linewidth]{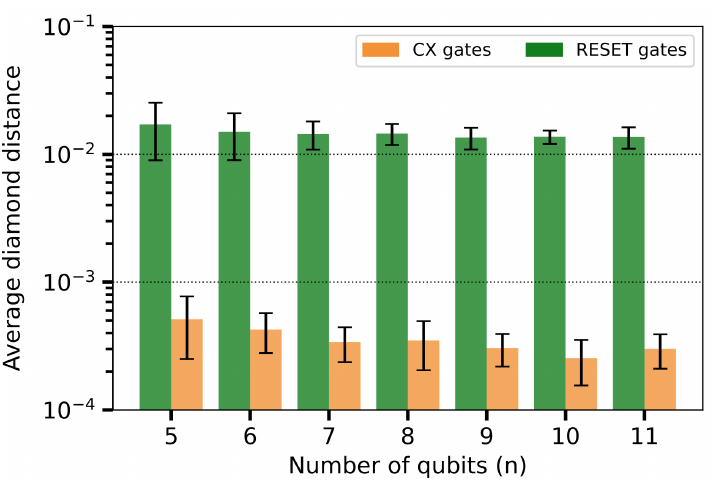}
  \caption{The average diamond distance between the original noisy $\CX$ and $\RESET$ gates and their respective estimations, for a fixed number of shots $N=10^6$, using the
  rotations $\{\Balpha_k,\psi_k\}$ from the map 2. Each data
  point being the average of 10 statistical noise realizations in
  the expectation values. The error bars indicate the uncertainty of
  two standard deviations.} \label{fig:RESU-numerics-2}
\end{figure}

The noise model we used in the simulations of the deterministic maps
consisted of generalized amplitude damping and non-uniform
depolarization. For each qubit this amounted to 5 Lindblad operators: 3
depolarization operators (one for each axis) and two operators for
the generalized amplitude damping\cc{ref:Srikanth2008, ref:Cafaro2014,
ref:Krantz2019}.  In addition, for each qubit we had two parameters
modeling its $\RESET$ gate according to
\Eq{eq:numerics-reset-noise}. As in the stochastic case, we assume
no coherent errors in the unitary gates. The full description of the
noise model is given in
\App{app-sec:deterministic-channel-noise-details}.

After running the simulations, we learned the noise model using the
optimization procedure of \Sec{sec:optim-routine}. To evaluate
the accuracy of the noise parameter estimation, we used the same
procedure as in the stochastic maps simulations, and compared the
noisy $\CX$ and $\RESET$ gates used in the simulations to the ones
obtained from the optimization routine.

In \Fig{fig:RESU-numerics-1} and \Fig{fig:RESU-numerics-2}, the average diamond distance between the original noisy $\CX$ and $\RESET$ gates and their respective estimations obtained through our method is presented. For a system of size $n$, the average is
taken over the $n-1$ possible $\RESET$ gates and $2(n-1)$ possible
$\CX$ gates. 

In \Fig{fig:RESU-numerics-1}, we present the average diamond
distance as a function of the number of shots per local Pauli
observable for a fixed system size of $n=8$ qubits, using three
randomly generated sets of rotation axes and angles,
$\{\Balpha_k,\psi_k\}_{k=1}^{4}$. Each of the three sets
resulted in a different steady state. The accuracy of the estimation
is robust to the different sets of rotation angles and axes (which
are given in \App{app-sec:deterministic-resu-angles}). For $N=10^6$
shots per local Pauli observable, the average diamond distance for
the $\CX$ gates is approximately $3\times 10^{-4}$, while for the
$\RESET$ gates it is approximately $1.4\times 10^{-2}$.

In \Fig{fig:RESU-numerics-2} we kept the number of shots per
Pauli fixed at $N=10^6$ and used the rotation angles and axes from
map 2. Increasing the system size from $n=5$ to $n=11$, we observe
that the accuracy of the parameter estimation improves slightly
between $n=5$ and $n=7$, and remains robust up to $n=11$. The
initial increase in accuracy for smaller systems in
\Fig{fig:RESU-numerics-2} may be explained by examining the
light cone shown in \Fig{fig:determinist-lightcone} in
\Sec{sec:deterministic}. For a smaller system size with open
boundary conditions, the light cone is truncated at each end of the
system, causing the expectation values at the edges of the system to
be 3-local instead of 4-local. As a result, the optimization routine
uses fewer statistics for the qubits located at the ends of the
system. However, for larger system sizes, this boundary effect
becomes negligible as the relative fraction of boundary qubits
decreases.

As in the case of the stochastic map in
\Sec{sec:stochastic-numerics}, the average diamond distance for the
$\CX$ gates is much smaller than the $\RESET$ gates.
This similar behavior arises because our method learns the underlying noise
parameters, which may appear in different gates within the circuit.
Consequently, the noise channel parameters in the $\CX$ gates are
present in multiple combinations within a single $\RESU$ gate, while
the channel parameters of the $\RESET$ gate only appear once per
$\RESU$ gate.

Comparing the numerical results depicted in \Fig{fig:kraus-numerics-1} and \Fig{fig:kraus-numerics-2} versus \Fig{fig:RESU-numerics-1} and \Fig{fig:RESU-numerics-2}, respectively, shows that, for a given number of shots $N$, the stochastic map achieves greater accuracy compared to the deterministic map. This disparity can be attributed to the stochastic map's additional "prior knowledge" regarding the circuits, given that the gate probabilities $p_k$ are known, consequently leading to more accurate results.

\section{Quantum device results}
\label{sec:ibmq-implementation}

In this section we describe the experiments we performed 
on actual quantum hardware to test our method. All experiments were
done on an IBMQ machine via the IBMQ cloud.

\subsection{Details of the experiment}
\label{sec:ibmq-measurement-protocol}

For the quantum hardware demonstration we chose the
\emph{ibm\_lagos} v1.0.32, which is one of the IBM Quantum Falcon
Processors\cc{ref:IBMQ} with $7$ qubits in the shape of a rotated H. Out
of the $7$ qubits, we chose $n=5$ qubits arranged in a line, which
we labeled by $0,1,2,3,4$. The IBMQ labels of these qubits were
$6,5,3,1,0$ in corresponding order.

We used the deterministic map-1 in \Sec{sec:deterministic} to
learn the noise of the quantum hardware. See
\App{app-sec:deterministic-resu-angles} for an exact description of the
map. Our noise model was the same model used in the deterministic
map simulations, which consisted of non-uniform depolarization errors,
generalized amplitude damping, and $\RESET$ errors (total of 7 free
parameters per qubit). In addition to these parameters we added a
modeling of coherent $\CX$ errors as free parameters in its
Hamiltonian --- see Appendices~\ref{app-sec:CX-details},
\ref{app-sec:deterministic-channel-noise-details} for a full description
of the model. 

After learning the hardware noise model using map-1,  
we ran deterministic map-2 and measured $2$-local Pauli
expectation values on its steady state. We used these measurements
to test the quality of the noise model we learned from map-1.

To reduce SPAM errors, we used a readout-error mitigation scheme,
which was partially based on \cRefs{ref:Geller_2020, ref:Geller_2021}. The
scheme relied on a preliminary step in which we estimated the
conditional probability of measuring a $4$-local bit string in the
computational basis, assuming the qubits were prepared in a
different computational basis. See \App{app-sec:ibmq-readout-scheme} for
an full description of our scheme.

Overall, our experiment consisted of three steps: (i) preliminary
readout-error mitigation measurements, (ii) running deterministic
map-1 and measuring the expectation values of $4$-local Paulis, and
(iii) running deterministic map-2 and measuring the expectation
values of $2$-local Paulis.

In all three parts of the experiment we needed to estimate the
expectation values of local Pauli strings. To that aim, we used the
overlapping local tomography method, described by Zubida \emph{et
al.} in \cRef{ref:Zubida2021}, which simultaneously estimates all
geometrically $k$-local Pauli strings using a set of $3^k$ different
measurement circuits. This means that using a total budget of $M$
shots, each Pauli string was estimated using $M/3^k$ shots.

We now describe the 3 parts of the experiment in more details.

{~}

\emph{(i) Readout error mitigation measurements}: Here we performed
preliminary readout-error mitigation measurements for the protocol
that is described in \App{app-sec:ibmq-readout-scheme}. Our measurements
were used to mitigate the readout errors of steps (ii),(iii). We
used the overlapping local tomography method (see Zubida \emph{et
al.} in \cRef{ref:Zubida2021}), to obtain $4$-local readout statistics
from a set of $2^4$ different measurement circuits of unit depth. On
each circuit we performed $2\times 10^5$ measurements (about
$M=3\times 10^6$ shots in total). 

{~}

\emph{(ii) $4$-local channel measurements}: In this step we ran the
deterministic map-1 on the quantum computer and prepared many copies
of the steady states of $\mcE_I$ and $\mcE_{II}$ channels. Following
the numerical simulations, we assumed that the steady state of each
channel was approximately reached after 10 steps. We then used the
overlapping local tomography to estimate the expectation values of
the $4$-local Pauli operators using $1.25\times 10^5$ measurements
for every $4$-local expectation value (and about $M=2\times 10^7$
shots in total).

{~}

\emph{(iii) $2$-local channel measurements}: In this step we ran the
deterministic map-2 on the quantum computer and performed $2$-local
Pauli measurements using the overlapping local tomography. As
explained in \Sec{sec:deterministic-numerics}, map-2 produces a
different steady state from the map-1. We used these $2$-local Pauli
measurements to asses the accuracy of our noise model we learned from the $4$-local expectation values of step (ii). As in part
(ii), we assumed that the steady states of the $\mcE_I$ and
$\mcE_{II}$ channels were approximately reached after 10 steps. The
$2$-local Pauli expectation values were obtained using overlapping
local tomography that used $1.25\times 10^5$ measurements for every
$2$-local expectation value (and about $M=2\times 10^6$ shots in
total).

{~}

Upon finishing the measurements in (ii), (iii), we used the results
of readout measurements (i) to correct these measurement
statistics (see \App{app-sec:ibmq-readout-scheme} for details).

\subsection{Results I: Noise model validation}
\label{sec:ibmq-loss-map}

The main application of our method is to \emph{validate} (benchmark) a
noise model of a QC by calculating its local cost functions $\Phi_q$
given a noise model that defines the noisy channels
$\{\tilde{\mcE}_k\}$, as explained in \Sec{sec:deterministic}.
The local cost function $\Phi_q$ for a qubit $q$ is the sum of all
the $A_j$ entries in $\Phi_I$ and $\Phi_{II}$ (see
\Eq{eq:cost-deterministic-noisy}) that overlap qubit $q$, divided by
the suitable normalization factor $C_{S_q}$ (see \Eq{eq:loss-Kraus-validation}). This happens when the light cone of $A_j$
includes qubit $q$, i.e., when $q\in \{j-1, \ldots, j+2\}$
(see \Fig{fig:determinist-lightcone} for an illustration).

\begin{figure}
  \includegraphics[width=\linewidth]{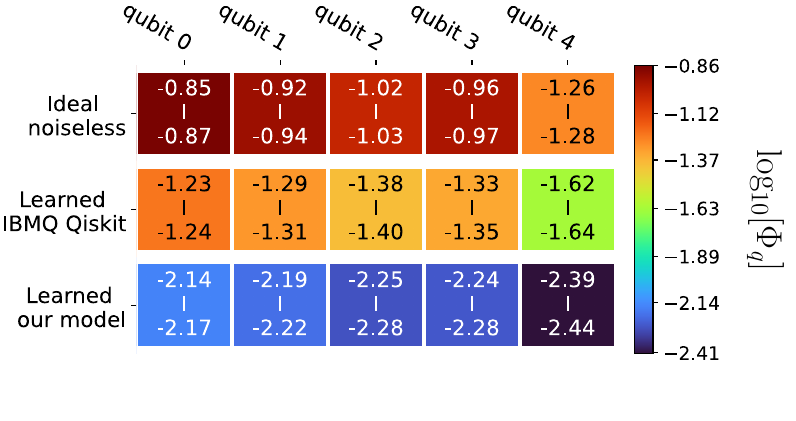}
  \caption{The cost map for the \emph{ibm\_lagos} device for three
  noise models: the ideal noiseless case (top row), the learned
  noise model from IBMQ Qiskit backend (middle row), and the learned
  noise model using our method (bottom row). The top number
  in each square corresponds to $\log_{10}[\Phi_q + \Delta\Phi_q]$,
  and the bottom number to $\log_{10}[\Phi_q - \Delta\Phi_q]$. Where
  $\Phi_q$ is cost function for the qubit $q$, and $\Delta\Phi_q$ is
  the uncertainty of one standard deviation in the cost function
  value due to the statistical noise present in the measurements of
  the local Paulis. The full details of the derivation for the $\Delta\Phi_q$ are given in the Appendix~\ref{app-sec:stat-error}. Each square is colored according to the value of
  $\log_{10}[\Phi_q]$.}\label{fig:ibmq-heatmap}
\end{figure}

We used our validation protocol to compare three noise models: (i)
the ideal noiseless model, (ii) our estimate of the IBMQ
Qiskit\cc{ref:Qiskit} noise model, imported from the device's backend
properties at the time of our experiment, and (iii) a noise model
that we learned from the IBMQ hardware using our optimization
method (see \Sec{sec:optim-and-noise-models}).

To calculate model (ii), we needed to know $\mcE^*_{odd},
\mcE^*_{even}$ of the IBMQ Qiskit noise model, from which the local
cost function is calculated (see Eqs.~(\ref{eq:determinist-vanilla},
\ref{eq:determinist-exp-val}, \ref{eq:cost-deterministic-noisy})).
However, as far as we could tell, IBMQ does not fully publish the
exact formulas it uses to model the local noise. Instead, it
publishes general noise parameters such as $T_1, T_2$, and provides
code that simulates noisy circuits. Therefore, in order to calculate
$\mcE^*_{odd}, \mcE^*_{even}$, we first \emph{learned} IBMQ Qiskit
noise model using the deterministic map-3, and used the resultant
model as an approximation to IBMQ Qiskit noise model. In more
details, we used IBMQ Qiskit simulator to simulate the steady state
of map-3, and calculated local Pauli expectations with vanishing
statistical noise. We then used our optimization routine to learn
Qiskit's noise model. We used the parameters described in
\App{app-sec:deterministic-channel-noise-details}, but without any coherent
errors, and with depolarization only along the $\hat{z}$ axis.  As a
sanity check, we used the model we learned to simulate the 5 qubits
steady states of maps-1,2,3 and compared them to Qiskit's steady
states. We noticed that in all cases the trace distance between our
steady state and Qiskit's steady state was less than $1\%$.

\Fig{fig:ibmq-heatmap} shows the single-qubit cost function, as
a heat map. The different rows represent $\log_{10}[\Phi_q \pm \Delta\Phi_q]$, for the
3 noise models, and $\Delta\Phi_q$ is the uncertainty of one standard deviation in the cost function (see Appendix~\ref{app-sec:stat-error} for detailed error analysis of the experiment). Unsurprisingly, it suggests that IBMQ Qiskit's noise
model provides a better description of the hardware than the ideal,
noiseless channel. It also suggests that our estimated noise model
performs significantly better than the two other models. This might
not be surprising, given the fact that we learned our noise model
using the same steady state that we used to validate it. In other
words, the learned noise model parameters are the ones that minimize
the global cost function, which is roughly the sum of the local cost
functions that appear in \Fig{fig:ibmq-heatmap}. In the machine
learning jargon, this might imply that our noise model estimation
procedure overfitted the particular steady state from which the
4-local measurements were sampled. To rule this out (at least to
some extent), we show in the next subsection that our estimated
noise model has a better predictive power than the other two: it is
able to better approximate the 2-local expectation values of a
steady state of map-2, which is different from the steady state of
map-1 with which we learned the noise model.

\subsection{Results II: Noise model characterization}
\label{sec:ibmq-characterization}

In this section we present the results of applying our optimization
algorithm to \emph{characterize} the noise within a quantum device.
The full learned noise model parameter details are presented in
\App{app-sec:ibmq-noise-model-results}, and the statistical error analysis is presented in \App{app-sec:stat-error}.

To evaluate the quality of our noise characterization, we took the 
noise model that we learned and used it to numerically simulate the
steady states of map-2 (of $\mcE_I$ and $\mcE_{II}$). On these
steady states we calculated the expectation values of $2$-local
Paulis in the bulk of the 5 qubits line (we used qubits $1,2$ for
$\mcE_I$ and $2,3$ for $\mcE_{II}$). We then compared these
simulated expectation values to the actual results we measured on
the quantum hardware when running map-2 (part (iii) of our
experiment described in \Sec{sec:ibmq-measurement-protocol}). We
used the same procedure to test the two other noise models from the
previous section, i.e., the ideal (noiseless) model, and IBMQ
Qiskit's model. In contrast to the previous section, to sample the
$2$-local Paulis from the IBMQ Qiskit noise model we don't need to
know $\mcE^*_{odd}, \mcE^*_{even}$. Hence, we calculated the
$2$-local Pauli expectation values directly from the IBMQ Qiskit
simulator using the noise model imported from the device's backend
properties at the time of our experiment.

\begin{figure}
  \includegraphics[width=\linewidth]{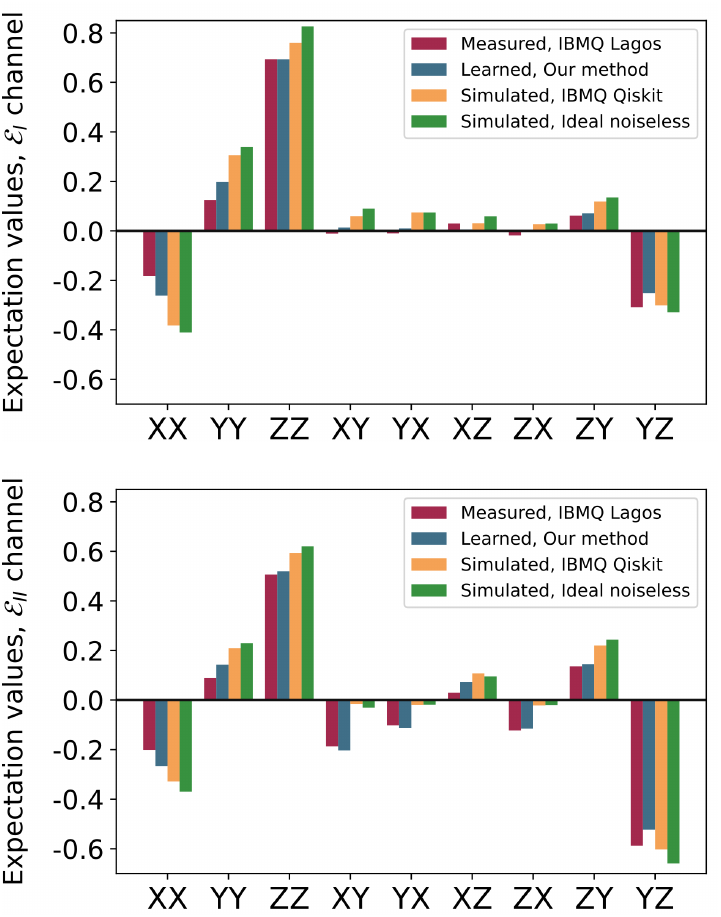}
  \caption{Expectation values of the 2-local Pauli operators
  obtained from: the \emph{ibm\_lagos} device (red), the learned
  noise model using our method (blue), the IBMQ Qiskit simulator
  (yellow) and the ideal noiseless case (green). We focus on qubits
  $1,2$ for $\mcE_{I}$ (top) and $2,3$ for $\mcE_{II}$ (bottom) to
  avoid including the boundary qubits $0$ and $4$, for which the
  measurement statistics become $3$-local instead of $4$-local (see
  \Sec{sec:deterministic-numerics}). See
  \autoref{table:trace-distances} for the corresponding trace
  distances between the measured RDM and the RDMs of the three
  models. } \label{fig:ibmq-two-local-exp-vals}
\end{figure}

Our results are presented as histograms in
\Fig{fig:ibmq-two-local-exp-vals}, where we show the expectation
values of the $9$ possible $2$-local Pauli expectation values for
the steady states of $\mcE_I, \mcE_{II}$ of map-2. The expectation
values from the quantum hardware are shown in red. As we used more
than $10^5$ measurements for each data point, the statistical error in the red bars in \Fig{fig:ibmq-two-local-exp-vals} is smaller than $0.003$, which is negligible comparing to the
differences with respect to other models. For the 3 analytical
models, we present the exact expectation values, without any
statistical error. Our model prediction is shown in blue, Qiskit's
model in yellow and the noiseless model in green. As can be seen in
the figure, while the predictions of all the models are in decent
agreement with the empirical results, our model is in better
agreement than the other two models. 

To compare our estimate in a quantitative way, we used the 2-local
Paulies expectation values of the IBMQ hardware to perform quantum
state tomography and estimate the underlying 2-local reduced density
matrices (RDMs) in the steady states of $\mcE_I, \mcE_{II}$ of
map-2. We compared it to the RDMs produced by simulations of our
model, the Qiskit model and the ideal model. For each case, we
calculated the trace distance $D(\rho, \sigma)\EqDef
\frac{1}{2}\norm{\rho-\sigma}_1$ between the simulation and the
estimated RDM of the IBMQ hardware. The results are given in
\autoref{table:trace-distances}, showing that indeed the trace
distance between IBMQ RDMs and our model RDMs are $50\%-70\%$ of the
distances from ideal model or the Qiskit model.

It is worth noting that the 2-local Pauli operators used for the
reliability assessment were collected after the 4-local Pauli
expectation values used for noise model characterization were
obtained. As a result, the noise model at the time of the 2-local
Pauli measurements may have undergone some drifting. This may
account for some of the variations seen in the 2-local expectation
values depicted in
\Fig{fig:ibmq-two-local-exp-vals}\cc{ref:Chow2009, ref:Fogarty2015,
ref:Klimov2018}.

\begin{table}[h]
    \centering
    \begin{tabular}{||l c c ||} 
    \hline
        & $\mcE_I$ & $\mcE_{II}$ \\ [0.5ex] 
    \hline\hline
    \ Ideal model  & 0.19 & 0.21 \\
    \ Qiskit model & 0.174 & 0.19 \\
    \ Learned model    & 0.09 & 0.13 \\
    \hline
    \end{tabular}
    \caption{The trace distance of the 2-local reduced density
    matrices obtained from the IBMQ hardware using quantum state
    tomography and the corresponding density matrices taken from
    simulations using the ideal model, Qiskit noise model, and our
    learned model. We calculated the distances on the steady states
    of $\mcE_I$ and $\mcE_{II}$ of map-2.}
    \label{table:trace-distances}
\end{table}

\section{Summary and outlook}
\label{sec:discussion}

We have presented and demonstrated numerically and experimentally a
new framework for learning a non-unital quantum channel from its
steady state, which can be used to learn the noise channel of a
quantum computer. It is inspired by the recent methods of
learning a local Hamiltonian or Lindbladian from their steady
states\cc{ref:Qi2019, ref:Bairey2019, ref:Bairey2020,
ref:Evans2019}. 

We have shown that when the underlying noise channel in a quantum
computer is Markovian and local, it can be learned using our method
by driving the system into an engineered dissipative steady state.
We have shown that there exist a set of linear constraints between
the expectation values of local observables in the steady state,
which can be used to learn or validate the underlying noise channel.
This steady state can be a correlated, entangled state, and might
even be classically inaccessible. Nevertheless, as we showed, the
constraints between its local expectation values are local and can
be efficiently verified once we measure these expectation values. We
can therefore apply our method to a system with a large number of
qubits and test how well a given noise model \emph{globally}
describes the system, which is in a state that might be classically
inaccessible.

For generic dissipative maps, the steady state is unique, and
consequently, our method is independent of the initial state and is
insensitive to state preparation errors. In addition, the fact that
we measure a steady state that was created using several iterations
allows us to test possible deviations from Markovianity. Finally, by
optimizing over a parametrized set of local Markovian noise models,
our method allows learning of the optimal noise model.

We have suggested two generic ways to engineer a dissipative map on
a quantum computer, both of which rely on non-unitary operations, such
as the $\RESET$ gate. The first is a stochastic map that is
implemented on a quantum computer by applying a set of gates
according to a prescribed probability distribution. The second uses
a brick-wall type circuit, whose building blocks are composed
$2$-local non-unitary gates and
can therefore be applied using a single circuit. We have
demonstrated numerically both methods using simulations of $5-11$
qubits, and showed that it can successfully learn an underlying
noise model. In our numerical tests we managed to learn the
underlying $\CNOT$ and $\RESET$ gates up to diamond distances of
$10^{-3}-10^{-5}$ and $10^{-1}-10^{-2}$ respectively, using
$10^4-10^6$ shots per observable.

We have also tested our method experimentally on the IBMQ
\textit{ibm\_lagos} machine using $5$ qubits. We have shown that
given a noise model, we can construct a `heat map' view of the
qubits showing how well the model describes each qubit. We have also
showed how our method can learn an underlying noise model for that
machine, and that the model it learned better predicted the outcomes
of another experiment than the noiseless model or IBMQ Qiskit's
model.

Our work leaves open several theoretical and practical questions. On
the theoretical side, it would be interesting to characterize the
complexity of the steady states of dissipative maps like the one we
used. Can they be sampled classically, or can they encode some
BQP-hard problems? More general engineered dissipative maps are
known to be BQP-complete\cc{ref:Verstraete2009}, but our maps are
much simpler. They also differ from the much studied class of
randomly generated circuits, for which there are several complexity
bounds (see, for example, the recent result in
\cRef{ref:Aharonov2022} and references within), in several aspects:
our maps are not completely random (we apply the same map over and
over), they are strongly dissipative and non-unital, and can be
applied for $\poly(n)$ time. 

It would also be interesting to find a systematic way of engineering
the dissipative maps (of both types) to have an optimal estimate of
the error parameters. This might be possible to do by studying the
quantum Fisher information\cc{ref:Meyer2021} of the steady states
with respect to the free parameters of the map. Related to that, it
would be interesting to understand how the quantum Fisher
information, or more general our ability to use the steady state for
learning, depends on the deviation of the channel from
unitality. Is
there a phase transition behavior, as seen in related measurement
induced phase transitions literature (see \cRef{ref:Li2022} and
references within), or is it a smooth transition?

On the more practical direction, it would be interesting to see if
there is a way to make our method completely insensitive to SPAM
errors, by removing its dependence on measurement errors. The
readout error mitigation procedure that we employed arguably removes
a large part of these errors, but not in a controlled, systematic
way, as done in methods like RB\cc{ref:Emerson2005, ref:Emerson2007,
ref:Knill2008, ref:Magesan2012} and GST\cc{ref:BlumeKohout2013,
ref:Kim2015, ref:Nielsen2021}.

\section*{Acknowledgements}

We thank Eyal Bairey and Netanel Lindner for enlightening
discussions. IA acknowledges the support of the Israel Science
Foundation (ISF) under the Research Grants in Quantum Technologies
and Science No. 2074/19, and the joint NRF-ISF Research Grant No.
3528/20. This research was supported by the National Research
Foundation, Singapore, under its Quantum Engineering Programme. We
are very grateful for the support of the National University of
Singapore (NUS) and the Centre of Quantum Technologies (CQT) for
their help in running the IBMQ experiments. We acknowledge the use
of IBM Quantum services for this work and to advanced services
provided by the IBM Quantum Researchers Program. The views expressed
are those of the authors, and do not reflect the official policy or
position of IBM or the IBM Quantum team. This research project was
partially supported by the Helen Diller Quantum Center at the
Technion – Israel Institute of Technology.


\bibliography{main.bib}

\begin{thebibliography}{64}%
\makeatletter
\providecommand \@ifxundefined [1]{%
 \@ifx{#1\undefined}
}%
\providecommand \@ifnum [1]{%
 \ifnum #1\expandafter \@firstoftwo
 \else \expandafter \@secondoftwo
 \fi
}%
\providecommand \@ifx [1]{%
 \ifx #1\expandafter \@firstoftwo
 \else \expandafter \@secondoftwo
 \fi
}%
\providecommand \natexlab [1]{#1}%
\providecommand \enquote  [1]{``#1''}%
\providecommand \bibnamefont  [1]{#1}%
\providecommand \bibfnamefont [1]{#1}%
\providecommand \citenamefont [1]{#1}%
\providecommand \href@noop [0]{\@secondoftwo}%
\providecommand \href [0]{\begingroup \@sanitize@url \@href}%
\providecommand \@href[1]{\@@startlink{#1}\@@href}%
\providecommand \@@href[1]{\endgroup#1\@@endlink}%
\providecommand \@sanitize@url [0]{\catcode `\\12\catcode `\$12\catcode `\&12\catcode `\#12\catcode `\^12\catcode `\_12\catcode `\%12\relax}%
\providecommand \@@startlink[1]{}%
\providecommand \@@endlink[0]{}%
\providecommand \url  [0]{\begingroup\@sanitize@url \@url }%
\providecommand \@url [1]{\endgroup\@href {#1}{\urlprefix }}%
\providecommand \urlprefix  [0]{URL }%
\providecommand \Eprint [0]{\href }%
\providecommand \doibase [0]{https://doi.org/}%
\providecommand \selectlanguage [0]{\@gobble}%
\providecommand \bibinfo  [0]{\@secondoftwo}%
\providecommand \bibfield  [0]{\@secondoftwo}%
\providecommand \translation [1]{[#1]}%
\providecommand \BibitemOpen [0]{}%
\providecommand \bibitemStop [0]{}%
\providecommand \bibitemNoStop [0]{.\EOS\space}%
\providecommand \EOS [0]{\spacefactor3000\relax}%
\providecommand \BibitemShut  [1]{\csname bibitem#1\endcsname}%
\let\auto@bib@innerbib\@empty
\bibitem [{\citenamefont {Burgarth}\ \emph {et~al.}(2009)\citenamefont {Burgarth}, \citenamefont {Maruyama},\ and\ \citenamefont {Nori}}]{ref:Burgarth2009}%
  \BibitemOpen
  \bibfield  {author} {\bibinfo {author} {\bibfnamefont {D.}~\bibnamefont {Burgarth}}, \bibinfo {author} {\bibfnamefont {K.}~\bibnamefont {Maruyama}},\ and\ \bibinfo {author} {\bibfnamefont {F.}~\bibnamefont {Nori}},\ }\bibfield  {title} {\bibinfo {title} {{Coupling strength estimation for spin chains despite restricted access}},\ }\bibfield  {journal} {\bibinfo  {journal} {Phys. Rev. A}\ }\textbf {\bibinfo {volume} {79}},\ \href {https://doi.org/10.1103/PhysRevA.79.020305} {10.1103/PhysRevA.79.020305} (\bibinfo {year} {2009})\BibitemShut {NoStop}%
\bibitem [{\citenamefont {{Di Franco}}\ \emph {et~al.}(2009)\citenamefont {{Di Franco}}, \citenamefont {Paternostro},\ and\ \citenamefont {Kim}}]{ref:DiFranco2009}%
  \BibitemOpen
  \bibfield  {author} {\bibinfo {author} {\bibfnamefont {C.}~\bibnamefont {{Di Franco}}}, \bibinfo {author} {\bibfnamefont {M.}~\bibnamefont {Paternostro}},\ and\ \bibinfo {author} {\bibfnamefont {M.~S.}\ \bibnamefont {Kim}},\ }\bibfield  {title} {\bibinfo {title} {{Hamiltonian tomography in an access-limited setting without state initialization}},\ }\bibfield  {journal} {\bibinfo  {journal} {Phys. Rev. Lett.}\ }\textbf {\bibinfo {volume} {102}},\ \href {https://doi.org/10.1103/PhysRevLett.102.187203} {10.1103/PhysRevLett.102.187203} (\bibinfo {year} {2009})\BibitemShut {NoStop}%
\bibitem [{\citenamefont {Shabani}\ \emph {et~al.}(2011)\citenamefont {Shabani}, \citenamefont {Mohseni}, \citenamefont {Lloyd}, \citenamefont {Kosut},\ and\ \citenamefont {Rabitz}}]{ref:Shabani2011a}%
  \BibitemOpen
  \bibfield  {author} {\bibinfo {author} {\bibfnamefont {A.}~\bibnamefont {Shabani}}, \bibinfo {author} {\bibfnamefont {M.}~\bibnamefont {Mohseni}}, \bibinfo {author} {\bibfnamefont {S.}~\bibnamefont {Lloyd}}, \bibinfo {author} {\bibfnamefont {R.~L.}\ \bibnamefont {Kosut}},\ and\ \bibinfo {author} {\bibfnamefont {H.}~\bibnamefont {Rabitz}},\ }\bibfield  {title} {\bibinfo {title} {{Estimation of many-body quantum hamiltonians via compressive sensing}},\ }\bibfield  {journal} {\bibinfo  {journal} {Phys. Rev. A}\ }\textbf {\bibinfo {volume} {84}},\ \href {https://doi.org/10.1103/PhysRevA.84.012107} {10.1103/PhysRevA.84.012107} (\bibinfo {year} {2011})\BibitemShut {NoStop}%
\bibitem [{\citenamefont {{Da Silva}}\ \emph {et~al.}(2011)\citenamefont {{Da Silva}}, \citenamefont {Landon-Cardinal},\ and\ \citenamefont {Poulin}}]{ref:DaSilva2011}%
  \BibitemOpen
  \bibfield  {author} {\bibinfo {author} {\bibfnamefont {M.~P.}\ \bibnamefont {{Da Silva}}}, \bibinfo {author} {\bibfnamefont {O.}~\bibnamefont {Landon-Cardinal}},\ and\ \bibinfo {author} {\bibfnamefont {D.}~\bibnamefont {Poulin}},\ }\bibfield  {title} {\bibinfo {title} {{Practical characterization of quantum devices without tomography}},\ }\bibfield  {journal} {\bibinfo  {journal} {Phys. Rev. Lett.}\ }\textbf {\bibinfo {volume} {107}},\ \href {https://doi.org/10.1103/PhysRevLett.107.210404} {10.1103/PhysRevLett.107.210404} (\bibinfo {year} {2011})\BibitemShut {NoStop}%
\bibitem [{\citenamefont {Zhang}\ and\ \citenamefont {Sarovar}(2014)}]{ref:Zhang2014}%
  \BibitemOpen
  \bibfield  {author} {\bibinfo {author} {\bibfnamefont {J.}~\bibnamefont {Zhang}}\ and\ \bibinfo {author} {\bibfnamefont {M.}~\bibnamefont {Sarovar}},\ }\bibfield  {title} {\bibinfo {title} {{Quantum Hamiltonian Identification from Measurement Time Traces}},\ }\href {https://doi.org/10.1103/PhysRevLett.113.080401} {\bibfield  {journal} {\bibinfo  {journal} {Phys. Rev. Lett.}\ }\textbf {\bibinfo {volume} {113}},\ \bibinfo {pages} {80401} (\bibinfo {year} {2014})}\BibitemShut {NoStop}%
\bibitem [{\citenamefont {{De Clercq}}\ \emph {et~al.}(2016)\citenamefont {{De Clercq}}, \citenamefont {Oswald}, \citenamefont {Fl{\"{u}}hmann}, \citenamefont {Keitch}, \citenamefont {Kienzler}, \citenamefont {Lo}, \citenamefont {Marinelli}, \citenamefont {Nadlinger}, \citenamefont {Negnevitsky},\ and\ \citenamefont {Home}}]{ref:DeClercq2016}%
  \BibitemOpen
  \bibfield  {author} {\bibinfo {author} {\bibfnamefont {L.~E.}\ \bibnamefont {{De Clercq}}}, \bibinfo {author} {\bibfnamefont {R.}~\bibnamefont {Oswald}}, \bibinfo {author} {\bibfnamefont {C.}~\bibnamefont {Fl{\"{u}}hmann}}, \bibinfo {author} {\bibfnamefont {B.}~\bibnamefont {Keitch}}, \bibinfo {author} {\bibfnamefont {D.}~\bibnamefont {Kienzler}}, \bibinfo {author} {\bibfnamefont {H.~Y.}\ \bibnamefont {Lo}}, \bibinfo {author} {\bibfnamefont {M.}~\bibnamefont {Marinelli}}, \bibinfo {author} {\bibfnamefont {D.}~\bibnamefont {Nadlinger}}, \bibinfo {author} {\bibfnamefont {V.}~\bibnamefont {Negnevitsky}},\ and\ \bibinfo {author} {\bibfnamefont {J.~P.}\ \bibnamefont {Home}},\ }\bibfield  {title} {\bibinfo {title} {{Estimation of a general time-dependent Hamiltonian for a single qubit}},\ }\bibfield  {journal} {\bibinfo  {journal} {Nat. Comm.}\ }\textbf {\bibinfo {volume} {7}},\ \href {https://doi.org/10.1038/ncomms11218} {10.1038/ncomms11218} (\bibinfo {year} {2016})\BibitemShut {NoStop}%
\bibitem [{\citenamefont {Wang}\ \emph {et~al.}(2018)\citenamefont {Wang}, \citenamefont {Dong}, \citenamefont {Qi}, \citenamefont {Zhang}, \citenamefont {Petersen},\ and\ \citenamefont {Yonezawa}}]{ref:Wang2018}%
  \BibitemOpen
  \bibfield  {author} {\bibinfo {author} {\bibfnamefont {Y.}~\bibnamefont {Wang}}, \bibinfo {author} {\bibfnamefont {D.}~\bibnamefont {Dong}}, \bibinfo {author} {\bibfnamefont {B.}~\bibnamefont {Qi}}, \bibinfo {author} {\bibfnamefont {J.}~\bibnamefont {Zhang}}, \bibinfo {author} {\bibfnamefont {I.~R.}\ \bibnamefont {Petersen}},\ and\ \bibinfo {author} {\bibfnamefont {H.}~\bibnamefont {Yonezawa}},\ }\bibfield  {title} {\bibinfo {title} {{A Quantum Hamiltonian Identification Algorithm: Computational Complexity and Error Analysis}},\ }\href {https://doi.org/10.1109/TAC.2017.2747507} {\bibfield  {journal} {\bibinfo  {journal} {IEEE Transactions on Automatic Control}\ }\textbf {\bibinfo {volume} {63}},\ \bibinfo {pages} {1388} (\bibinfo {year} {2018})}\BibitemShut {NoStop}%
\bibitem [{\citenamefont {{Zubida}}\ \emph {et~al.}(2021)\citenamefont {{Zubida}}, \citenamefont {{Yitzhaki}}, \citenamefont {{Lindner}},\ and\ \citenamefont {{Bairey}}}]{ref:Zubida2021}%
  \BibitemOpen
  \bibfield  {author} {\bibinfo {author} {\bibfnamefont {A.}~\bibnamefont {{Zubida}}}, \bibinfo {author} {\bibfnamefont {E.}~\bibnamefont {{Yitzhaki}}}, \bibinfo {author} {\bibfnamefont {N.~H.}\ \bibnamefont {{Lindner}}},\ and\ \bibinfo {author} {\bibfnamefont {E.}~\bibnamefont {{Bairey}}},\ }\bibfield  {title} {\bibinfo {title} {{Optimal short-time measurements for Hamiltonian learning}},\ }\href {https://doi.org/10.48550/arXiv.2108.08824} {\bibfield  {journal} {\bibinfo  {journal} {arXiv}\ ,\ \bibinfo {eid} {arXiv:2108.08824}} (\bibinfo {year} {2021})}\BibitemShut {NoStop}%
\bibitem [{\citenamefont {Kokail}\ \emph {et~al.}(2021)\citenamefont {Kokail}, \citenamefont {van Bijnen}, \citenamefont {Elben}, \citenamefont {Vermersch},\ and\ \citenamefont {Zoller}}]{ref:Kokail2021}%
  \BibitemOpen
  \bibfield  {author} {\bibinfo {author} {\bibfnamefont {C.}~\bibnamefont {Kokail}}, \bibinfo {author} {\bibfnamefont {R.}~\bibnamefont {van Bijnen}}, \bibinfo {author} {\bibfnamefont {A.}~\bibnamefont {Elben}}, \bibinfo {author} {\bibfnamefont {B.}~\bibnamefont {Vermersch}},\ and\ \bibinfo {author} {\bibfnamefont {P.}~\bibnamefont {Zoller}},\ }\bibfield  {title} {\bibinfo {title} {Entanglement hamiltonian tomography in quantum simulation},\ }\href {https://doi.org/10.1038/s41567-021-01260-w} {\bibfield  {journal} {\bibinfo  {journal} {Nature Physics}\ }\textbf {\bibinfo {volume} {17}},\ \bibinfo {pages} {936} (\bibinfo {year} {2021})}\BibitemShut {NoStop}%
\bibitem [{\citenamefont {{Yu}}\ \emph {et~al.}(2022)\citenamefont {{Yu}}, \citenamefont {{Sun}}, \citenamefont {{Han}},\ and\ \citenamefont {{Yuan}}}]{ref:Yu2022}%
  \BibitemOpen
  \bibfield  {author} {\bibinfo {author} {\bibfnamefont {W.}~\bibnamefont {{Yu}}}, \bibinfo {author} {\bibfnamefont {J.}~\bibnamefont {{Sun}}}, \bibinfo {author} {\bibfnamefont {Z.}~\bibnamefont {{Han}}},\ and\ \bibinfo {author} {\bibfnamefont {X.}~\bibnamefont {{Yuan}}},\ }\bibfield  {title} {\bibinfo {title} {{Practical and Efficient Hamiltonian Learning}},\ }\href {https://doi.org/10.48550/arXiv.2201.00190} {\bibfield  {journal} {\bibinfo  {journal} {arXiv}\ ,\ \bibinfo {eid} {arXiv:2201.00190}} (\bibinfo {year} {2022})}\BibitemShut {NoStop}%
\bibitem [{\citenamefont {Wilde}\ \emph {et~al.}(2022)\citenamefont {Wilde}, \citenamefont {Kshetrimayum}, \citenamefont {Roth}, \citenamefont {Hangleiter}, \citenamefont {Sweke},\ and\ \citenamefont {Eisert}}]{ref:Wilde2022}%
  \BibitemOpen
  \bibfield  {author} {\bibinfo {author} {\bibfnamefont {F.}~\bibnamefont {Wilde}}, \bibinfo {author} {\bibfnamefont {A.}~\bibnamefont {Kshetrimayum}}, \bibinfo {author} {\bibfnamefont {I.}~\bibnamefont {Roth}}, \bibinfo {author} {\bibfnamefont {D.}~\bibnamefont {Hangleiter}}, \bibinfo {author} {\bibfnamefont {R.}~\bibnamefont {Sweke}},\ and\ \bibinfo {author} {\bibfnamefont {J.}~\bibnamefont {Eisert}},\ }\bibfield  {title} {\bibinfo {title} {{Scalably learning quantum many-body Hamiltonians from dynamical data}},\ }\href {http://arxiv.org/abs/2209.14328} {\bibfield  {journal} {\bibinfo  {journal} {arXiv}\ } (\bibinfo {year} {2022})},\ \Eprint {https://arxiv.org/abs/2209.14328} {arXiv:2209.14328} \BibitemShut {NoStop}%
\bibitem [{\citenamefont {Kappen}(2020)}]{ref:Kappen2020}%
  \BibitemOpen
  \bibfield  {author} {\bibinfo {author} {\bibfnamefont {H.~J.}\ \bibnamefont {Kappen}},\ }\bibfield  {title} {\bibinfo {title} {Learning quantum models from quantum or classical data},\ }\href {https://doi.org/10.1088/1751-8121/ab7df6} {\bibfield  {journal} {\bibinfo  {journal} {Journal of Physics A: Mathematical and Theoretical}\ }\textbf {\bibinfo {volume} {53}},\ \bibinfo {pages} {214001} (\bibinfo {year} {2020})}\BibitemShut {NoStop}%
\bibitem [{\citenamefont {Anshu}\ \emph {et~al.}(2021)\citenamefont {Anshu}, \citenamefont {Arunachalam}, \citenamefont {Kuwahara},\ and\ \citenamefont {Soleimanifar}}]{ref:Anshu2021}%
  \BibitemOpen
  \bibfield  {author} {\bibinfo {author} {\bibfnamefont {A.}~\bibnamefont {Anshu}}, \bibinfo {author} {\bibfnamefont {S.}~\bibnamefont {Arunachalam}}, \bibinfo {author} {\bibfnamefont {T.}~\bibnamefont {Kuwahara}},\ and\ \bibinfo {author} {\bibfnamefont {M.}~\bibnamefont {Soleimanifar}},\ }\bibfield  {title} {\bibinfo {title} {Sample-efficient learning of interacting quantum systems},\ }\href {https://doi.org/10.1038/s41567-021-01232-0} {\bibfield  {journal} {\bibinfo  {journal} {Nature Physics}\ }\textbf {\bibinfo {volume} {17}},\ \bibinfo {pages} {931} (\bibinfo {year} {2021})}\BibitemShut {NoStop}%
\bibitem [{\citenamefont {{Haah}}\ \emph {et~al.}(2021)\citenamefont {{Haah}}, \citenamefont {{Kothari}},\ and\ \citenamefont {{Tang}}}]{ref:Haah2021}%
  \BibitemOpen
  \bibfield  {author} {\bibinfo {author} {\bibfnamefont {J.}~\bibnamefont {{Haah}}}, \bibinfo {author} {\bibfnamefont {R.}~\bibnamefont {{Kothari}}},\ and\ \bibinfo {author} {\bibfnamefont {E.}~\bibnamefont {{Tang}}},\ }\bibfield  {title} {\bibinfo {title} {{Optimal learning of quantum Hamiltonians from high-temperature Gibbs states}},\ }\href {https://doi.org/10.48550/arXiv.2108.04842} {\bibfield  {journal} {\bibinfo  {journal} {arXiv}\ ,\ \bibinfo {eid} {arXiv:2108.04842}} (\bibinfo {year} {2021})}\BibitemShut {NoStop}%
\bibitem [{\citenamefont {{Lifshitz}}\ \emph {et~al.}(2021)\citenamefont {{Lifshitz}}, \citenamefont {{Bairey}}, \citenamefont {{Arbel}}, \citenamefont {{Aleksandrowicz}}, \citenamefont {{Landa}},\ and\ \citenamefont {{Arad}}}]{ref:Lifshitz2021}%
  \BibitemOpen
  \bibfield  {author} {\bibinfo {author} {\bibfnamefont {Y.~Y.}\ \bibnamefont {{Lifshitz}}}, \bibinfo {author} {\bibfnamefont {E.}~\bibnamefont {{Bairey}}}, \bibinfo {author} {\bibfnamefont {E.}~\bibnamefont {{Arbel}}}, \bibinfo {author} {\bibfnamefont {G.}~\bibnamefont {{Aleksandrowicz}}}, \bibinfo {author} {\bibfnamefont {H.}~\bibnamefont {{Landa}}},\ and\ \bibinfo {author} {\bibfnamefont {I.}~\bibnamefont {{Arad}}},\ }\bibfield  {title} {\bibinfo {title} {{Practical Quantum State Tomography for Gibbs states}},\ }\href {https://doi.org/10.48550/arXiv.2112.10418} {\bibfield  {journal} {\bibinfo  {journal} {arXiv}\ ,\ \bibinfo {eid} {arXiv:2112.10418}} (\bibinfo {year} {2021})}\BibitemShut {NoStop}%
\bibitem [{\citenamefont {Granade}\ \emph {et~al.}(2012)\citenamefont {Granade}, \citenamefont {Ferrie}, \citenamefont {Wiebe},\ and\ \citenamefont {Cory}}]{ref:Granade2012}%
  \BibitemOpen
  \bibfield  {author} {\bibinfo {author} {\bibfnamefont {C.~E.}\ \bibnamefont {Granade}}, \bibinfo {author} {\bibfnamefont {C.}~\bibnamefont {Ferrie}}, \bibinfo {author} {\bibfnamefont {N.}~\bibnamefont {Wiebe}},\ and\ \bibinfo {author} {\bibfnamefont {D.~G.}\ \bibnamefont {Cory}},\ }\bibfield  {title} {\bibinfo {title} {{Robust online Hamiltonian learning}},\ }\bibfield  {journal} {\bibinfo  {journal} {New Journal of Physics}\ }\textbf {\bibinfo {volume} {14}},\ \href {https://doi.org/10.1088/1367-2630/14/10/103013} {10.1088/1367-2630/14/10/103013} (\bibinfo {year} {2012})\BibitemShut {NoStop}%
\bibitem [{\citenamefont {Wiebe}\ \emph {et~al.}(2014{\natexlab{a}})\citenamefont {Wiebe}, \citenamefont {Granade}, \citenamefont {Ferrie},\ and\ \citenamefont {Cory}}]{ref:Wiebe2014}%
  \BibitemOpen
  \bibfield  {author} {\bibinfo {author} {\bibfnamefont {N.}~\bibnamefont {Wiebe}}, \bibinfo {author} {\bibfnamefont {C.}~\bibnamefont {Granade}}, \bibinfo {author} {\bibfnamefont {C.}~\bibnamefont {Ferrie}},\ and\ \bibinfo {author} {\bibfnamefont {D.~G.}\ \bibnamefont {Cory}},\ }\bibfield  {title} {\bibinfo {title} {{Hamiltonian learning and certification using quantum resources}},\ }\bibfield  {journal} {\bibinfo  {journal} {Phys. Rev. Lett.}\ }\textbf {\bibinfo {volume} {112}},\ \href {https://doi.org/10.1103/PhysRevLett.112.190501} {10.1103/PhysRevLett.112.190501} (\bibinfo {year} {2014}{\natexlab{a}})\BibitemShut {NoStop}%
\bibitem [{\citenamefont {Wiebe}\ \emph {et~al.}(2014{\natexlab{b}})\citenamefont {Wiebe}, \citenamefont {Granade}, \citenamefont {Ferrie},\ and\ \citenamefont {Cory}}]{ref:Wiebe2014a}%
  \BibitemOpen
  \bibfield  {author} {\bibinfo {author} {\bibfnamefont {N.}~\bibnamefont {Wiebe}}, \bibinfo {author} {\bibfnamefont {C.}~\bibnamefont {Granade}}, \bibinfo {author} {\bibfnamefont {C.}~\bibnamefont {Ferrie}},\ and\ \bibinfo {author} {\bibfnamefont {D.}~\bibnamefont {Cory}},\ }\bibfield  {title} {\bibinfo {title} {{Quantum Hamiltonian learning using imperfect quantum resources}},\ }\bibfield  {journal} {\bibinfo  {journal} {Phys. Rev. A}\ }\textbf {\bibinfo {volume} {89}},\ \href {https://doi.org/10.1103/PhysRevA.89.042314} {10.1103/PhysRevA.89.042314} (\bibinfo {year} {2014}{\natexlab{b}})\BibitemShut {NoStop}%
\bibitem [{\citenamefont {Wiebe}\ \emph {et~al.}(2015)\citenamefont {Wiebe}, \citenamefont {Granade},\ and\ \citenamefont {Cory}}]{ref:Wiebe2015}%
  \BibitemOpen
  \bibfield  {author} {\bibinfo {author} {\bibfnamefont {N.}~\bibnamefont {Wiebe}}, \bibinfo {author} {\bibfnamefont {C.}~\bibnamefont {Granade}},\ and\ \bibinfo {author} {\bibfnamefont {D.~G.}\ \bibnamefont {Cory}},\ }\bibfield  {title} {\bibinfo {title} {{Quantum bootstrapping via compressed quantum Hamiltonian learning}},\ }\bibfield  {journal} {\bibinfo  {journal} {New Journal of Physics}\ }\textbf {\bibinfo {volume} {17}},\ \href {https://doi.org/10.1088/1367-2630/17/2/022005} {10.1088/1367-2630/17/2/022005} (\bibinfo {year} {2015})\BibitemShut {NoStop}%
\bibitem [{\citenamefont {Wang}\ \emph {et~al.}(2017)\citenamefont {Wang}, \citenamefont {Paesani}, \citenamefont {Santagati}, \citenamefont {Knauer}, \citenamefont {Gentile}, \citenamefont {Wiebe}, \citenamefont {Petruzzella}, \citenamefont {O'brien}, \citenamefont {Rarity}, \citenamefont {Laing},\ and\ \citenamefont {Thompson}}]{ref:Wang2017}%
  \BibitemOpen
  \bibfield  {author} {\bibinfo {author} {\bibfnamefont {J.}~\bibnamefont {Wang}}, \bibinfo {author} {\bibfnamefont {S.}~\bibnamefont {Paesani}}, \bibinfo {author} {\bibfnamefont {R.}~\bibnamefont {Santagati}}, \bibinfo {author} {\bibfnamefont {S.}~\bibnamefont {Knauer}}, \bibinfo {author} {\bibfnamefont {A.~A.}\ \bibnamefont {Gentile}}, \bibinfo {author} {\bibfnamefont {N.}~\bibnamefont {Wiebe}}, \bibinfo {author} {\bibfnamefont {M.}~\bibnamefont {Petruzzella}}, \bibinfo {author} {\bibfnamefont {J.~L.}\ \bibnamefont {O'brien}}, \bibinfo {author} {\bibfnamefont {J.~G.}\ \bibnamefont {Rarity}}, \bibinfo {author} {\bibfnamefont {A.}~\bibnamefont {Laing}},\ and\ \bibinfo {author} {\bibfnamefont {M.~G.}\ \bibnamefont {Thompson}},\ }\bibfield  {title} {\bibinfo {title} {{Experimental quantum Hamiltonian learning}},\ }\href {https://doi.org/10.1038/nphys4074} {\bibfield  {journal} {\bibinfo  {journal} {Nat. Phys.}\ }\textbf {\bibinfo {volume} {13}},\ \bibinfo {pages} {551} (\bibinfo {year} {2017})}\BibitemShut
  {NoStop}%
\bibitem [{\citenamefont {Sone}\ and\ \citenamefont {Cappellaro}(2017)}]{ref:Sone2017}%
  \BibitemOpen
  \bibfield  {author} {\bibinfo {author} {\bibfnamefont {A.}~\bibnamefont {Sone}}\ and\ \bibinfo {author} {\bibfnamefont {P.}~\bibnamefont {Cappellaro}},\ }\bibfield  {title} {\bibinfo {title} {{Hamiltonian identifiability assisted by a single-probe measurement}},\ }\href {https://doi.org/10.1103/PhysRevA.95.022335} {\bibfield  {journal} {\bibinfo  {journal} {Physical Review A}\ }\textbf {\bibinfo {volume} {95}},\ \bibinfo {pages} {1} (\bibinfo {year} {2017})}\BibitemShut {NoStop}%
\bibitem [{\citenamefont {Greiter}\ \emph {et~al.}(2018)\citenamefont {Greiter}, \citenamefont {Schnells},\ and\ \citenamefont {Thomale}}]{ref:Greiter2018}%
  \BibitemOpen
  \bibfield  {author} {\bibinfo {author} {\bibfnamefont {M.}~\bibnamefont {Greiter}}, \bibinfo {author} {\bibfnamefont {V.}~\bibnamefont {Schnells}},\ and\ \bibinfo {author} {\bibfnamefont {R.}~\bibnamefont {Thomale}},\ }\bibfield  {title} {\bibinfo {title} {Method to identify parent hamiltonians for trial states},\ }\href {https://doi.org/10.1103/PhysRevB.98.081113} {\bibfield  {journal} {\bibinfo  {journal} {Phys. Rev. B}\ }\textbf {\bibinfo {volume} {98}},\ \bibinfo {pages} {081113} (\bibinfo {year} {2018})}\BibitemShut {NoStop}%
\bibitem [{\citenamefont {Chertkov}\ and\ \citenamefont {Clark}(2018)}]{ref:Chertkov2018}%
  \BibitemOpen
  \bibfield  {author} {\bibinfo {author} {\bibfnamefont {E.}~\bibnamefont {Chertkov}}\ and\ \bibinfo {author} {\bibfnamefont {B.~K.}\ \bibnamefont {Clark}},\ }\bibfield  {title} {\bibinfo {title} {Computational inverse method for constructing spaces of quantum models from wave functions},\ }\href {https://doi.org/10.1103/PhysRevX.8.031029} {\bibfield  {journal} {\bibinfo  {journal} {Phys. Rev. X}\ }\textbf {\bibinfo {volume} {8}},\ \bibinfo {pages} {031029} (\bibinfo {year} {2018})}\BibitemShut {NoStop}%
\bibitem [{\citenamefont {Qi}\ and\ \citenamefont {Ranard}(2019)}]{ref:Qi2019}%
  \BibitemOpen
  \bibfield  {author} {\bibinfo {author} {\bibfnamefont {X.-L.}\ \bibnamefont {Qi}}\ and\ \bibinfo {author} {\bibfnamefont {D.}~\bibnamefont {Ranard}},\ }\bibfield  {title} {\bibinfo {title} {Determining a local {H}amiltonian from a single eigenstate},\ }\href {https://doi.org/10.22331/q-2019-07-08-159} {\bibfield  {journal} {\bibinfo  {journal} {{Quantum}}\ }\textbf {\bibinfo {volume} {3}},\ \bibinfo {pages} {159} (\bibinfo {year} {2019})}\BibitemShut {NoStop}%
\bibitem [{\citenamefont {Bairey}\ \emph {et~al.}(2019)\citenamefont {Bairey}, \citenamefont {Arad},\ and\ \citenamefont {Lindner}}]{ref:Bairey2019}%
  \BibitemOpen
  \bibfield  {author} {\bibinfo {author} {\bibfnamefont {E.}~\bibnamefont {Bairey}}, \bibinfo {author} {\bibfnamefont {I.}~\bibnamefont {Arad}},\ and\ \bibinfo {author} {\bibfnamefont {N.~H.}\ \bibnamefont {Lindner}},\ }\bibfield  {title} {\bibinfo {title} {Learning a local hamiltonian from local measurements},\ }\href {https://doi.org/10.1103/PhysRevLett.122.020504} {\bibfield  {journal} {\bibinfo  {journal} {Phys. Rev. Lett.}\ }\textbf {\bibinfo {volume} {122}},\ \bibinfo {pages} {020504} (\bibinfo {year} {2019})}\BibitemShut {NoStop}%
\bibitem [{\citenamefont {Bairey}\ \emph {et~al.}(2020)\citenamefont {Bairey}, \citenamefont {Guo}, \citenamefont {Poletti}, \citenamefont {Lindner},\ and\ \citenamefont {Arad}}]{ref:Bairey2020}%
  \BibitemOpen
  \bibfield  {author} {\bibinfo {author} {\bibfnamefont {E.}~\bibnamefont {Bairey}}, \bibinfo {author} {\bibfnamefont {C.}~\bibnamefont {Guo}}, \bibinfo {author} {\bibfnamefont {D.}~\bibnamefont {Poletti}}, \bibinfo {author} {\bibfnamefont {N.~H.}\ \bibnamefont {Lindner}},\ and\ \bibinfo {author} {\bibfnamefont {I.}~\bibnamefont {Arad}},\ }\bibfield  {title} {\bibinfo {title} {{Learning the dynamics of open quantum systems from their steady states}},\ }\bibfield  {journal} {\bibinfo  {journal} {New Journal of Physics}\ }\textbf {\bibinfo {volume} {22}},\ \href {https://doi.org/10.1088/1367-2630/ab73cd} {10.1088/1367-2630/ab73cd} (\bibinfo {year} {2020})\BibitemShut {NoStop}%
\bibitem [{\citenamefont {{Evans}}\ \emph {et~al.}(2019)\citenamefont {{Evans}}, \citenamefont {{Harper}},\ and\ \citenamefont {{Flammia}}}]{ref:Evans2019}%
  \BibitemOpen
  \bibfield  {author} {\bibinfo {author} {\bibfnamefont {T.~J.}\ \bibnamefont {{Evans}}}, \bibinfo {author} {\bibfnamefont {R.}~\bibnamefont {{Harper}}},\ and\ \bibinfo {author} {\bibfnamefont {S.~T.}\ \bibnamefont {{Flammia}}},\ }\bibfield  {title} {\bibinfo {title} {{Scalable Bayesian Hamiltonian learning}},\ }\href {https://doi.org/10.48550/arXiv.1912.07636} {\bibfield  {journal} {\bibinfo  {journal} {arXiv}\ ,\ \bibinfo {eid} {arXiv:1912.07636}} (\bibinfo {year} {2019})},\ \Eprint {https://arxiv.org/abs/1912.07636} {arXiv:1912.07636 [quant-ph]} \BibitemShut {NoStop}%
\bibitem [{Note1()}]{Note1}%
  \BibitemOpen
  \bibinfo {note} {Strictly speaking, this is true up to an overall factor.}\BibitemShut {Stop}%
\bibitem [{Note2()}]{Note2}%
  \BibitemOpen
  \bibinfo {note} {While it would also be more natural to normalize by $\protect \frac {1}{2}({\| \rho \|}_S^2 + {\| \protect \mathcal {E}(\rho ) \|}_S^2)$, such normalization substantially complicates the optimization routine because ${\| \protect \mathcal {E}(\rho ) \|}_S^2$ depends on the optimization parameters. In turn, this normalization does not introduce a significant change in the accuracy of the optimization as close to the convergence point $\protect \mathcal {E}(\rho )$ has to be close to $\rho $.}\BibitemShut {Stop}%
\bibitem [{\citenamefont {Head-Marsden}\ \emph {et~al.}(2021)\citenamefont {Head-Marsden}, \citenamefont {Krastanov}, \citenamefont {Mazziotti},\ and\ \citenamefont {Narang}}]{ref:Head-Marsden2021}%
  \BibitemOpen
  \bibfield  {author} {\bibinfo {author} {\bibfnamefont {K.}~\bibnamefont {Head-Marsden}}, \bibinfo {author} {\bibfnamefont {S.}~\bibnamefont {Krastanov}}, \bibinfo {author} {\bibfnamefont {D.~A.}\ \bibnamefont {Mazziotti}},\ and\ \bibinfo {author} {\bibfnamefont {P.}~\bibnamefont {Narang}},\ }\bibfield  {title} {\bibinfo {title} {{Capturing non-Markovian dynamics on near-term quantum computers}},\ }\href {https://doi.org/10.1103/PhysRevResearch.3.013182} {\bibfield  {journal} {\bibinfo  {journal} {Physical Review Research}\ }\textbf {\bibinfo {volume} {3}},\ \bibinfo {pages} {13182} (\bibinfo {year} {2021})}\BibitemShut {NoStop}%
\bibitem [{\citenamefont {Fratus}\ \emph {et~al.}(2022)\citenamefont {Fratus}, \citenamefont {Bark}, \citenamefont {Vogt}, \citenamefont {Leppakangas}, \citenamefont {Zanker}, \citenamefont {Marthaler},\ and\ \citenamefont {Reiner}}]{ref:Fratus2022}%
  \BibitemOpen
  \bibfield  {author} {\bibinfo {author} {\bibfnamefont {K.~R.}\ \bibnamefont {Fratus}}, \bibinfo {author} {\bibfnamefont {K.}~\bibnamefont {Bark}}, \bibinfo {author} {\bibfnamefont {N.}~\bibnamefont {Vogt}}, \bibinfo {author} {\bibfnamefont {J.}~\bibnamefont {Leppakangas}}, \bibinfo {author} {\bibfnamefont {S.}~\bibnamefont {Zanker}}, \bibinfo {author} {\bibfnamefont {M.}~\bibnamefont {Marthaler}},\ and\ \bibinfo {author} {\bibfnamefont {J.-M.}\ \bibnamefont {Reiner}},\ }\bibfield  {title} {\bibinfo {title} {{Describing Trotterized Time Evolutions on Noisy Quantum Computers via Static Effective Lindbladians}},\ }\href {https://arxiv.org/abs/2210.11371} {\bibfield  {journal} {\bibinfo  {journal} {arXiv}\ } (\bibinfo {year} {2022})},\ \Eprint {https://arxiv.org/abs/2210.11371} {arXiv:2210.11371} \BibitemShut {NoStop}%
\bibitem [{\citenamefont {van~den Berg}\ \emph {et~al.}(2022)\citenamefont {van~den Berg}, \citenamefont {Minev}, \citenamefont {Kandala},\ and\ \citenamefont {Temme}}]{ref:Berg2022}%
  \BibitemOpen
  \bibfield  {author} {\bibinfo {author} {\bibfnamefont {E.}~\bibnamefont {van~den Berg}}, \bibinfo {author} {\bibfnamefont {Z.~K.}\ \bibnamefont {Minev}}, \bibinfo {author} {\bibfnamefont {A.}~\bibnamefont {Kandala}},\ and\ \bibinfo {author} {\bibfnamefont {K.}~\bibnamefont {Temme}},\ }\bibfield  {title} {\bibinfo {title} {{Probabilistic error cancellation with sparse Pauli-Lindblad models on noisy quantum processors}},\ }\href {http://arxiv.org/abs/2201.09866} {\bibfield  {journal} {\bibinfo  {journal} {arXiv}\ } (\bibinfo {year} {2022})},\ \Eprint {https://arxiv.org/abs/2201.09866} {arXiv:2201.09866} \BibitemShut {NoStop}%
\bibitem [{\citenamefont {Kjaergaard}\ \emph {et~al.}(2020)\citenamefont {Kjaergaard}, \citenamefont {Schwartz}, \citenamefont {Braumuller}, \citenamefont {Krantz}, \citenamefont {Wang}, \citenamefont {Gustavsson},\ and\ \citenamefont {Oliver}}]{ref:Kjaergaard2020}%
  \BibitemOpen
  \bibfield  {author} {\bibinfo {author} {\bibfnamefont {M.}~\bibnamefont {Kjaergaard}}, \bibinfo {author} {\bibfnamefont {M.~E.}\ \bibnamefont {Schwartz}}, \bibinfo {author} {\bibfnamefont {J.}~\bibnamefont {Braumuller}}, \bibinfo {author} {\bibfnamefont {P.}~\bibnamefont {Krantz}}, \bibinfo {author} {\bibfnamefont {J.~I.-J.}\ \bibnamefont {Wang}}, \bibinfo {author} {\bibfnamefont {S.}~\bibnamefont {Gustavsson}},\ and\ \bibinfo {author} {\bibfnamefont {W.~D.}\ \bibnamefont {Oliver}},\ }\bibfield  {title} {\bibinfo {title} {Superconducting qubits: Current state of play},\ }\href {https://doi.org/10.1146/annurev-conmatphys-031119-050605} {\bibfield  {journal} {\bibinfo  {journal} {Annual Review of Condensed Matter Physics}\ }\textbf {\bibinfo {volume} {11}},\ \bibinfo {pages} {369} (\bibinfo {year} {2020})},\ \Eprint {https://arxiv.org/abs/https://doi.org/10.1146/annurev-conmatphys-031119-050605} {https://doi.org/10.1146/annurev-conmatphys-031119-050605} \BibitemShut {NoStop}%
\bibitem [{\citenamefont {Jurcevic}\ \emph {et~al.}(2021)\citenamefont {Jurcevic}, \citenamefont {Javadi-Abhari}, \citenamefont {Bishop}, \citenamefont {Lauer}, \citenamefont {Bogorin}, \citenamefont {Brink}, \citenamefont {Capelluto}, \citenamefont {Günlük}, \citenamefont {Itoko},\ and\ \citenamefont {Kanazawa}}]{ref:Jurcevic_2021}%
  \BibitemOpen
  \bibfield  {author} {\bibinfo {author} {\bibfnamefont {P.}~\bibnamefont {Jurcevic}}, \bibinfo {author} {\bibfnamefont {A.}~\bibnamefont {Javadi-Abhari}}, \bibinfo {author} {\bibfnamefont {L.~S.}\ \bibnamefont {Bishop}}, \bibinfo {author} {\bibfnamefont {I.}~\bibnamefont {Lauer}}, \bibinfo {author} {\bibfnamefont {D.~F.}\ \bibnamefont {Bogorin}}, \bibinfo {author} {\bibfnamefont {M.}~\bibnamefont {Brink}}, \bibinfo {author} {\bibfnamefont {L.}~\bibnamefont {Capelluto}}, \bibinfo {author} {\bibfnamefont {O.}~\bibnamefont {Günlük}}, \bibinfo {author} {\bibfnamefont {T.}~\bibnamefont {Itoko}},\ and\ \bibinfo {author} {\bibfnamefont {N.}~\bibnamefont {Kanazawa}},\ }\bibfield  {title} {\bibinfo {title} {Demonstration of quantum volume 64 on a superconducting quantum computing system},\ }\href {https://doi.org/10.1088/2058-9565/abe519} {\bibfield  {journal} {\bibinfo  {journal} {Quantum Science and Technology}\ }\textbf {\bibinfo {volume} {6}},\ \bibinfo {pages} {025020} (\bibinfo {year} {2021})}\BibitemShut
  {NoStop}%
\bibitem [{\citenamefont {{Rost}}\ \emph {et~al.}(2021)\citenamefont {{Rost}}, \citenamefont {{Del Re}}, \citenamefont {{Earnest}}, \citenamefont {{Kemper}}, \citenamefont {{Jones}},\ and\ \citenamefont {{Freericks}}}]{ref:Rost2021}%
  \BibitemOpen
  \bibfield  {author} {\bibinfo {author} {\bibfnamefont {B.}~\bibnamefont {{Rost}}}, \bibinfo {author} {\bibfnamefont {L.}~\bibnamefont {{Del Re}}}, \bibinfo {author} {\bibfnamefont {N.}~\bibnamefont {{Earnest}}}, \bibinfo {author} {\bibfnamefont {A.~F.}\ \bibnamefont {{Kemper}}}, \bibinfo {author} {\bibfnamefont {B.}~\bibnamefont {{Jones}}},\ and\ \bibinfo {author} {\bibfnamefont {J.~K.}\ \bibnamefont {{Freericks}}},\ }\bibfield  {title} {\bibinfo {title} {{Demonstrating robust simulation of driven-dissipative problems on near-term quantum computers}},\ }\href {https://doi.org/10.48550/arXiv.2108.01183} {\bibfield  {journal} {\bibinfo  {journal} {arXiv}\ ,\ \bibinfo {eid} {arXiv:2108.01183}} (\bibinfo {year} {2021})}\BibitemShut {NoStop}%
\bibitem [{\citenamefont {Kingma}\ and\ \citenamefont {Ba}(2015)}]{ref:Kingma2015}%
  \BibitemOpen
  \bibfield  {author} {\bibinfo {author} {\bibfnamefont {D.~P.}\ \bibnamefont {Kingma}}\ and\ \bibinfo {author} {\bibfnamefont {J.~L.}\ \bibnamefont {Ba}},\ }\bibfield  {title} {\bibinfo {title} {{Adam: A method for stochastic optimization}},\ }\href@noop {} {\bibfield  {journal} {\bibinfo  {journal} {3rd International Conference on Learning Representations, ICLR 2015 - Conference Track Proceedings}\ ,\ \bibinfo {pages} {1}} (\bibinfo {year} {2015})},\ \Eprint {https://arxiv.org/abs/1412.6980} {arXiv:1412.6980} \BibitemShut {NoStop}%
\bibitem [{\citenamefont {Paszke}\ \emph {et~al.}(2019)\citenamefont {Paszke}, \citenamefont {Gross}, \citenamefont {Massa}, \citenamefont {Lerer}, \citenamefont {Bradbury}, \citenamefont {Chanan}, \citenamefont {Killeen}, \citenamefont {Lin}, \citenamefont {Gimelshein}, \citenamefont {Antiga}, \citenamefont {Desmaison}, \citenamefont {Kopf}, \citenamefont {Yang}, \citenamefont {DeVito}, \citenamefont {Raison}, \citenamefont {Tejani}, \citenamefont {Chilamkurthy}, \citenamefont {Steiner}, \citenamefont {Fang}, \citenamefont {Bai},\ and\ \citenamefont {Chintala}}]{ref:NEURIPS2019_9015}%
  \BibitemOpen
  \bibfield  {author} {\bibinfo {author} {\bibfnamefont {A.}~\bibnamefont {Paszke}}, \bibinfo {author} {\bibfnamefont {S.}~\bibnamefont {Gross}}, \bibinfo {author} {\bibfnamefont {F.}~\bibnamefont {Massa}}, \bibinfo {author} {\bibfnamefont {A.}~\bibnamefont {Lerer}}, \bibinfo {author} {\bibfnamefont {J.}~\bibnamefont {Bradbury}}, \bibinfo {author} {\bibfnamefont {G.}~\bibnamefont {Chanan}}, \bibinfo {author} {\bibfnamefont {T.}~\bibnamefont {Killeen}}, \bibinfo {author} {\bibfnamefont {Z.}~\bibnamefont {Lin}}, \bibinfo {author} {\bibfnamefont {N.}~\bibnamefont {Gimelshein}}, \bibinfo {author} {\bibfnamefont {L.}~\bibnamefont {Antiga}}, \bibinfo {author} {\bibfnamefont {A.}~\bibnamefont {Desmaison}}, \bibinfo {author} {\bibfnamefont {A.}~\bibnamefont {Kopf}}, \bibinfo {author} {\bibfnamefont {E.}~\bibnamefont {Yang}}, \bibinfo {author} {\bibfnamefont {Z.}~\bibnamefont {DeVito}}, \bibinfo {author} {\bibfnamefont {M.}~\bibnamefont {Raison}}, \bibinfo {author} {\bibfnamefont {A.}~\bibnamefont {Tejani}}, \bibinfo
  {author} {\bibfnamefont {S.}~\bibnamefont {Chilamkurthy}}, \bibinfo {author} {\bibfnamefont {B.}~\bibnamefont {Steiner}}, \bibinfo {author} {\bibfnamefont {L.}~\bibnamefont {Fang}}, \bibinfo {author} {\bibfnamefont {J.}~\bibnamefont {Bai}},\ and\ \bibinfo {author} {\bibfnamefont {S.}~\bibnamefont {Chintala}},\ }\bibfield  {title} {\bibinfo {title} {Pytorch:an imperative style, high-performance deep learning library},\ }in\ \href {http://papers.neurips.cc/paper/9015-pytorch-an-imperative-style-high-performance-deep-learning-library.pdf} {\emph {\bibinfo {booktitle} {Advances in Neural Information Processing Systems 32}}}\ (\bibinfo  {publisher} {Curran Associates, Inc.},\ \bibinfo {year} {2019})\ pp.\ \bibinfo {pages} {8024--8035}\BibitemShut {NoStop}%
\bibitem [{\citenamefont {Srikanth}\ and\ \citenamefont {Banerjee}(2008)}]{ref:Srikanth2008}%
  \BibitemOpen
  \bibfield  {author} {\bibinfo {author} {\bibfnamefont {R.}~\bibnamefont {Srikanth}}\ and\ \bibinfo {author} {\bibfnamefont {S.}~\bibnamefont {Banerjee}},\ }\bibfield  {title} {\bibinfo {title} {{Squeezed generalized amplitude damping channel}},\ }\bibfield  {journal} {\bibinfo  {journal} {Physical Review A - Atomic, Molecular, and Optical Physics}\ }\textbf {\bibinfo {volume} {77}},\ \href {https://doi.org/10.1103/PhysRevA.77.012318} {10.1103/PhysRevA.77.012318} (\bibinfo {year} {2008})\BibitemShut {NoStop}%
\bibitem [{\citenamefont {Cafaro}\ and\ \citenamefont {{Van Loock}}(2014)}]{ref:Cafaro2014}%
  \BibitemOpen
  \bibfield  {author} {\bibinfo {author} {\bibfnamefont {C.}~\bibnamefont {Cafaro}}\ and\ \bibinfo {author} {\bibfnamefont {P.}~\bibnamefont {{Van Loock}}},\ }\bibfield  {title} {\bibinfo {title} {{Approximate quantum error correction for generalized amplitude-damping errors}},\ }\href {https://doi.org/10.1103/PhysRevA.89.022316} {\bibfield  {journal} {\bibinfo  {journal} {Physical Review A - Atomic, Molecular, and Optical Physics}\ }\textbf {\bibinfo {volume} {89}},\ \bibinfo {pages} {1} (\bibinfo {year} {2014})}\BibitemShut {NoStop}%
\bibitem [{\citenamefont {Krantz}\ \emph {et~al.}(2019)\citenamefont {Krantz}, \citenamefont {Kjaergaard}, \citenamefont {Yan}, \citenamefont {Orlando}, \citenamefont {Gustavsson},\ and\ \citenamefont {Oliver}}]{ref:Krantz2019}%
  \BibitemOpen
  \bibfield  {author} {\bibinfo {author} {\bibfnamefont {P.}~\bibnamefont {Krantz}}, \bibinfo {author} {\bibfnamefont {M.}~\bibnamefont {Kjaergaard}}, \bibinfo {author} {\bibfnamefont {F.}~\bibnamefont {Yan}}, \bibinfo {author} {\bibfnamefont {T.~P.}\ \bibnamefont {Orlando}}, \bibinfo {author} {\bibfnamefont {S.}~\bibnamefont {Gustavsson}},\ and\ \bibinfo {author} {\bibfnamefont {W.~D.}\ \bibnamefont {Oliver}},\ }\bibfield  {title} {\bibinfo {title} {{A quantum engineer's guide to superconducting qubits}},\ }\bibfield  {journal} {\bibinfo  {journal} {Applied Physics Reviews}\ }\textbf {\bibinfo {volume} {6}},\ \href {https://doi.org/10.1063/1.5089550} {10.1063/1.5089550} (\bibinfo {year} {2019})\BibitemShut {NoStop}%
\bibitem [{ref(2022{\natexlab{a}})}]{ref:IBMQ}%
  \BibitemOpen
  \href@noop {} {\bibinfo {title} {Ibm quantum. https://quantum-computing.ibm.com/, 2022}} (\bibinfo {year} {2022}{\natexlab{a}})\BibitemShut {NoStop}%
\bibitem [{\citenamefont {Geller}(2020)}]{ref:Geller_2020}%
  \BibitemOpen
  \bibfield  {author} {\bibinfo {author} {\bibfnamefont {M.~R.}\ \bibnamefont {Geller}},\ }\bibfield  {title} {\bibinfo {title} {Rigorous measurement error correction},\ }\href {https://doi.org/10.1088/2058-9565/ab9591} {\bibfield  {journal} {\bibinfo  {journal} {Quantum Science and Technology}\ }\textbf {\bibinfo {volume} {5}},\ \bibinfo {pages} {03LT01} (\bibinfo {year} {2020})}\BibitemShut {NoStop}%
\bibitem [{\citenamefont {Geller}\ and\ \citenamefont {Sun}(2021)}]{ref:Geller_2021}%
  \BibitemOpen
  \bibfield  {author} {\bibinfo {author} {\bibfnamefont {M.~R.}\ \bibnamefont {Geller}}\ and\ \bibinfo {author} {\bibfnamefont {M.}~\bibnamefont {Sun}},\ }\bibfield  {title} {\bibinfo {title} {Toward efficient correction of multiqubit measurement errors: pair correlation method},\ }\href {https://doi.org/10.1088/2058-9565/abd5c9} {\bibfield  {journal} {\bibinfo  {journal} {Quantum Science and Technology}\ }\textbf {\bibinfo {volume} {6}},\ \bibinfo {pages} {025009} (\bibinfo {year} {2021})}\BibitemShut {NoStop}%
\bibitem [{ref(2022{\natexlab{b}})}]{ref:Qiskit}%
  \BibitemOpen
  \href {https://doi.org/10.5281/zenodo.2573505} {\bibinfo {title} {Qiskit: An open-source framework for quantum computing}} (\bibinfo {year} {2022}{\natexlab{b}})\BibitemShut {NoStop}%
\bibitem [{\citenamefont {Chow}\ \emph {et~al.}(2009)\citenamefont {Chow}, \citenamefont {Gambetta}, \citenamefont {Tornberg}, \citenamefont {Koch}, \citenamefont {Bishop}, \citenamefont {Houck}, \citenamefont {Johnson}, \citenamefont {Frunzio}, \citenamefont {Girvin},\ and\ \citenamefont {Schoelkopf}}]{ref:Chow2009}%
  \BibitemOpen
  \bibfield  {author} {\bibinfo {author} {\bibfnamefont {J.~M.}\ \bibnamefont {Chow}}, \bibinfo {author} {\bibfnamefont {J.~M.}\ \bibnamefont {Gambetta}}, \bibinfo {author} {\bibfnamefont {L.}~\bibnamefont {Tornberg}}, \bibinfo {author} {\bibfnamefont {J.}~\bibnamefont {Koch}}, \bibinfo {author} {\bibfnamefont {L.~S.}\ \bibnamefont {Bishop}}, \bibinfo {author} {\bibfnamefont {A.~A.}\ \bibnamefont {Houck}}, \bibinfo {author} {\bibfnamefont {B.~R.}\ \bibnamefont {Johnson}}, \bibinfo {author} {\bibfnamefont {L.}~\bibnamefont {Frunzio}}, \bibinfo {author} {\bibfnamefont {S.~M.}\ \bibnamefont {Girvin}},\ and\ \bibinfo {author} {\bibfnamefont {R.~J.}\ \bibnamefont {Schoelkopf}},\ }\bibfield  {title} {\bibinfo {title} {{Randomized benchmarking and process tomography for gate errors in a solid-state qubit}},\ }\href {https://doi.org/10.1103/PhysRevLett.102.090502} {\bibfield  {journal} {\bibinfo  {journal} {Physical Review Letters}\ }\textbf {\bibinfo {volume} {102}},\ \bibinfo {pages} {1} (\bibinfo {year}
  {2009})}\BibitemShut {NoStop}%
\bibitem [{\citenamefont {Fogarty}\ \emph {et~al.}(2015)\citenamefont {Fogarty}, \citenamefont {Veldhorst}, \citenamefont {Harper}, \citenamefont {Yang}, \citenamefont {Bartlett}, \citenamefont {Flammia},\ and\ \citenamefont {Dzurak}}]{ref:Fogarty2015}%
  \BibitemOpen
  \bibfield  {author} {\bibinfo {author} {\bibfnamefont {M.~A.}\ \bibnamefont {Fogarty}}, \bibinfo {author} {\bibfnamefont {M.}~\bibnamefont {Veldhorst}}, \bibinfo {author} {\bibfnamefont {R.}~\bibnamefont {Harper}}, \bibinfo {author} {\bibfnamefont {C.~H.}\ \bibnamefont {Yang}}, \bibinfo {author} {\bibfnamefont {S.~D.}\ \bibnamefont {Bartlett}}, \bibinfo {author} {\bibfnamefont {S.~T.}\ \bibnamefont {Flammia}},\ and\ \bibinfo {author} {\bibfnamefont {A.~S.}\ \bibnamefont {Dzurak}},\ }\bibfield  {title} {\bibinfo {title} {{Nonexponential fidelity decay in randomized benchmarking with low-frequency noise}},\ }\href {https://doi.org/10.1103/PhysRevA.92.022326} {\bibfield  {journal} {\bibinfo  {journal} {Physical Review A - Atomic, Molecular, and Optical Physics}\ }\textbf {\bibinfo {volume} {92}},\ \bibinfo {pages} {1} (\bibinfo {year} {2015})}\BibitemShut {NoStop}%
\bibitem [{\citenamefont {Klimov}\ \emph {et~al.}(2018)\citenamefont {Klimov}, \citenamefont {Kelly}, \citenamefont {Chen}, \citenamefont {Neeley}, \citenamefont {Megrant}, \citenamefont {Burkett}, \citenamefont {Barends}, \citenamefont {Arya}, \citenamefont {Chiaro},\ and\ \citenamefont {Chen}}]{ref:Klimov2018}%
  \BibitemOpen
  \bibfield  {author} {\bibinfo {author} {\bibfnamefont {P.~V.}\ \bibnamefont {Klimov}}, \bibinfo {author} {\bibfnamefont {J.}~\bibnamefont {Kelly}}, \bibinfo {author} {\bibfnamefont {Z.}~\bibnamefont {Chen}}, \bibinfo {author} {\bibfnamefont {M.}~\bibnamefont {Neeley}}, \bibinfo {author} {\bibfnamefont {A.}~\bibnamefont {Megrant}}, \bibinfo {author} {\bibfnamefont {B.}~\bibnamefont {Burkett}}, \bibinfo {author} {\bibfnamefont {R.}~\bibnamefont {Barends}}, \bibinfo {author} {\bibfnamefont {K.}~\bibnamefont {Arya}}, \bibinfo {author} {\bibfnamefont {B.}~\bibnamefont {Chiaro}},\ and\ \bibinfo {author} {\bibfnamefont {Y.}~\bibnamefont {Chen}},\ }\bibfield  {title} {\bibinfo {title} {{Fluctuations of Energy-Relaxation Times in Superconducting Qubits}},\ }\href {https://doi.org/10.1103/PhysRevLett.121.090502} {\bibfield  {journal} {\bibinfo  {journal} {Physical Review Letters}\ }\textbf {\bibinfo {volume} {121}},\ \bibinfo {pages} {90502} (\bibinfo {year} {2018})}\BibitemShut {NoStop}%
\bibitem [{\citenamefont {Verstraete}\ \emph {et~al.}(2009)\citenamefont {Verstraete}, \citenamefont {Wolf},\ and\ \citenamefont {Ignacio~Cirac}}]{ref:Verstraete2009}%
  \BibitemOpen
  \bibfield  {author} {\bibinfo {author} {\bibfnamefont {F.}~\bibnamefont {Verstraete}}, \bibinfo {author} {\bibfnamefont {M.~M.}\ \bibnamefont {Wolf}},\ and\ \bibinfo {author} {\bibfnamefont {J.}~\bibnamefont {Ignacio~Cirac}},\ }\bibfield  {title} {\bibinfo {title} {Quantum computation and quantum-state engineering driven by dissipation},\ }\href {https://doi.org/10.1038/nphys1342} {\bibfield  {journal} {\bibinfo  {journal} {Nature Physics}\ }\textbf {\bibinfo {volume} {5}},\ \bibinfo {pages} {633} (\bibinfo {year} {2009})}\BibitemShut {NoStop}%
\bibitem [{\citenamefont {{Aharonov}}\ \emph {et~al.}(2022)\citenamefont {{Aharonov}}, \citenamefont {{Gao}}, \citenamefont {{Landau}}, \citenamefont {{Liu}},\ and\ \citenamefont {{Vazirani}}}]{ref:Aharonov2022}%
  \BibitemOpen
  \bibfield  {author} {\bibinfo {author} {\bibfnamefont {D.}~\bibnamefont {{Aharonov}}}, \bibinfo {author} {\bibfnamefont {X.}~\bibnamefont {{Gao}}}, \bibinfo {author} {\bibfnamefont {Z.}~\bibnamefont {{Landau}}}, \bibinfo {author} {\bibfnamefont {Y.}~\bibnamefont {{Liu}}},\ and\ \bibinfo {author} {\bibfnamefont {U.}~\bibnamefont {{Vazirani}}},\ }\bibfield  {title} {\bibinfo {title} {{A polynomial-time classical algorithm for noisy random circuit sampling}},\ }\href {https://doi.org/10.48550/arXiv.2211.03999} {\bibfield  {journal} {\bibinfo  {journal} {arXiv}\ ,\ \bibinfo {eid} {arXiv:2211.03999}} (\bibinfo {year} {2022})}\BibitemShut {NoStop}%
\bibitem [{\citenamefont {Meyer}(2021)}]{ref:Meyer2021}%
  \BibitemOpen
  \bibfield  {author} {\bibinfo {author} {\bibfnamefont {J.~J.}\ \bibnamefont {Meyer}},\ }\bibfield  {title} {\bibinfo {title} {Fisher {I}nformation in {N}oisy {I}ntermediate-{S}cale {Q}uantum {A}pplications},\ }\href {https://doi.org/10.22331/q-2021-09-09-539} {\bibfield  {journal} {\bibinfo  {journal} {{Quantum}}\ }\textbf {\bibinfo {volume} {5}},\ \bibinfo {pages} {539} (\bibinfo {year} {2021})}\BibitemShut {NoStop}%
\bibitem [{\citenamefont {{Li}}\ \emph {et~al.}(2022)\citenamefont {{Li}}, \citenamefont {{Zou}}, \citenamefont {{Glorioso}}, \citenamefont {{Altman}},\ and\ \citenamefont {{Fisher}}}]{ref:Li2022}%
  \BibitemOpen
  \bibfield  {author} {\bibinfo {author} {\bibfnamefont {Y.}~\bibnamefont {{Li}}}, \bibinfo {author} {\bibfnamefont {Y.}~\bibnamefont {{Zou}}}, \bibinfo {author} {\bibfnamefont {P.}~\bibnamefont {{Glorioso}}}, \bibinfo {author} {\bibfnamefont {E.}~\bibnamefont {{Altman}}},\ and\ \bibinfo {author} {\bibfnamefont {M.~P.~A.}\ \bibnamefont {{Fisher}}},\ }\bibfield  {title} {\bibinfo {title} {{Cross Entropy Benchmark for Measurement-Induced Phase Transitions}},\ }\href {https://doi.org/10.48550/arXiv.2209.00609} {\bibfield  {journal} {\bibinfo  {journal} {arXiv}\ ,\ \bibinfo {eid} {arXiv:2209.00609}} (\bibinfo {year} {2022})}\BibitemShut {NoStop}%
\bibitem [{\citenamefont {Emerson}\ \emph {et~al.}(2005)\citenamefont {Emerson}, \citenamefont {Alicki},\ and\ \citenamefont {Życzkowski}}]{ref:Emerson2005}%
  \BibitemOpen
  \bibfield  {author} {\bibinfo {author} {\bibfnamefont {J.}~\bibnamefont {Emerson}}, \bibinfo {author} {\bibfnamefont {R.}~\bibnamefont {Alicki}},\ and\ \bibinfo {author} {\bibfnamefont {K.}~\bibnamefont {Życzkowski}},\ }\bibfield  {title} {\bibinfo {title} {Scalable noise estimation with random unitary operators},\ }\href {https://doi.org/10.1088/1464-4266/7/10/021} {\bibfield  {journal} {\bibinfo  {journal} {Journal of Optics B: Quantum and Semiclassical Optics}\ }\textbf {\bibinfo {volume} {7}},\ \bibinfo {pages} {S347} (\bibinfo {year} {2005})}\BibitemShut {NoStop}%
\bibitem [{\citenamefont {Emerson}\ \emph {et~al.}(2007)\citenamefont {Emerson}, \citenamefont {Silva}, \citenamefont {Moussa}, \citenamefont {Ryan}, \citenamefont {Laforest}, \citenamefont {Baugh}, \citenamefont {Cory},\ and\ \citenamefont {Laflamme}}]{ref:Emerson2007}%
  \BibitemOpen
  \bibfield  {author} {\bibinfo {author} {\bibfnamefont {J.}~\bibnamefont {Emerson}}, \bibinfo {author} {\bibfnamefont {M.}~\bibnamefont {Silva}}, \bibinfo {author} {\bibfnamefont {O.}~\bibnamefont {Moussa}}, \bibinfo {author} {\bibfnamefont {C.}~\bibnamefont {Ryan}}, \bibinfo {author} {\bibfnamefont {M.}~\bibnamefont {Laforest}}, \bibinfo {author} {\bibfnamefont {J.}~\bibnamefont {Baugh}}, \bibinfo {author} {\bibfnamefont {D.~G.}\ \bibnamefont {Cory}},\ and\ \bibinfo {author} {\bibfnamefont {R.}~\bibnamefont {Laflamme}},\ }\bibfield  {title} {\bibinfo {title} {Symmetrized characterization of noisy quantum processes},\ }\href {https://doi.org/10.1126/science.1145699} {\bibfield  {journal} {\bibinfo  {journal} {Science}\ }\textbf {\bibinfo {volume} {317}},\ \bibinfo {pages} {1893} (\bibinfo {year} {2007})},\ \Eprint {https://arxiv.org/abs/https://www.science.org/doi/pdf/10.1126/science.1145699} {https://www.science.org/doi/pdf/10.1126/science.1145699} \BibitemShut {NoStop}%
\bibitem [{\citenamefont {Knill}\ \emph {et~al.}(2008)\citenamefont {Knill}, \citenamefont {Leibfried}, \citenamefont {Reichle}, \citenamefont {Britton}, \citenamefont {Blakestad}, \citenamefont {Jost}, \citenamefont {Langer}, \citenamefont {Ozeri}, \citenamefont {Seidelin},\ and\ \citenamefont {Wineland}}]{ref:Knill2008}%
  \BibitemOpen
  \bibfield  {author} {\bibinfo {author} {\bibfnamefont {E.}~\bibnamefont {Knill}}, \bibinfo {author} {\bibfnamefont {D.}~\bibnamefont {Leibfried}}, \bibinfo {author} {\bibfnamefont {R.}~\bibnamefont {Reichle}}, \bibinfo {author} {\bibfnamefont {J.}~\bibnamefont {Britton}}, \bibinfo {author} {\bibfnamefont {R.~B.}\ \bibnamefont {Blakestad}}, \bibinfo {author} {\bibfnamefont {J.~D.}\ \bibnamefont {Jost}}, \bibinfo {author} {\bibfnamefont {C.}~\bibnamefont {Langer}}, \bibinfo {author} {\bibfnamefont {R.}~\bibnamefont {Ozeri}}, \bibinfo {author} {\bibfnamefont {S.}~\bibnamefont {Seidelin}},\ and\ \bibinfo {author} {\bibfnamefont {D.~J.}\ \bibnamefont {Wineland}},\ }\bibfield  {title} {\bibinfo {title} {Randomized benchmarking of quantum gates},\ }\href {https://doi.org/10.1103/PhysRevA.77.012307} {\bibfield  {journal} {\bibinfo  {journal} {Phys. Rev. A}\ }\textbf {\bibinfo {volume} {77}},\ \bibinfo {pages} {012307} (\bibinfo {year} {2008})}\BibitemShut {NoStop}%
\bibitem [{\citenamefont {Magesan}\ \emph {et~al.}(2012)\citenamefont {Magesan}, \citenamefont {Gambetta},\ and\ \citenamefont {Emerson}}]{ref:Magesan2012}%
  \BibitemOpen
  \bibfield  {author} {\bibinfo {author} {\bibfnamefont {E.}~\bibnamefont {Magesan}}, \bibinfo {author} {\bibfnamefont {J.~M.}\ \bibnamefont {Gambetta}},\ and\ \bibinfo {author} {\bibfnamefont {J.}~\bibnamefont {Emerson}},\ }\bibfield  {title} {\bibinfo {title} {Characterizing quantum gates via randomized benchmarking},\ }\href {https://doi.org/10.1103/PhysRevA.85.042311} {\bibfield  {journal} {\bibinfo  {journal} {Phys. Rev. A}\ }\textbf {\bibinfo {volume} {85}},\ \bibinfo {pages} {042311} (\bibinfo {year} {2012})}\BibitemShut {NoStop}%
\bibitem [{\citenamefont {Blume-Kohout}\ \emph {et~al.}(2013)\citenamefont {Blume-Kohout}, \citenamefont {Gamble}, \citenamefont {Nielsen}, \citenamefont {Mizrahi}, \citenamefont {Sterk},\ and\ \citenamefont {Maunz}}]{ref:BlumeKohout2013}%
  \BibitemOpen
  \bibfield  {author} {\bibinfo {author} {\bibfnamefont {R.}~\bibnamefont {Blume-Kohout}}, \bibinfo {author} {\bibfnamefont {J.~K.}\ \bibnamefont {Gamble}}, \bibinfo {author} {\bibfnamefont {E.}~\bibnamefont {Nielsen}}, \bibinfo {author} {\bibfnamefont {J.}~\bibnamefont {Mizrahi}}, \bibinfo {author} {\bibfnamefont {J.~D.}\ \bibnamefont {Sterk}},\ and\ \bibinfo {author} {\bibfnamefont {P.}~\bibnamefont {Maunz}},\ }\bibfield  {title} {\bibinfo {title} {Robust, self-consistent, closed-form tomography of quantum logic gates on a trapped ion qubit},\ }\bibfield  {journal} {\bibinfo  {journal} {arXiv}\ }\href {https://doi.org/10.48550/ARXIV.1310.4492} {10.48550/ARXIV.1310.4492} (\bibinfo {year} {2013})\BibitemShut {NoStop}%
\bibitem [{\citenamefont {Kim}\ \emph {et~al.}(2015)\citenamefont {Kim}, \citenamefont {Ward}, \citenamefont {Simmons}, \citenamefont {Gamble}, \citenamefont {Blume-Kohout}, \citenamefont {Nielsen}, \citenamefont {Savage}, \citenamefont {Lagally}, \citenamefont {Friesen}, \citenamefont {Coppersmith},\ and\ \citenamefont {Eriksson}}]{ref:Kim2015}%
  \BibitemOpen
  \bibfield  {author} {\bibinfo {author} {\bibfnamefont {D.}~\bibnamefont {Kim}}, \bibinfo {author} {\bibfnamefont {D.~R.}\ \bibnamefont {Ward}}, \bibinfo {author} {\bibfnamefont {C.~B.}\ \bibnamefont {Simmons}}, \bibinfo {author} {\bibfnamefont {J.~K.}\ \bibnamefont {Gamble}}, \bibinfo {author} {\bibfnamefont {R.}~\bibnamefont {Blume-Kohout}}, \bibinfo {author} {\bibfnamefont {E.}~\bibnamefont {Nielsen}}, \bibinfo {author} {\bibfnamefont {D.~E.}\ \bibnamefont {Savage}}, \bibinfo {author} {\bibfnamefont {M.~G.}\ \bibnamefont {Lagally}}, \bibinfo {author} {\bibfnamefont {M.}~\bibnamefont {Friesen}}, \bibinfo {author} {\bibfnamefont {S.~N.}\ \bibnamefont {Coppersmith}},\ and\ \bibinfo {author} {\bibfnamefont {M.~A.}\ \bibnamefont {Eriksson}},\ }\bibfield  {title} {\bibinfo {title} {Microwave-driven coherent operation of a semiconductor quantum dot charge qubit},\ }\href {https://doi.org/10.1038/nnano.2014.336} {\bibfield  {journal} {\bibinfo  {journal} {Nature Nanotechnology}\ }\textbf {\bibinfo {volume}
  {10}},\ \bibinfo {pages} {243} (\bibinfo {year} {2015})}\BibitemShut {NoStop}%
\bibitem [{\citenamefont {Nielsen}\ \emph {et~al.}(2021)\citenamefont {Nielsen}, \citenamefont {Gamble}, \citenamefont {Rudinger}, \citenamefont {Scholten}, \citenamefont {Young},\ and\ \citenamefont {Blume-Kohout}}]{ref:Nielsen2021}%
  \BibitemOpen
  \bibfield  {author} {\bibinfo {author} {\bibfnamefont {E.}~\bibnamefont {Nielsen}}, \bibinfo {author} {\bibfnamefont {J.~K.}\ \bibnamefont {Gamble}}, \bibinfo {author} {\bibfnamefont {K.}~\bibnamefont {Rudinger}}, \bibinfo {author} {\bibfnamefont {T.}~\bibnamefont {Scholten}}, \bibinfo {author} {\bibfnamefont {K.}~\bibnamefont {Young}},\ and\ \bibinfo {author} {\bibfnamefont {R.}~\bibnamefont {Blume-Kohout}},\ }\bibfield  {title} {\bibinfo {title} {Gate {S}et {T}omography},\ }\href {https://doi.org/10.22331/q-2021-10-05-557} {\bibfield  {journal} {\bibinfo  {journal} {{Quantum}}\ }\textbf {\bibinfo {volume} {5}},\ \bibinfo {pages} {557} (\bibinfo {year} {2021})}\BibitemShut {NoStop}%
\bibitem [{\citenamefont {Steffen}\ \emph {et~al.}(2011)\citenamefont {Steffen}, \citenamefont {DiVincenzo}, \citenamefont {Chow}, \citenamefont {Theis},\ and\ \citenamefont {Ketchen}}]{ref:SteffenIBM2011}%
  \BibitemOpen
  \bibfield  {author} {\bibinfo {author} {\bibfnamefont {M.}~\bibnamefont {Steffen}}, \bibinfo {author} {\bibfnamefont {D.~P.}\ \bibnamefont {DiVincenzo}}, \bibinfo {author} {\bibfnamefont {J.~M.}\ \bibnamefont {Chow}}, \bibinfo {author} {\bibfnamefont {T.~N.}\ \bibnamefont {Theis}},\ and\ \bibinfo {author} {\bibfnamefont {M.~B.}\ \bibnamefont {Ketchen}},\ }\bibfield  {title} {\bibinfo {title} {Quantum computing: An ibm perspective},\ }\href {https://doi.org/10.1147/JRD.2011.2165678} {\bibfield  {journal} {\bibinfo  {journal} {IBM Journal of Research and Development}\ }\textbf {\bibinfo {volume} {55}},\ \bibinfo {pages} {13:1} (\bibinfo {year} {2011})}\BibitemShut {NoStop}%
\bibitem [{\citenamefont {Rigetti}\ and\ \citenamefont {Devoret}(2010)}]{ref:Rigetti2010}%
  \BibitemOpen
  \bibfield  {author} {\bibinfo {author} {\bibfnamefont {C.}~\bibnamefont {Rigetti}}\ and\ \bibinfo {author} {\bibfnamefont {M.}~\bibnamefont {Devoret}},\ }\bibfield  {title} {\bibinfo {title} {{Fully microwave-tunable universal gates in superconducting qubits with linear couplings and fixed transition frequencies}},\ }\href {https://doi.org/10.1103/PhysRevB.81.134507} {\bibfield  {journal} {\bibinfo  {journal} {Physical Review B - Condensed Matter and Materials Physics}\ }\textbf {\bibinfo {volume} {81}},\ \bibinfo {pages} {1} (\bibinfo {year} {2010})}\BibitemShut {NoStop}%
\bibitem [{\citenamefont {Malekakhlagh}\ \emph {et~al.}(2020)\citenamefont {Malekakhlagh}, \citenamefont {Magesan},\ and\ \citenamefont {McKay}}]{ref:Malekakhlagh2020}%
  \BibitemOpen
  \bibfield  {author} {\bibinfo {author} {\bibfnamefont {M.}~\bibnamefont {Malekakhlagh}}, \bibinfo {author} {\bibfnamefont {E.}~\bibnamefont {Magesan}},\ and\ \bibinfo {author} {\bibfnamefont {D.~C.}\ \bibnamefont {McKay}},\ }\bibfield  {title} {\bibinfo {title} {{First-principles analysis of cross-resonance gate operation}},\ }\href {https://doi.org/10.1103/PhysRevA.102.042605} {\bibfield  {journal} {\bibinfo  {journal} {Physical Review A}\ }\textbf {\bibinfo {volume} {102}},\ \bibinfo {pages} {1} (\bibinfo {year} {2020})}\BibitemShut {NoStop}%
\bibitem [{\citenamefont {Magesan}\ and\ \citenamefont {Gambetta}(2020)}]{ref:Magesan2020}%
  \BibitemOpen
  \bibfield  {author} {\bibinfo {author} {\bibfnamefont {E.}~\bibnamefont {Magesan}}\ and\ \bibinfo {author} {\bibfnamefont {J.~M.}\ \bibnamefont {Gambetta}},\ }\bibfield  {title} {\bibinfo {title} {{Effective Hamiltonian models of the cross-resonance gate}},\ }\bibfield  {journal} {\bibinfo  {journal} {Physical Review A}\ }\textbf {\bibinfo {volume} {101}},\ \href {https://doi.org/10.1103/PhysRevA.101.052308} {10.1103/PhysRevA.101.052308} (\bibinfo {year} {2020})\BibitemShut {NoStop}%
\bibitem [{\citenamefont {Kandala}\ \emph {et~al.}(2021)\citenamefont {Kandala}, \citenamefont {Wei}, \citenamefont {Srinivasan}, \citenamefont {Magesan}, \citenamefont {Carnevale}, \citenamefont {Keefe}, \citenamefont {Klaus}, \citenamefont {Dial},\ and\ \citenamefont {McKay}}]{ref:Kandala2021}%
  \BibitemOpen
  \bibfield  {author} {\bibinfo {author} {\bibfnamefont {A.}~\bibnamefont {Kandala}}, \bibinfo {author} {\bibfnamefont {K.~X.}\ \bibnamefont {Wei}}, \bibinfo {author} {\bibfnamefont {S.}~\bibnamefont {Srinivasan}}, \bibinfo {author} {\bibfnamefont {E.}~\bibnamefont {Magesan}}, \bibinfo {author} {\bibfnamefont {S.}~\bibnamefont {Carnevale}}, \bibinfo {author} {\bibfnamefont {G.~A.}\ \bibnamefont {Keefe}}, \bibinfo {author} {\bibfnamefont {D.}~\bibnamefont {Klaus}}, \bibinfo {author} {\bibfnamefont {O.}~\bibnamefont {Dial}},\ and\ \bibinfo {author} {\bibfnamefont {D.~C.}\ \bibnamefont {McKay}},\ }\bibfield  {title} {\bibinfo {title} {{Demonstration of a High-Fidelity cnot Gate for Fixed-Frequency Transmons with Engineered Suppression}},\ }\href {https://doi.org/10.1103/PhysRevLett.127.130501} {\bibfield  {journal} {\bibinfo  {journal} {Physical Review Letters}\ }\textbf {\bibinfo {volume} {127}},\ \bibinfo {pages} {130501} (\bibinfo {year} {2021})}\BibitemShut {NoStop}%
\bibitem [{\citenamefont {James}\ \emph {et~al.}(2013)\citenamefont {James}, \citenamefont {Witten}, \citenamefont {Hastie}, \citenamefont {Tibshirani} \emph {et~al.}}]{ref:James2013-Stat-intro}%
  \BibitemOpen
  \bibfield  {author} {\bibinfo {author} {\bibfnamefont {G.}~\bibnamefont {James}}, \bibinfo {author} {\bibfnamefont {D.}~\bibnamefont {Witten}}, \bibinfo {author} {\bibfnamefont {T.}~\bibnamefont {Hastie}}, \bibinfo {author} {\bibfnamefont {R.}~\bibnamefont {Tibshirani}}, \emph {et~al.},\ }\href@noop {} {\emph {\bibinfo {title} {An introduction to statistical learning}}},\ Vol.\ \bibinfo {volume} {112}\ (\bibinfo  {publisher} {Springer},\ \bibinfo {year} {2013})\BibitemShut {NoStop}%
\end{thebibliography}%

\appendix

\newpage 

{~}


\section{Modeling the noisy CX}
\label{app-sec:CX-details}

We modeled the noisy $\CX$ gate to approximate the actual $\CX$ gate
available on the IBMQ hardware, which is based on the supercoducting
qubits\cc{ref:SteffenIBM2011}. In IBMQ machines, a two-qubit $\CX(i
\to j)$ gate is realized as a microwave-activated two-qubit
cross-resonance ($\CR$) gate that consists of driving the control
$i$ qubit at the frequency of the target qubit $j$\cc{ref:Rigetti2010}. 

Based on the discussions in \cRefs{ref:Malekakhlagh2020, ref:Magesan2020,
ref:Kandala2021}, we modeled the effective Hamiltonian 
$\CX(i\to j)$ by:
\begin{align}
\label{eq:cx-parameterized}
  H_{\CX(i\to j)} \EqDef \theta^{(ij)}_{1} Z_i \otimes X_{j} 
    - \theta^{(ij)}_{2} \Id_i \otimes X_{j} 
    - \theta^{(ij)}_{3} Z_i \otimes \Id_{j},
\end{align}
where $\theta^{(ij)}_{m=1,2,3}$ are
variational parameters that represent the strength for each of the
interaction terms. The ideal value for all these parameters is $1$,
which leads to a $\CX$ gate when using dimensionless execution time
$T/T_0 = \pi / 4$. In reality, these parameters can be different
from $1$, leading to coherent $\CX$ errors.  We refer the reader to
the \cRefs{ref:Rigetti2010, ref:Malekakhlagh2020, ref:Magesan2020,
ref:Kandala2021} for more details on the $\CR$ gate implementation on
transmons.

\section{Details of the stochastic maps}
\label{app-sec:stochastic-channel-details}

\subsection{Gates and probabilities}
\label{app-sec:stochastic-probabilities} 

As discussed in \Sec{sec:stochastic-numerics}, our stochastic
maps were defined using the gates $\RESCX$, $X^{1/2}$, $H$,
$R_\Balpha(\phi)$. We used 3 stochastic maps, defined by 3
probability distributions. In each map, the probability of acting
with certain gate was independent of the qubit number. The 3 probability
distributions are given in the following table. The parameters of
the single-qubit rotations $R_\Balpha(\phi)$ were chosen randomly.

\begin{table}[h]
    \centering
    \begin{tabular}{||c c c c c||} 
    \hline
    Gate & $RESCX$ & $X^{1/2}$ & $H$ & $R_\Balpha(\phi)$ \\ [0.5ex] 
    \hline\hline
    Probability set 1 & 0.3 & 0.2 & 0.2 & 0.2 \\ 
    Probability set 2 & 0.5 & 0.1 & 0.2 & 0.2 \\
    Probability set 3 & 0.7 & 0.1 & 0.1 & 0.1 \\ [1ex] 
    \hline
    \end{tabular}
    \caption{Three different probability distributions $\{p_k\}$
    used in the numerical simulations. The rotation axis
    $\Balpha$ and the rotation angle $\psi$ were defined by
    three rotation angles $\{\phi_i\}_{i=1}^{3}$ which we randomly
    generated.} \label{eq:kraus-probs}
\end{table}

\subsection{Noise model in the stochastic case}
\label{app-sec:stochastic-noise-model}

In the stochastic case, we used 4 parameters for each qubit to model
its noise. We used $\theta_z, \theta_{amp}$ to model dephasing and
amplitude damping, which are given by the Lindbladians
\begin{align}
\label{eq:lindblad-stochastic}
    \mcL_z(\rho) &\EqDef Z\rho Z - \rho, \\
    \mcL_{amp}(\rho) &\EqDef \sigma_+ \rho \sigma_- -
    \frac{1}{2}\{\sigma_-\sigma_+,\rho\},
\end{align}
where
\begin{align}
\label{def:sigma-pm}
  \sigma_{\pm} &\EqDef (X \pm i Y) / 2.
\end{align}
In addition, we used $\theta_0,\theta_1$ to model the noisy RESET
gate, as defined in \Eq{eq:numerics-reset-noise}.

We note that in all the simulations of the stochastic maps, we
assumed that the $\CX$ gate had no coherent errors; all its errors
came from the single-qubit errors of its two-qubits.

For every qubit, the dimensionless noise parameters $\theta_z,
\theta_{amp}$ that we used in our simulations were randomly picked
between the values of $0.01$ and $0.07$. In terms of the coherence
times (given by $T_0 / \{\theta_m\}$), our choices correspond
approximately to a range of $20\,[\mu sec]$ to $150\,[\mu sec]$.

The parameters for the noisy $\RESET$ were randomly generated
between the values $0.87$ and $0.98$. Following the discussion in
\cRef{ref:Rost2021}, we required $\theta_0\geq\theta_1$ to match the
typical values on the IBMQ hardware. However, we note that the
experimentally determined noise parameters of the $\RESET$ gates did
not follow this assumption.

\subsection{Gate durations}
\label{app-sec:stochastic-durations}

The noisy versions of all single-qubit unitary gates were simulated
using Suzuki-Trotter expansion, as outlined in
\Eq{eq:numerics-unitary-trotter}. For the noisy version of the $\CX$
we used \Eq{eq:numerics-unitary-gate-noise} with the Hamiltonian
defined in \App{app-sec:CX-details} with vanishing coherent errors.

To define the gate channels in the stochastic map, we used the
execution times that resemble the execution times of actual quantum
gates in superconducting qubits\cc{ref:Kjaergaard2020, ref:Jurcevic_2021}.
For the $\RESET$ gates, duration times were generated between $0.8\,[\mu sec]$
and $1.1\,[\mu sec]$. For the $\CX$ gates, durations were
generated between $0.3\,[\mu sec]$ and $0.45\,[\mu sec]$. The duration time
of the $X^{1/2}$ gate was set to $15\,[nsec]$, and for the $H$
gate the duration was set to $20\,[nsec]$. As stated in
\Sec{sec:stochastic-numerics}, the $R_\Balpha(\psi)$ gate
is a product of two $X^{1/2}$ and three $R_{\bm{z}}(\phi_i)$ gates. The
$R_{\bm{z}}(\phi_i)$ gates are implemented on the superconducting
platforms, such as IBMQ, with zero error and duration
\cc{ref:Kjaergaard2020}. Therefore, we took the duration of the
general single-qubit rotation gate $R_\Balpha(\psi)$, defined
in \Eq{eq:rotation-gate-parameterized}, to be twice the duration of
the $X^{1/2}$ gate.

In our simulations, the time scale $T_0$ was dictated by 
duration of the $\RESCX$ gate, and was roughly $1.5\,[\mu sec]$.

\section{Details of the deterministic map}
\label{app-sec:deterministic-map-details}

\subsection{Rotation angles in the RESU gates}
\label{app-sec:deterministic-resu-angles}

We used the deterministic map \emph{both} in numerical simulations
and real quantum hardware. Consequently, it was important that the
maps we defined will converge to the fixed point in at most 10
steps, which was the maximal number of steps that we could run of
these types of circuits on the IBMQ hardware. To that aim, we first
ran few numerical simulations to find a combination of rotation
angles that will guarantee fast convergence. 

Recall from \Fig{fig:determinist-example} that the $\RESU$ gate
is defined by 4 single-qubit rotations, which amounts to 12 rotation
parameters per $\RESU$ gate (3 parameters for every single-qubit
rotation). To facilitate the search for optimal rotation angles
(which will give rapid convergence), we reduced this number by half
by using the same parameters in the two single-qubit rotations in
$U_1$ (the 2-local unitary before the $\RESET$ gate), and similarly
in $U_2$ (the 2-local unitary after the $\RESET$ gate). In other
words, we set $R_\Balpha=R_{\bm{\beta}}$ and $R_{\bm{\delta}}
=R_{\bm{\gamma}}$. It's worth mentioning that even when each $\RESU$ gate utilizes 12 rotation parameters, convergence is still achieved within a few tens of steps. However, due to classical hardware constraints on IBMQ machines, we were compelled to approximate the steady state within a maximum of 10 steps.

The parameters we used for map-1,2,3 are presented in 
\autoref{table:deterministic-angles-even} and
\autoref{table:deterministic-angles-odd} for the even and odd
layers, respectively.

\begin{table}[h]
    \centering
    \begin{tabular}{||c c c c||} 
    \hline
                    & map-1 & map-2 & map-3 \\ [0.5ex] 
    \hline\hline
    $U_1$, $\phi_1$ & -2.63 & 2.13 & 0.77 \\ 
    $U_1$, $\phi_2$ & 1.71 & -1.31 & -2.46 \\
    $U_1$, $\phi_3$ & 0.21 & -0.71 & 2.59 \\ 
    $U_2$, $\phi_1$ & -1.07 & -2.07 & -0.79\\ 
    $U_2$, $\phi_2$ & -2.12 & -1.12 & 1.09 \\ 
    $U_2$, $\phi_3$ & 2.73 & -2.73 & -1.80\\ [1ex]
    \hline
    \end{tabular}
    \caption{Rotation angles used to define the $\RESU$ gates in the
    three simulated deterministic maps, for the even layer. The
    angles are given in radians.}
    \label{table:deterministic-angles-even}
\end{table}

\begin{table}[h]
    \centering
    \begin{tabular}{||c c c c||} 
    \hline
                    & map-1 & map-2 & map-3 \\ [0.5ex] 
    \hline\hline
    $U_1$, $\phi_1$ & 1.42 & -2.32 & 2.57 \\ 
    $U_1$, $\phi_2$ & -2.92 & 1.92 & 3.04 \\
    $U_1$, $\phi_3$ & -0.33 & -0.73 & 2.67 \\ 
    $U_2$, $\phi_1$ & 2.83 & -2.83 & -2.72\\ 
    $U_2$, $\phi_2$ & 1.36 & 1.87 & -1.79 \\ 
    $U_2$, $\phi_3$ & -1.08 & -1.98 & 0.21\\ [1ex]
    \hline
    \end{tabular}
    \caption{Rotation angles used to define the $\RESU$ gates in the
    three simulated deterministic maps, for the odd layer. The
    angles are given in radians.}
    \label{table:deterministic-angles-odd}
\end{table}

\subsection{Noise model in the deterministic case}
\label{app-sec:deterministic-channel-noise-details}

In the deterministic case, we used a more expressive noise model
than was used in the stochastic case, since we wanted the same model
to describe also the real quantum hardware experiments of
\Sec{sec:ibmq-implementation}. For each qubit, we used 7 noise
parameters: $\theta_x, \theta_y, \theta_z$, to model 
depolarization in the $\hat{x},\hat{y},\hat{z}$ directions,
$\theta_{amp}, \theta_{ex}$ to model general amplitude damping
noise, and $\theta_0,\theta_1$ to model the noisy RESET gate (as was
done in the stochastic case).

The depolarization noise channels in the different directions were
modeled by the Lindbladians
\begin{align}
\label{eq:XYZ-dephasing}
    \mcL_\sigma(\rho) &\EqDef \sigma\rho \sigma - \rho, &
    \sigma &\in \{X,Y, Z\}, 
\end{align}

For the generalized amplitude damping channel\cc{ref:Srikanth2008, ref:Cafaro2014}, we used the Lindbladian
\begin{align*}
  \theta_{amp}\big[ (1-\theta_{ex})\mcL_{amp,+} +
  \theta_{ex}\mcL_{amp,-}\big] ,
\end{align*}
where 
\begin{align}
  \mcL_{amp,\pm}(\rho) \EqDef \sigma_\pm \rho \sigma_\mp
   - \frac{1}{2}\{\sigma_\mp\sigma_\pm, \rho\} .
\end{align}
Physically, the $\theta_{amp}$ models the strength of the general 
amplitude damping process, and $\theta_{ex}$ models the occupation
of the qubit's state $\ket{1}$ at thermal equilibrium. The simple
amplitude damping model is described by $\mcL_{amp,+}$, which
corresponds $\theta_{ex}=0$, where the equilibrium state of the
qubit is assumed to be $\ketbra{0}{0}$ (which correspond to a qubit
at zero temperature).

In the numerical simulations we chose the dimensionless noise 
parameters $\{\theta_x,\theta_y,\theta_z,\theta_{amp},\theta_{ex}\}$
randomly between the values of $0.01$ and $0.1$. The
$\theta_0,\theta_1$ parameters for the noisy $\RESET$ were randomly
generated between the values $0.87$ and $0.98$, as in the stochastic
case.

In our numerical simulations we have used several randomly generated
gate durations using the same bounds as for the stochastic map,
given in \App{app-sec:stochastic-durations}. The resulting time scale
$T_0$ was approximately $2\,[\mu sec]$.

\section{Details of the quantum device noise model characterization
results} \label{app-sec:ibmq-noise-model-params}

\subsection{Details of the readout error mitigation scheme for 
  k-local Paulis}
\label{app-sec:ibmq-readout-scheme}

For quantum readout error mitigation (QREM), we followed the
approach of \cRef{ref:Geller_2020}, but adapted it to $k$-local readout
statistics obtained using the overlapping local tomography from
\cRef{ref:Zubida2021}. The method empirically evaluates the conditional
probability distribution $P(\ux|\uy)$ of measuring
$\ux=(x_1,\ldots,x_k)$ classical bit string in standard basis after
the system is prepared the quantum state $\ketbra{\uy}{\uy}$ for
$\uy=(y_1, \ldots, y_k)$. In our case, we used $k=4$ for learning
the noise channel (part (ii) in
\Sec{sec:ibmq-measurement-protocol}) and $k=2$ for testing our
characterization (part (iii) in
\Sec{sec:ibmq-measurement-protocol}). In the ideal case,
$P(\cdot|\cdot)$ is a Kronecker delta-function. Upon obtaining the
$P(\cdot|\cdot)$, we used it to `fix' an empirical probability
$P_{noisy}(\ux)$ of measuring $\ux$ in the computational basis, as
follows. Assuming that $P_{noisy}(\ux) = \sum_{\uy} P(\ux|\uy)\cdot
P_{ideal}(\uy) $, we extracted $P_{ideal}(\ux)$ by
\begin{align*}
  P_{ideal}(\ux) = \sum_{\uy} P^{-1}(\ux|\uy)\cdot P_{noisy}(\uy) ,
\end{align*}
where $P^{-1}(\ux|\uy)$ is the matrix inverse of $P(\ux|\uy)$.

The readout error mitigation process precedes the numerical optimization, as outlined in \Sec{sec:optim-routine}, where a matrix $P(\ux|\uy)$ with $O(n^2)$ values is inverted with pre-processing computational cost of $O(n^3)$.

\subsection{Estimated noise parameters of the ibm\_lagos v1.0.32 device}
\label{app-sec:ibmq-noise-model-results}

The experimental results presented in \Sec{sec:ibmq-loss-map}
and \Sec{sec:ibmq-characterization} were obtained from the
\emph{ibm\_lagos} v1.0.32 quantum device on the IBMQ platform on
September 11, 2022. The time scale based on the data from the device
backend was $T_0=1.792\,[\mu sec]$.

In \autoref{table:ibmq-lindblad-params}, we give the estimated
values of the dimensionless noise (coupling strength) parameters
$\{\theta_x, \theta_y, \theta_z, \theta_{amp}, \theta_{ex}\}$, 
which are described in
\App{app-sec:deterministic-channel-noise-details}. The values of
$10^{-5}$ and $0.1$ are the lower and the upper bounds on the
parameters in the numerical optimization, at which the gradient with
respect to the specific parameter is forced to zero as an overall
stability measure.  

\begin{table*}[p]
    \centering
    \setlength{\tabcolsep}{12pt}
    \begin{tabular}{||c | c | c | c | c | c||} 
    \hline
     & qubit 0 & qubit 1 & qubit 2 & qubit 3 & qubit 4 \\ [0.5ex] 
    \hline\hline
    \ $\theta_x$ & $0.1 $ & $10^{-5}$ & $10^{-5}$ & $10^{-5}$ & $10^{-5}$ \\ 
    \ $\theta_y$ & $10^{-5}$ & $0.1$ & $0.1$ & $10^{-5}$ & $10^{-5}$ \\
    \ $\theta_z$ & $10^{-5}$ & $0.1$ & $0.1$ & $10^{-5}$ & $10^{-5}$ \\
    \ $\theta_{amp}$ & $0.0598\pm (0.011)$ & $0.0377\pm (0.019)$ & $0.1\pm (0.0013)$ & $10^{-5}\pm (10^{-4})$ & $0.0749\pm (0.0052)$ \\
    \ $\theta_{ex}$ & $10^{-5}$ & $0.0961\pm (0.017)$ & $0.0705$ & $0.000279\pm (10^{-5})$ & $10^{-5}$ \\ [1ex]
    \hline
    \end{tabular}
    \caption{Estimated values of the dimensionless noise parameters.
    The error values represent one standard deviation and are
    determined using the bootstrapping method outlined in
    Appendix~\ref{app-sec:stat-error-bootstrap}. Entries without an error estimate
    correspond to ``run away'' optimization parameters, which
    consistently saturated the allowed parameter range $[10^{-5},
    10^{-1}]$ in the optimization for all instances in the
    bootstrapping ensemble.} \label{table:ibmq-lindblad-params}
\end{table*}

The estimated values of the $\theta_0,\theta_1$ parameters of the
noisy $\RESET$ gates are given in \autoref{table:ibmq-reset-params}
(see \Eq{eq:numerics-reset-noise} in \Sec{sec:noise-models}). In
the optimization, we used $0.99$ as an upper bound to
$\theta_0,\theta_1$.  We note that for the $\RESET$ gate on the
qubit $4$ the optimization was not performed. This
is because in the definition the $\RESU$ gate (see
\Fig{fig:determinist-example}) the $\RESET$ gate acts on the
first qubit in the pair, and so we did not use $\RESET$ on qubit
$4$.

\begin{table*}[p]
    \centering
    \setlength{\tabcolsep}{12pt}
    \begin{tabular}{||c | c | c | c | c | c||} 
    \hline
     & qubit 0 & qubit 1 & qubit 2 & qubit 3 & qubit 4 \\ [0.5ex] 
    \hline\hline
    $\theta_0$ & $0.926\pm (0.0047)$ & $0.941\pm (0.0071)$ & $0.905\pm (0.0047)$ & $0.99\pm (0.023)$ & n/a \\ 
    $\theta_1$ & $0.99$ & $0.927\pm (0.0059)$ & $0.929\pm (0.0051)$ & $0.895\pm (0.022)$ & n/a \\ [1ex]
    \hline
    \end{tabular}
    \caption{Estimated values of the $\RESET$ gate parameters.
    The error values represent one standard deviation and are
    determined using the bootstrapping method outlined in
    Appendix~\ref{app-sec:stat-error-bootstrap}. Entries without an error estimate
    correspond to ``run away'' optimization parameters, which
    consistently saturated the allowed upper parameter bound of $0.99$ in the optimization for all instances in the
    bootstrapping ensemble. Note that for the qubit 4, the $\RESET$ gate was
    never applied, and therefore, the relevant parameters were not estimated.}
    \label{table:ibmq-reset-params}
\end{table*}

In \autoref{table:ibmq-cx-params}, we give the estimated
values of $\CX$ gate noise parameters, representing the coherent
errors in the two-qubit gate (see \Eq{eq:cx-parameterized} in
\App{app-sec:CX-details}). We note that the $\theta_3$ parameter in the
gates $\CX(0\to 1), \CX(1\to 2), \CX(2\to 3), \CX(3\to 4)$  was not affected by the
optimization (i.e., the gradient with respect to $\theta_3$ vanished for these gates), and remained in its ideal inital value of $1.0$. In a
hindsight, this can be explained by the fact that $\theta_3$ is
equivalent to adding a $\hat{z}$ rotation after $\CX$. In the gates
$\CX(0\to 1), \CX(1\to 2), \CX(2\to 3), \CX(3\to 4)$, this rotation is
followed by a $\RESET$ gate, which is insensitive to phases in the
standard basis, and as a result the $\theta_3$ parameter does not
affect $\RESU$ gate (and therefore it does not affect the steady
state). 

\begin{table*}[p]
    \setlength{\tabcolsep}{18pt}
    \centering
    \begin{tabular}{||c | c | c | c||} 
    \hline
     & $\theta_{1}$ & $\theta_{2}$ & $\theta_{3}$ \\ [0.5ex] 
    \hline\hline
    $\CX(0\to 1)$ & $0.976\pm (0.021)$ & $1.352\pm (0.031)$ & $1.0$ \\
    $\CX(1\to 0)$ & $1.053\pm (0.0022)$ & $1.029\pm (0.0024)$ & $0.587\pm (0.0061)$ \\
    
    $\CX(1\to 2)$ & $0.791\pm (0.0036)$ & $0.711\pm (0.031)$ & $1.0$ \\
    $\CX(2\to 1)$ & $0.987\pm (0.0018)$ & $1.020\pm (0.0021)$ & $0.936\pm (0.0091)$ \\
        
    $\CX(2\to 3)$ & $1.082\pm (0.0077)$ & $0.973\pm (0.027)$ & $1.0$ \\
    $\CX(3\to 2)$ & $0.970\pm (0.0018)$ & $1.011\pm (0.0022)$ & $0.627\pm (0.011)$ \\
    
    $\CX(3\to 4)$ & $0.958\pm (0.014)$ & $0.930\pm (0.0069)$ & $1.0$ \\
    $\CX(4\to 3)$ & $1.062\pm (0.0017)$ & $1.008\pm (0.0018)$ & $1.030\pm (0.022)$ \\ [1ex]
    \hline
    \end{tabular}
    \caption{Estimated values of the $\CX$ gate parameters. The
    notation $\CX(i\to j)$ corresponds to the gate with control
    qubit $i$ and target qubit $j$. The error values,
    representing one standard deviation, are determined using the
    bootstrapping method outlined in Appendix~\ref{app-sec:stat-error-bootstrap}.
    Error values of precisely $0$ for $\theta_3$ parameter in the
    gates $\CX(0\to 1), \CX(1\to 2), \CX(2\to 3), \CX(3\to 4)$ are a
    result of the vanishing gradient, as discussed in
    \App{app-sec:ibmq-noise-model-results}.}
    \label{table:ibmq-cx-params}
\end{table*}

The notable fluctuations in learning
precision across different families of noise parameters, such as
between single qubit depolarization parameters and two-qubit $\CX$
gate parameters, can be attributed to the differences in the
gradients of the cost function $\Phi$ with respect to the parameters
of the $\CX$ and $\RESET$ gates compared to single qubit parameters.
This difference arises from the fact that the quantum circuit
implementing the deterministic map primarily comprises $\CX$ gates
along with $\RESET$ gates. The single qubit parameters come into
play mostly during idle states of the qubit or during the action of
single qubit rotation gates. As a result, the influence of single
qubit parameters on the cost function is smaller, leading to lower
learning accuracy for these parameters in presence of the
statistical noise.

{~}

\section{Bootstrapping method and statistical error analysis via error propagation} 
\label{app-sec:stat-error}

\subsection{Bootstrapping method assuming independent sampling}
\label{app-sec:stat-error-bootstrap}

To estimate the statistical errors in the cost function
$\Phi$ and the fitted parameters $\vtheta$ we applied a simple bootstrapping
method\cc{ref:James2013-Stat-intro}. Given a set of Pauli
expectation values obtained from the experiment using $N$ statistical shots, we augmented each expectation value by
adding a realization of a Gaussian random variable with zero mean
and variance of $1/N$. Subsequently, our optimization routine was
executed using these ``noisy" Pauli expectation values as input. All
other hyperparameters, such as the optimization step size, number of
iterations, and stopping conditions, remained consistent.

By employing this approach, the error bars on the fitted parameters, as illustrated in \autoref{table:ibmq-lindblad-params}, \autoref{table:ibmq-reset-params}, and \autoref{table:ibmq-cx-params}, were determined as one standard deviation of the fitted parameters acquired from 150 separate realizations of Gaussian noise applied to the expectation values obtained from the experiment. Consequently, the error in the cost function $\Phi$, presented in \Fig{fig:ibmq-heatmap}, for a fixed set of parameters $\vtheta$, was likewise calculated as one standard deviation of the cost function values derived from 150 distinct realizations of Gaussian noise affecting the expectation values obtained from the experiment.

The primary assumption underlying this method hinges on the notion that each Pauli expectation value was independently sampled. In practice, however, measurements were conducted utilizing the overlapping local tomography method (as elucidated in Section \ref{sec:ibmq-measurement-protocol}), wherein all geometrically $k$-local Pauli strings are simultaneously estimated using a set of $3^k$ distinct measurement circuits. Consequently, while considering Pauli expectation values as random variables, it's important to acknowledge that these variables may indeed exhibit correlations. Nonetheless, conducting a comprehensive analysis of such correlations would necessitate measuring $k+1$-local Pauli strings, a task that would substantially increase the execution time on the IBMQ machine.

Below, we provide an analytical description of the errors in both the cost function and the parameters, assuming independent sampling. Additionally, we explore a scenario where independent sampling is not assumed, leading to an upper bound on the error bars associated with the estimated parameters.

\subsection{Statistical error analysis}
\label{app-sec:stat-error-analysis}

The errors in cost function $\Phi$ can be estimated analytically using simple error
propagation analysis. From the functional form of the cost function
in \Eq{eq:cost-Pauli-Kraus}, we get that for a fixed set of
parameters $\vtheta$, the error in $\Phi$ due to statistical errors
in the estimation of $\av{P_i}$ is given by
\begin{align}
  \Delta\Phi = \left(\sum_j
  \left(\frac{\partial\Phi(\vtheta)}{\partial P_j}\right)^2
    \cdot(\Delta P_j)^2\right)^{1/2} ,
\end{align}
where we assumed that the Pauli operators were sampled independently.

To derive the analytical expression for the errors in the fitted parameters, recall from \Eq{eq:cost-Pauli-Kraus} that the cost function is given
by 
\begin{align*}
\label{}
  \Phi(\vtheta) = \frac{1}{C_S} 
    \sum_i \Big( \sum_j C_{ij}(\vtheta)\av{P_j}_\infty\Big)^2 .
\end{align*}

In order not to overload the notation, we shall drop the $\infty$
subscript from $\av{P_j}_\infty$, and in addition we will use
$\av{P}$ to denote the vector of all $\av{P_j}$ expectation values.

Let then $\av{P_j}$ denote the exact expectation value of the Pauli
operator, and by $\bP_j$ the empirical expectation value we obtain
using $N$ independent shots. Then we may write $\bP_j\EqDef \av{P_j} 
+ \delta P_j$, where $\delta P_j$ can be modeled by a Gaussian
variable with zero mean and standard deviation $\Delta P_j =
\sqrt{\av{(\delta P_j)^2}} = \sqrt{(1-\av{P_j}^2)/N}\le \frac{1}{\sqrt{N}}$. During the
optimization step, we find $\vtheta$ that minimizes $\Phi$ at
the set of empirical expectation values $\{\av{\bP_j}\}$. This is of
course a random variable by itself, which fluctuates around the
exact minima that we would have obtained had we minimized $\Phi$ at
$\{\av{P_j}\}$. Let $\vtheta_0$ denote the exact minima, then we
define the random variable $\delta\vtheta$ by 
$\vtheta \EqDef \vtheta_0 + \delta\vtheta$. Finally, define $J_\alpha$
to be the gradient of the cost function: $J_\alpha = \frac{\partial
\Phi}{\partial \theta_\alpha}$. Then
\begin{align}
  0 &= J_\alpha(\vtheta;\ \bP) =
    J_\alpha(\vtheta_0 + \delta\vtheta;\  \av{P} + \delta P)\\
  &\simeq  J_\alpha(\vtheta_0;\ \av{P})
   + \sum_\beta \frac{\partial J_\alpha}{\partial \theta_\beta}
   \delta \theta_\beta +
   \sum_j \frac{\partial J_\alpha}{\partial \av{P_j}}
   \delta P_j .
\end{align}
By definition, also $J_\alpha(\vtheta_0;\  \av{P})=0$, since
it is the minimum of the cost function at the exact expectation
values, and therefore we get a linear equation for $\delta\theta$,
whose solution is
\begin{align*}
  \delta\theta_\beta = \sum_j M_{\beta j}\cdot \delta P_j, 
\end{align*}
where the matrix $M$ is given by
\begin{align*}
  M_{\beta j} = -\sum_\alpha (H^{-1})_{\beta \alpha}\cdot
    \frac{\partial J_\alpha}{\partial \av{P_j}},
\end{align*}
and $H$ is the Hessian of the cost function,
\begin{align*}
  H_{\beta\alpha} \EqDef \frac{\partial^2 \Phi}
    {\partial \theta_\alpha\partial\theta_\beta} .
\end{align*}
The matrix $M$ is evaluated at $(\vtheta_0, \av{P})$, and is therefore
not a random variable. Taking the $\delta P_j$ to be independent
Gaussian variables with variance $(\Delta P_j)^2$, we may estimate
\begin{align}
\label{eq:theta-err}
  \Delta \theta_\beta = 
    \left(\sum_j |M_{\beta j}|^2\cdot (\Delta P_j)^2 \right)^{1/2} .
\end{align}
In practice, we do not know $M$ exactly at $(\vtheta_0, \av{P})$,
but as a zero order approximation, we can use its value at the
empirical values $(\vtheta, \bP)$.

We note that the bound in \Eq{eq:theta-err} was derived by assuming
that $\{\delta P_j\}$ are independenet Gaussian variables, which is
justified when using independent samples to estimate different
expectation values. When this is not the case, one can use the
weaker bound
\begin{align}
\label{eq:theta-err-weak}
  \Delta \theta_\beta \le \frac{1}{\sqrt{N}}\cdot 
    \sum_j |M_{\beta j}|, 
\end{align}
which does not assume statistical independence, and in fact provides the worst-case bound on the error which can be attributed to the case where all Pauli expectation values are correlated.

During the analysis of the error, we computed the error bars for the fitted parameters using two approaches: the bootstrapping method assuming independent sampling (refer to Appendix~\ref{app-sec:stat-error-bootstrap}) and the upper bound derived from \Eq{eq:theta-err-weak}. However, we find that the calculated upper bound offers limited insight into the actual errors in the estimation due to its substantial magnitude. To illustrate this point, we compare the errors obtained from the bootstrapping method described in Appendix~\ref{app-sec:stat-error-bootstrap} with those derived from Equation \Eq{eq:theta-err-weak} for the $\CX(i\rightarrow j)$ gates positioned in the middle of the five-qubit system utilized in our experiment.

This comparison of error estimates is presented in \autoref{table:ibmq-cx-error-comparison}. Notably, the bound provided by Equation \Eq{eq:theta-err-weak} greatly exceeds the estimate generated by our bootstrapping method. However, if the actual errors in the fitted parameters were indeed close to the worst-case bound, our method would likely fail to accurately reproduce the dynamics of the learned quantum channel. In such a scenario, we would not observe a satisfactory agreement between the estimated and experimentally obtained $2$-local Pauli expectation values, as demonstrated in \autoref{sec:ibmq-characterization}, particularly in \autoref{fig:ibmq-two-local-exp-vals}. Therefore, our findings suggest that the independence assumption is more likely to approximate the actual error propagation scenario than the upper bound.

\begin{table*}[b]
\centering
\setlength{\tabcolsep}{6pt}
\begin{tabular}{||c|ll|ll|ll||}
\hline
\multirow{2}{*}{} & \multicolumn{2}{c|}{$\Delta\theta_1$}      & \multicolumn{2}{c|}{$\Delta\theta_2$}       & \multicolumn{2}{c|}{$\Delta\theta_3$}        \\ \cline{2-7} 
                  & \multicolumn{1}{l|}{Bootstrapping}  & Worst-case  & \multicolumn{1}{l|}{Bootstrapping}  & Worst-case   & \multicolumn{1}{l|}{Bootstrapping}   & Worst-case   \\ \hline
$\CX(1\to 2)$     & \multicolumn{1}{c|}{0.0036} & 0.2375 & \multicolumn{1}{c|}{0.0310} & 0.5696  & \multicolumn{1}{c|}{n/a}  & n/a  \\ \hline
$\CX(2\to 1)$     & \multicolumn{1}{c|}{0.0018} & 0.0244 & \multicolumn{1}{c|}{0.0021} & 0.0311 & \multicolumn{1}{c|}{0.0091} & 0.0505 \\ \hline
$\CX(2\to 3)$     & \multicolumn{1}{c|}{0.0077} & 0.2663 & \multicolumn{1}{c|}{0.0271} & 1.0973 & \multicolumn{1}{c|}{n/a} & n/a \\ \hline
$\CX(3\to 2)$     & \multicolumn{1}{c|}{0.0018} & 0.0159 & \multicolumn{1}{c|}{0.0022} & 0.0564 & \multicolumn{1}{c|}{0.0112} & 0.2311 \\ \hline
\end{tabular}
\caption{Comparison between the error estimates $\Delta\theta_i$ in the fitted parameters $\theta_i$ for the $\CX(i\to j)$ gates located in the middle of the five-qubit system used in the experiment. The bootstrapping column is taken from \autoref{table:ibmq-cx-params}, and corresponds to the error analysis given in Appendix~\ref{app-sec:stat-error-bootstrap}. The worst-case column is the error bound obtained from \Eq{eq:theta-err-weak}.}
\label{table:ibmq-cx-error-comparison}
\end{table*}

\end{document}